\documentclass[fleqn,usenatbib]{mnras}
\usepackage[T1]{fontenc}
\usepackage{ae,aecompl}
\usepackage{hyperref}
\usepackage{graphicx}	
\usepackage{amsmath}	
\usepackage{amssymb}	
\usepackage{rotating}
\usepackage{txfonts}
\usepackage{rotating}
\usepackage{accents}
\usepackage{float}
\usepackage{epsfig}
\usepackage{textcomp}
\usepackage{upgreek}
\usepackage{epstopdf}
\usepackage{color}
\usepackage{pdflscape}
\usepackage{}
\usepackage{tabularx}
\newcommand{\orcid}[1]{\href{https://orcid.org/#1}{\includegraphics[width=10pt]{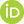}}}
\begin{document}
\title[{MEGASTAR DR1 late-type stars: stellar parameters}]{MEGASTAR (III). Stellar parameters and data products for DR1 late-type stars}

\author[Moll\'{a} et al. ] 
{M. Moll\'{a}$^{1}$\thanks{E-mail:mercedes.molla@ciemat.es}\orcid{0000-0003-0817-581X},
M.L. Garc{\'\i}a-Vargas$^{2}$\orcid{0000-0002-2058-3528}, 
I. Mill{\'a}n-Irigoyen$^{1}$\orcid{0000-0003-4115-5140}, 
N. Cardiel$^{3, 4}$\orcid{0000-0002-9334-2979},
E. Carrasco$^{5}$\orcid{0000-0002-9174-5491},
\newauthor
A. Gil de Paz$^{3, 4}$\orcid{0000-0001-6150-2854}, 
S. R. Berlanas$^{6,7}$\orcid{0000-0002-2613-8564},
and P. G{\'o}mez-{\'A}lvarez$^{2}$\orcid{0000-0002-8594-5358}
\\
$^{1}$ Dpto. de Investigaci\'{o}n B\'{a}sica, CIEMAT, Avda. Complutense 40, E-28040 Madrid, Spain\\
$^{2}$ FRACTAL S.L.N.E., Calle Tulip{\'a}n 2, portal 13, 1A, E-28231 Las Rozas de Madrid, Spain \\
$^{3}$ Dpto. de F{\'\i}sica de la Tierra y Astrof{\'\i}sica, Fac. CC. F{\'\i}sicas, Universidad Complutense de Madrid, Plaza de las Ciencias, 1, E-28040 Madrid, Spain \\
$^{4}$ Instituto de F{\'\i}sica de Part{\'\i}culas y del Cosmos, IPARCOS, Fac. CC. F{\'\i}sicas, Universidad Complutense de Madrid, Plaza de las Ciencias 1, E-28040 Madrid, Spain\\
$^{5}$ Instituto Nacional de Astrof{\'\i}sica, {\'O}ptica y Electr{\'o}nica, INAOE, Calle Luis Enrique Erro 1, C.P. 72840 Santa Mar{\'\i}a Tonantzintla, Puebla, Mexico\\
$^{6}$ Departamento de Física Aplicada, Universidad de Alicante, 03690 San Vicente del Raspeig, Alicante, Spain\\
$^{7}$ Astrophysics Group, Keele University, Keele ST5 5BG, Staffordshire,  UK
}

\date{Accepted Received ; in original form}
\pagerange{\pageref{firstpage}--\pageref{lastpage}} \pubyear{}

\maketitle
\label{firstpage}

\begin{abstract}
MEGARA is the optical integral field and multi-object spectrograph at the Gran Telescopio Canarias. We have created MEGASTAR, an empirical library of stellar spectra obtained using MEGARA at high resolution $R=20\,000$ (FWHM), available in two wavelength ranges: one centered in H${\alpha}$, from 6420 to 6790\,\AA\  and the other centered in the \ion{Ca}{ii} triplet, from 8370 to 8885\,\AA\ (\mbox{HR-R} and \mbox{HR-I} VPH-grating configurations). In this work, we use MEGASTAR spectra, combination of these two short wavelength intervals, to estimate the stellar parameters namely effective temperature, surface gravity and metallicity (and their associated errors) for a sample of 351 MEGASTAR members with spectral types earlier than B2. We have applied a $\chi^2$ technique by comparing MEGASTAR data to theoretical stellar models. For those stars with stellar parameters derived in the literature, we have obtained a good agreement between those published parameters and ours.  Besides the stellar parameters, we also provide several products like the rectified spectra, radial velocities and stellar indices for this sample of stars.  In a near future, we will use MEGASTAR spectra and their derived stellar parameters to compute stellar population evolutionary synthesis models, which will contribute to a better interpretation of star clusters and galaxies spectra obtained with MEGARA.
\end{abstract}


\section{Introduction}
\label{Introduction}

MEGARA (Multi Espectr{\'o}grafo en GTC de Alta Resoluci{\'o}n para Astronom{\'\i}a) is the optical integral-field and multi-object fibre fed spectrograph for the Gran Telescopio Canarias (GTC)\footnote{\url{http://www.gtc.iac.es}}, the 10.4m telescope located in La Palma, Canary Islands, Spain. The instrument offers 18 spectral configurations: six in low resolutions (LR) of 6,000, ten in medium resolution (MR) of 12,000, both covering the complete visible wavelength interval, and two in high resolution (HR) of 20,000, one centered in H${\alpha}$ (HR-R) and the other in the \ion{Ca}{ii} triplet, CaT, (HR-I). The wavelength intervals of  these high resolution configurations are  6420\ -- 6790\,\AA\ and  8370\ -- 8885\,\AA, respectively. The 18 spectral configurations are achieved via volume phase holographic gratings (VPHs) placed at the pupil of the spectrograph the collimated beam. For a detailed description of the instrument and its scientific validation see \citet{carspie18}, \citet{gilspie18}, \citet{dullo19} and \citet{gilAAS22}.

Population Synthesis models have proven to be crucial for the interpretation of galaxy spectra in terms of combination of Simple Stellar Populations (SSPs) or building blocks that give their star formation histories. There have been in the past many studies devoted to the computation of integrated properties or SSPs SEDs \citep{CER1994,Fioc_Rocca_Volmerange1997,lei99,bc03,Gonzalez+2005,Maraston2005, fritze06, Coelho+2007, elst09, Conroy_Gunn_White2009, Maraston+2009, Conroy_VanDokkum2012, Maraston_Stromback2011, lei14, vazdekis2015, vaz16, Fioc_Rocca_Volmerange2019, Maraston+2020, Coelho+2020}, including our own {\sc PopStar} model \citep[][] {popstar} and recently its update for high resolution, {\sc HR-pyPopStar} \citep[][hereinafter MI21]{Millan2021}. Important differences between these SSP models arise from the use of different stellar tracks (so different isochrones), stellar libraries, spectral coverage, inclusion or not of nebular emission, different input physics or even computational algorithms. In particular, the spectral resolution of the models, coming from the stellar libraries, most affects the SSPs.

Ideally, a set of high-resolution synthetic Spectral Energy Distributions (SEDs) is the best tool to interpret data from the MEGARA HR set-ups. At the time of starting MEGASTAR, there was not other theoretical or empirical library suitable for the characteristics of MEGARA high resolution configurations. In MI21, the authors developed a new version of the classical {\sc PopStar} model, by incorporating stellar atmosphere models with high spectral resolution\footnote{The grid of {\sc HR-pyPopStar} models can be found in \url{https://www.fractal-es.com/PopStar} and in \url{https://cdsarc.cds.unistra.fr/viz-bin/cat/J/MNRAS/506/4781}.}, is ideal for interpreting MEGARA high resolution observations as it  provides a good coverage of the stellar parameters. However, the contributions responsible for the spectral lines intrinsic broadening are still lacking. The idea of creating the library was precisely to use MEGASTAR spectra as input to this evolutionary synthesis code for the correct interpretation of data obtained with MEGARA high resolution set-ups. 

This work is the third of a series of papers. In \citet[][hereinafter, Paper I]{gvargas2020} the library rationale is presented,  the almost 3000 targets selection criteria are described, and a pilot observations program carried out during MEGARA commissioning and the data reduction pipeline are also explained. Additionally, we show main features of the public and accessible data base developed to manage the library.

In \citet[][hereinafter, Paper II]{carrasco2021}, the first data release (DR1) of the MEGASTAR library is presented. DR1 is composed by 414 stars and 838 spectra\footnote{These 838 reduced and calibrated spectra are available at the page web of the MEGASTAR project \url{https://www.fractal-es.com/megaragtc-stellarlibrary}, username: {\it{public}}, password: {\it{Q50ybAZm}}.} observed in both \mbox{HR-R} and HR-I set-ups through a GTC filler program awarded with 250\,h of observing time during three semesters (2018B, 219A and 2019B). Since then, the number of observed stars has increased with data obtained in subsequent semesters as the work is in progress. 

In this new piece of work, we present the determination of the stellar parameters: effective temperature, $T_{\mathrm{eff}}$, , surface gravity, $\log{g}$, and metallicity, [M/H], for the MEGASTAR DR1 sample, excluding the hottest stars with spectral types earlier than B2. Our aim is to obtain the parameters with the same method for all the stars of the library with these spectral types. To validate the method, we compare our results to those from the literature whenever available. This is particularly important as we are using two very narrow spectral windows given by the wavelength intervals of the \mbox{HR-R} and \mbox{HR-I} spectral configurations.  
We will also provide to the community with some relevant scientific data products: stellar radial velocities, rectified spectra and measurements of some spectral absorption lines.

For estimating the stellar parameters of FGKM type stars, there are two broadly accepted techniques. The first method is based on Equivalent Widths (EW) measurements. The strengths of some spectral lines, in general \ion{Fe}{i} and \ion{Fe}{ii}, are calculated by taking into account the balance between ionization and excitation and, from the measurement of their EWs, the stellar parameters are inferred. The second technique is based on the comparison of the observed and synthetic spectra in a given spectral window, assuming the stellar parameters from the best fitting model. Moreover, some  authors combine both methods \citep[e.g.][for iSpec or FASMA]{blanco-cuaresma2014,andreasen2017}. There are numerous works in the literature devoted to these methods.  \citet{tabernero2022}, review thoroughly the existing methodologies and include a list of references dedicated to the stellar parameters determination in recent stellar large surveys (APOGEE, GALAH, LAMOST, LEGUE, RAVE, SEGUE, GAIA, WEAVE, 4MOST and MANGA, among others). We highlight that each project uses the tool or technique most adequate to the characteristics of their observational data. Thus, these methods differ in wavelength range, spectral resolution, theoretical models used, numerical methodology and so on. As an example, \citet{tabernero2022} present the code {\sc SteParSyn}, which is only applicable to FGKM stars and, therefore, limited to $T_{\mathrm{eff}}$,  in the range [3500 -- 7000]\,K. The authors apply a $\chi^{2}$-technique plus a Montecarlo  (MCMC {\tt emcee}) code based on the MARCS atmosphere models from \citet{gustafsson2008}. They compute synthetic spectra in regions around 261 \ion{Fe}{i} and 31 \ion{Fe}{ii} lines, with spectral windows of 3\,\AA, while in their previous version of the code, {\sc StePar} \citep{tabernero2019}, Kurucz models and the CaT spectral range (8400 -- 8800\,\AA) were used for AFGKM stars. 

Other authors take advantage of all the information available of the stars. For example,  \citet{gent2022} beside  the spectroscopic information  use parallaxes,  photometry and even asterosysmology constraints to  build their own code based on Bayesian inference to determine the stellar parameters that will be used in the PLATO mission. These authors have tested the code with 19 FGK stars, 2 GK sub-giants and 2 red giants with high-resolution, high signal-to-noise (above 300) data in the 5520 -- 5600\,\AA\ wavelength range.  In \citet{imig2022}, with spectral resolution R $\sim$ 1800, obtain the stellar parameters for the library of the SDSS project, Ma-Star. The model is based on a neural network, which in turn uses models from \citet{allende-prieto2018}. The objective is to apply this technique to 59266 stars, the prototype being used for stars in common with the APOGEE survey (which uses MARCS models and the FERRE code). They determine $T_{\mathrm{eff}}$,  and $\log{g}$ from a $\chi^{2}$ technique applied to stellar models, and then vary the elemental abundances until finding the best fit to the spectra. 

Our scientific objective is to derive the stellar parameters in a similar way to the one presented by \citet{tabernero2022} in their Fig$.$\,1 (first part of the diagram). Our wavelength windows are, however, wider than the one (3\,\AA) used by these authors, being similar to the one from {\sc SterPar} for the CaT window. Our method will be applied to the stellar spectra obtained with R$\sim 20\,000$ in the two wavelength ranges 6420--6790\,\AA\ and 8370--8885\,\AA.

We point out that the methodology for deriving stellar parameters demands very different tools and techniques depending on the effective temperature of the stars. Modelling the atmospheres of hot stars is very different from standard techniques applied to cool stars \citep[see e.g.][]{Langer1997,kudri2006}, since it is necessary to consider phenomena such as rotational velocity or stellar winds. Therefore, we have divided the analysis of the full sample of MEGASTAR DR1 into two different works: 1) this paper for the coolest stars,  spectral types from B2 to S, with a limiting upper $T_{\mathrm{eff}}$ of 25\,000\,K;  and 2) a forthcoming paper (Berlanas et al. in preparation) for the hottest stars of the atlas. 

This paper is organised as follows: the spectral sample, the rectification process and a description of the adopted theoretical stellar models are in Section~\ref{spectra}. The derived stellar parameters are given in Section~\ref{parest}, explaining also the calculation process and comparing our results with those from the literature. The summary and conclusions are given in Section~\ref{conclusions}. 
Additionally, in Appendix~\ref{indices} we include an initial compilation of relevant line-strength measurements.  In Appendix B (Supplementary Material), we include the MEGASTAR spectra of this sample of stars  with their theoretical model fits.  Finally, in Appendix C (Supplementary Material), we describe a first prototype of the evolutionary synthesis code {\sc MegaPopStar} that uses this sample spectra as input and  provide some resulting spectra for solar metallicity and ages older than 10\,Myr, as an example.  Appendices B  and C will be given as supplementary material.

\begin{figure*}
\centering
\includegraphics[width=\textwidth,angle=0]{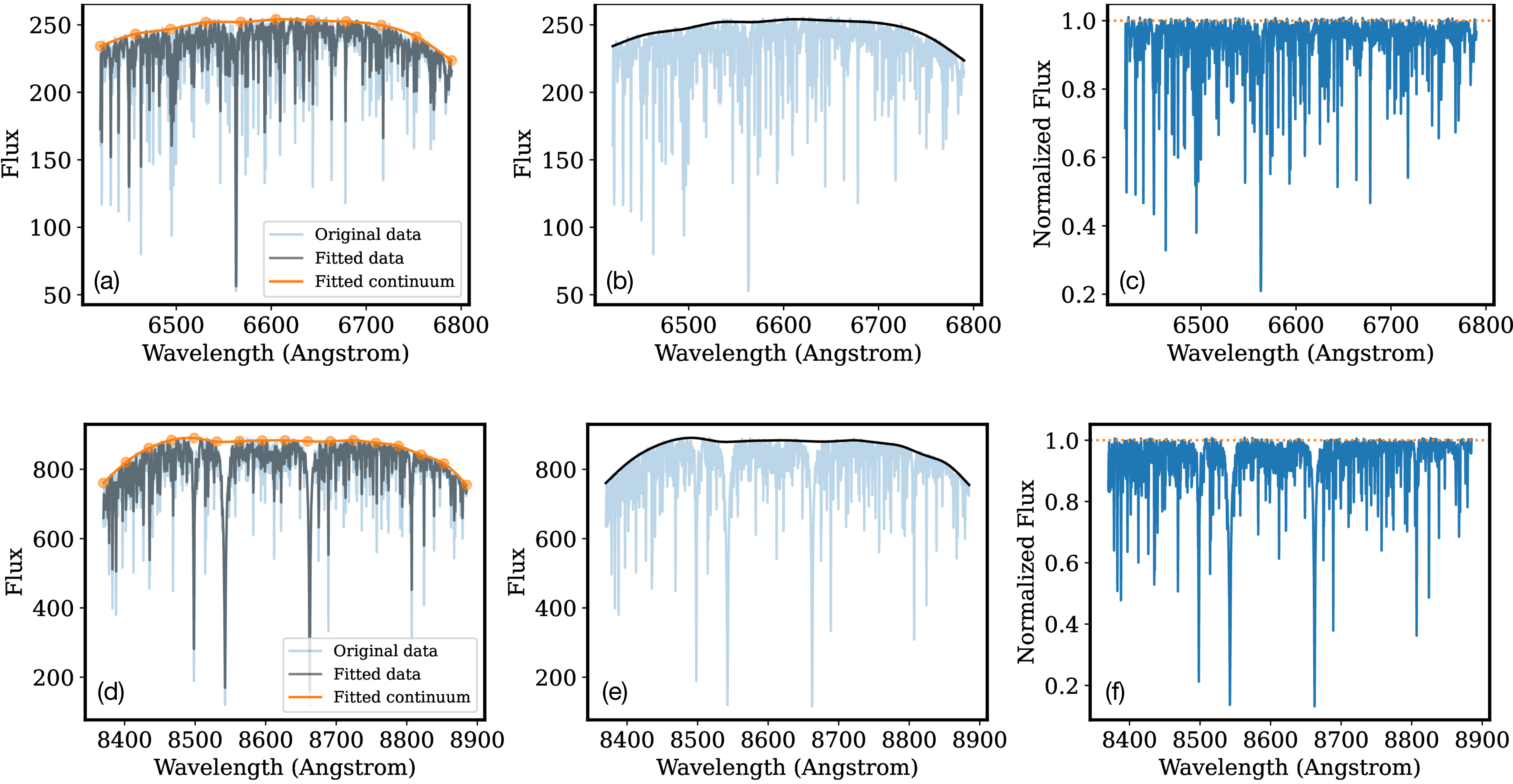}
\caption{Example of boundary fits employed to automatically rectify MEGASTAR spectra for the star HD~115136 (see subsection~\ref{rectify} for details). Each row represents the procedure followed with the spectrum corresponding to HR-R [panels~(a), (b) and~(c)] and HR-I [panels~(d), (e) and~(f)] spectra. The first column [panels~(a) and~(d)] shows the original spectra prior to the fitting procedure (light blue line), the result of applying a median filter of 5 pixels (dark grey line), and the spline fit to the median-filtered data (orange line, with knots displayed as filled dots). The central column [(b) and~(e)] simply displays the final fit (black line) over the original spectra (light blue line) to better assess the result. The last column [(c) and~(f)] shows the rectified spectra obtained in each case.} 
\label{boundary-fit}
\end{figure*}

\section{Stellar spectra}
\label{spectra}

\subsection{Rectified spectra}
\label{rectify}

As a preliminary step to facilitate the estimation of the atmospheric stellar parameters, we have performed an individual continuum rectification\footnote{We note that rectifying a spectrum is not the same as normalising it. By normalising, the whole spectrum is divided by the value at a selected point or wavelength, still maintaining its shape, while to rectify a spectrum implies to divide it by its continuum, obtained by a polynomial fit or any other method \citep[see][for a detailed explanation and figures]{gray}.} to the 351 DR1 stellar spectra\footnote{These spectra correspond to 349 different stars, since two of them have been observed twice.}, with spectral types range from B2 to S, for both wavelength ranges, \mbox{HR-R} and \mbox{HR-I}. 

We have applied our method of rectification to both the observed MEGASTAR spectra and to the theoretical models used in this work (see next section). Although the method was briefly described in Paper~I, we are providing a more detailed explanation, including some auxiliary plots to illustrate the procedure.

The rectification of any spectrum is typically based on the identification of spectral regions free from absorption spectral lines. Although a simple polynomial fit to the flux in those selected spectral windows provides the normalising continuum level, some problems arise: 1)~the result unavoidably depends on the location of the continuum regions (which is something that is commonly performed interactively through the visual examination of the spectrum); and 2)~the continuum level is a fit that, by definition, leaves points at both sides of the fitted data. In order to avoid these problems and to automatise as much as possible the procedure, we use the generalised least-squares method described in \citet{cardiel09}, which automatically provides boundary functions (simple polynomials or splines) for arbitrary data sets. The computation of these fits has been performed by using the public software {\sc boundfit}\footnote{\url{https://boundfit.readthedocs.io/en/latest/}}, which is based on the asymmetric treatment of the data at both sides of the fitted function: after computing an initial ordinary least-squares fit, an asymmetric weight is assigned to the data points above and below the computed fit. By iterating the fitting procedure, the result shifts towards the upper or the lower boundary of the data set. 

In our case, the upper boundary provides an automatic determination of the continuum of any spectrum. It is important to highlight that this method does not require the identification of regions free from absorption features, since the data points corresponding to those wavelengths automatically receive a much lower weight in the fit as the iterative process progresses. In addition, the upper boundary tends to leave all the fitted data below the boundary itself, which helps to obtain a more reliable continuum estimate when the spectra exhibit a high Signal to Noise Ratio, SNR, and fainter absorption features can bias the fitted continuum level. For each stellar spectrum, corresponding to either the MEGASTAR spectra or theoretical models, we have visually inspected all the resulting fits, performed using splines with a number of knots adapted to the spectral type. In the case of the MEGASTAR spectra, a median filter of a few pixels has been applied to each spectrum before the fitting process in order to minimise the bias that the data random noise may introduce in this kind of asymmetric fit.  We display an example of the method applied to the star HD\,115136 in Fig.~\ref{boundary-fit}.

\begin{figure*}
\centering
\includegraphics[width=0.35\textwidth,angle=-90]{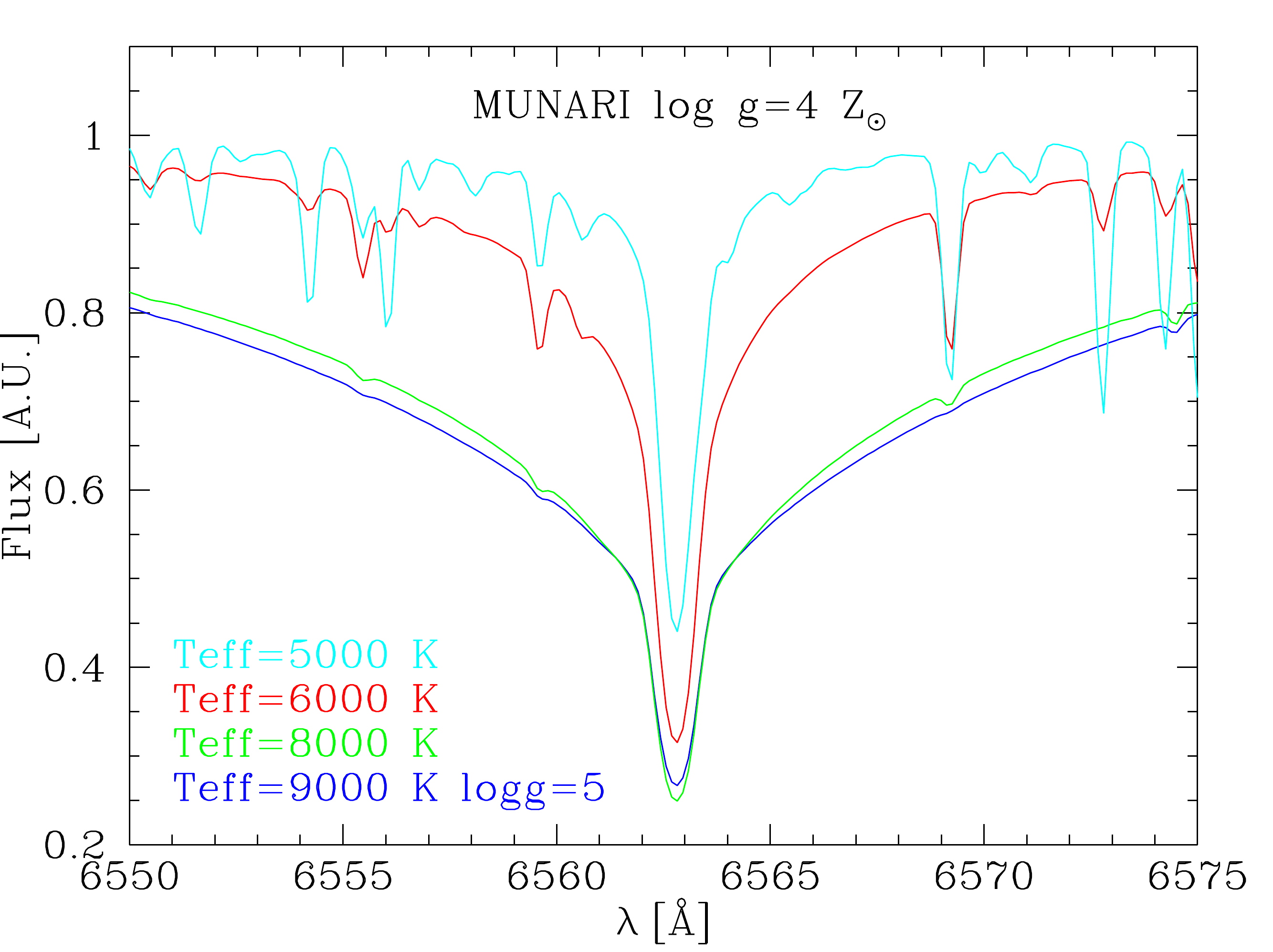}
\includegraphics[width=0.35\textwidth,angle=-90]{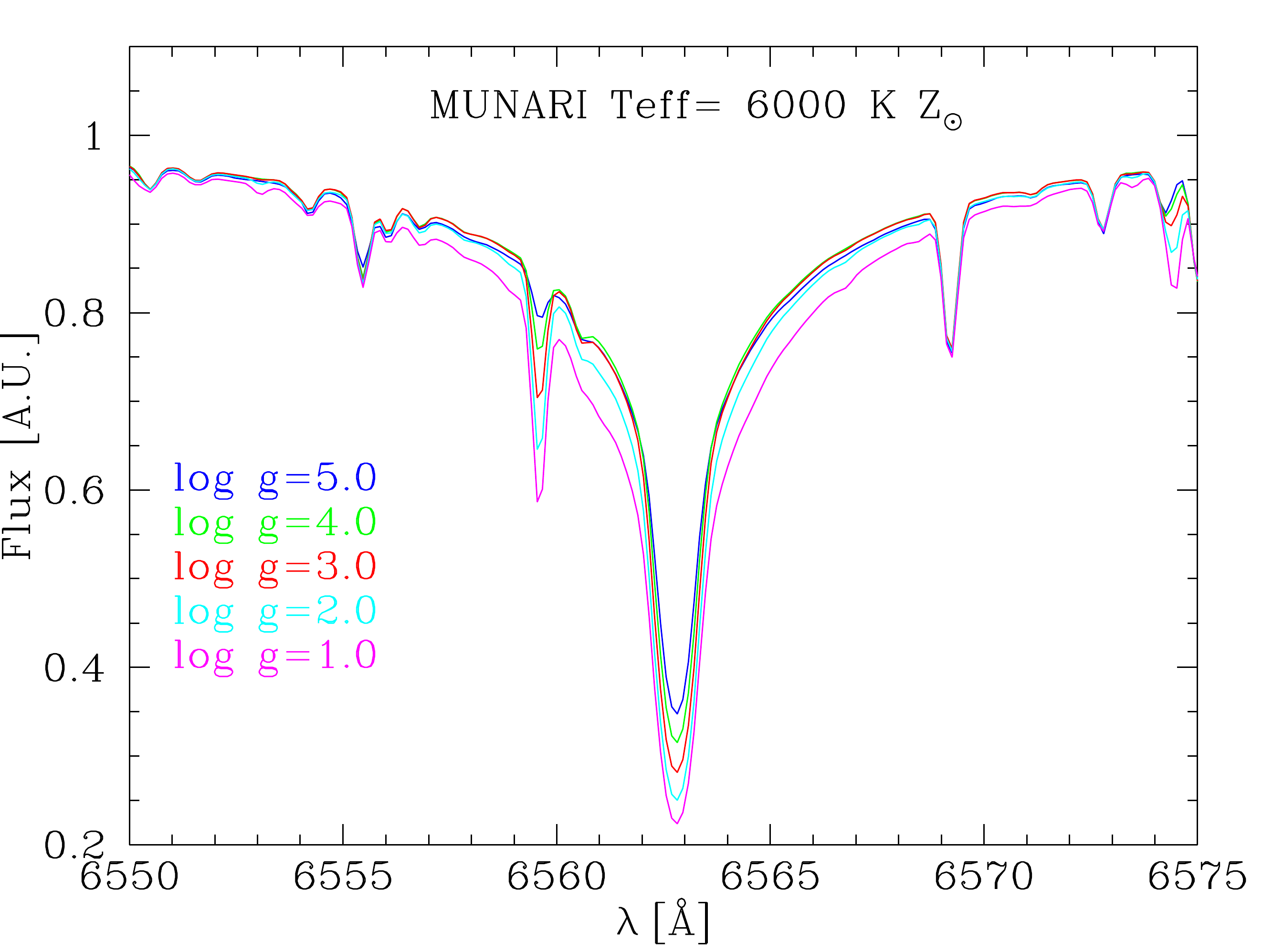}
\caption{Some MUN05 solar metallicity stellar models for cool stars, rectified as explained in sub-section~\ref{rectify}. Left panel: comparison of spectra with different effective temperature and a constant surface gravity $\log{g} = 4$ (except for the case T$_\mathrm{eff}$\,=\,9\,000\,$\mathrm{K}$, where $\log{g} = 5$ is used because $\log{g} = 4$ model is not available). Right panel: comparison of spectra with different values of surface gravity, and a constant effective temperature of 6\,000\,$\mathrm{K}$.}
\label{stellar-models}
\end{figure*}

\subsection{Theoretical models for cool stars}
\label{cool-models}

We have compared our observed spectra with the stellar models by \citet[][hereinafter MUN05]{munari05}.
MUN05 presented a library of synthetic spectra based on Kurucz's code that covers the range from 2\,500 to 10\,500\,\AA. These models are available for different combinations of stellar parameters, in particular, effective temperature in the range  \mbox{$500\le T_{\mathrm{eff}} \le 47\,500$\,K}, (with steps of \mbox{250\,K} up to \mbox{10\,000\,K} and progressively larger step for hotter stars), stellar surface gravity in the range \mbox{$0.0\le \log{g}\le 5.0$}, metallicity in the interval \mbox{$-2.5\le[\mathrm{M/H}]\le 0.5$}, $\alpha$-enhancement abundance for \mbox{$[\alpha/\mathrm{Fe}]=0.0$ and $+0.4$}, three values of micro-turbulence velocity, \mbox{$\xi=1$, 2, and\,4\,km\,s$^{-1}$}, and stellar rotation in the range \mbox{$0\le V_{\mathrm{rot}}\le 500$\,km\,s$^{-1}$}. We have selected models with \mbox{$\mathrm [\alpha/\mathrm{Fe}]=0.0$}\footnote{We have used the \mbox{$\mathrm [\alpha/\mathrm{Fe}]=0.0$} models as the stars in our sample have low [$\alpha$/Fe] values. We have identified in our sample 67 cool stars with \mbox{$T_{\mathrm{eff}} \le 7\,000$\,K} for which GAIA data give \mbox{$\mathrm [\alpha/\mathrm{Fe}] > +0.2$}\,dex. We have checked that their stellar parameters either using \mbox{$\mathrm [\alpha/\mathrm{Fe}]=+0.0$} or \mbox{$\mathrm [\alpha/\mathrm{Fe}]=+0.4$} models are similar, the $\chi^{2}$ being worse when using this set  than obtained with \mbox{$\mathrm [\alpha/\mathrm{Fe}]=0.0$}}, \mbox{$\xi\,=\,2$\,km\,s$^{-1}$} and \mbox{$V_{\mathrm{rot}}\,=\,0$\,km\,s$^{-1}$}. As we have excluded the hottest stars (see Section~\ref{Introduction}), we consider models with \mbox{$T_{\mathrm{eff}} \le 25\,000$\,K}.

MUN05 were originally computed at several resolving powers to simulate different survey data: R$=11\,500$ and 20\,000, for GAIA \citep{Gaia18}; R$=8\,500$ for RAVE \citep{bundy15} and R$=2\,000$ for MaNGA SDSS \citep{steinmetz06}, using an uniform dispersion of 1 and \mbox{10\,\AA\,pix$^{-1}$}. For this work, we have used those with R\,=\,20\,000, appropiate for MEGASTAR. MUN05 spectra are given either with flux units or normalized. We have employed the former format to perform the same rectification method (see subsection~\ref{rectify}) for both MUN05 and MEGASTAR spectra.

Fig.~\ref{stellar-models} shows some examples of the rectified models at solar abundance (Z=$Z_{\sun}$) around H${\alpha}$ line. As expected for cool stars, $T_{\mathrm{eff}}$ variations have an important effect on the amplitude of the line wings, while oscillations in $\log{g}$ mainly change the line depth.

\subsection{Comparison of observed and modelled spectra: radial velocities}
\label{spectra-models}

\begin{figure}
\centering
\includegraphics[width=0.48\textwidth, angle=0]{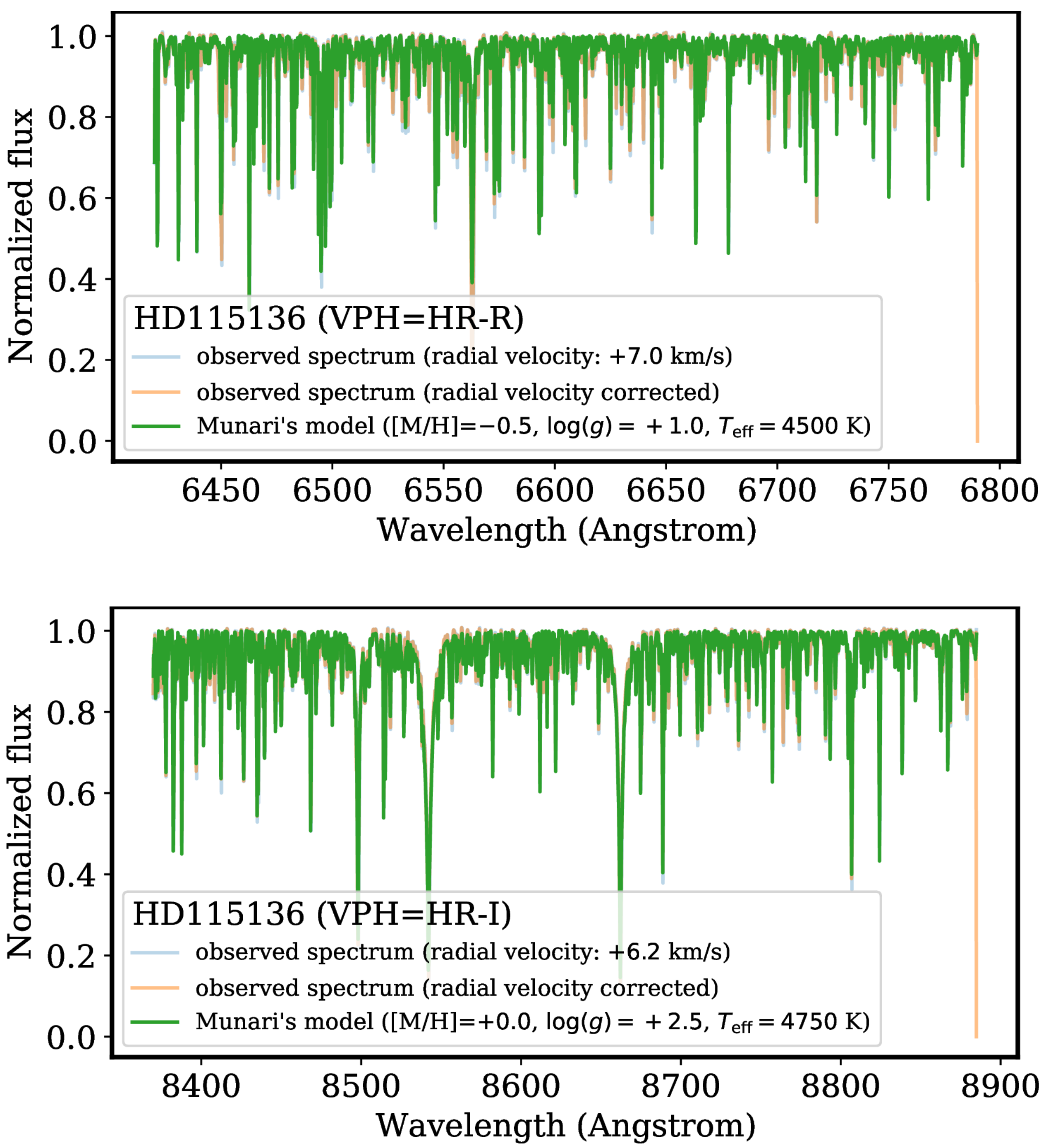}
\caption{Example of radial velocity estimate using the rectified spectra of the star HD~115136 obtained as explained in Fig.~\ref{boundary-fit}. Top panel shows \mbox{HR-R} spectrum while bottom panel displays \mbox{HR-I} range. The two independent radial velocity estimates agree with a difference of \mbox{$0.8\,$km\,s$^{-1}$}.} \label{rvshift}
\end{figure}

As a final step before deriving the atmospheric stellar parameters, we have corrected each rectified MEGARA spectrum of its observed (topocentric) radial velocity. The task was performed individually in each available spectrum, corresponding to either the \mbox{HR-R} or the \mbox{HR-I} VPH set-ups. The radial velocity was determined through the cross-correlation of the MEGASTAR spectrum with all the MUN05 model rectified spectra. The stellar spectra were first logarithmically re-sampled in wavelength scale, and the peak of the cross-correlation function was stored for each combination of stellar spectra and atmospheric parameters given by the modelled spectra. Interestingly, the radial velocity estimates for each observed spectrum were quite robust for a wide range of atmospheric stellar parameters of the models. This result is also confirmed when we compare the radial velocity values independently obtained from both \mbox{HR-R} and \mbox{HR-I} spectra: the median value of the standard deviation in radial velocity for the whole star sample presented in this paper is \mbox{$0.6\,$km\,s$^{-1}$}. This value corresponds to $\sim 1/7$ of one pixel in the wavelength scale for both \mbox{HR-R} and \mbox{HR-I} VPH settings. Fig.~\ref{rvshift} illustrates the result of this procedure for the star HD~115136. In this example, the measured radial velocities in the two spectral ranges differ by \mbox{$0.8\,$km\,s$^{-1}$}. 
The radial velocities of this sample will be available in the MEGASTAR project web page, where the rectified spectra in both \mbox{HR-R} and \mbox{HR-I} set-ups, will also be given.

The robustness of the cross-correlation method to determine reliable radial velocity corrections implies that this technique is not that suitable to estimate accurate atmospheric stellar parameters, as the peak of the cross-correlation function is dominated by the presence of conspicuous spectroscopic features, not being sensitive enough to discriminate among similar modelled spectra covering a relatively wide range in atmospheric stellar parameters. For this reason, we need to apply an independent method to determine more accurate stellar parameters, as described in Section ~\ref{parest}.

\subsection{Spectral lines and measurements of stellar indices}

The {\it classical} definition of a stellar index is the measurement of the strength of one or several spectral lines by establishing a central bandpass, covering the spectral feature/s of interest, plus one or several continuum bandpasses (at both blue and red sides of the central region), used to adjust a local continuum reference level through a fit (linear, polynomial etc.) to the mean values of the continuum bands. The indices can be defined by spectral slopes (like colours), by the measurement of EW of the line in the pre-defined window after the fitting of the pseudo-continuum, or as the ratio of the measurement of the central line depth of two neighbouring spectral stellar features. The spectral indices are usually defined to measure one single spectral line (although this will depend on the spectral resolution of the instrument). Sometimes, however, the combination of indices of several lines (usually of the same atomic specie) has become a new index itself.  

There are many important works related to indices definition, measurement and applications and it is out of the scope of this paper to give the bibliographic references to all of them and to explain the differences. We cite here few papers on indices falling in MEGASTAR spectral windows. In the \mbox{HR-R} wavelength range, the most studied index has been H${\alpha}$, but with different definitions of the line and pseudo-continuum windows for hot \citep[][]{Gonzalez+2005} and cool stars \citep{sou10}. A good discussion and numerous references are given in these two papers. In the \mbox{HR-I} spectral region, the most widely popular features are the Ca triplet lines, CaT, which have been used as indicators of stellar parameters for more than 60 years, starting with the pioneer work by \citet{s56} for M-stars and followed by others \citep[ etc.]{b81,khm91,g94,jaj84,cvp86,ab89}, finding a strong anti-correlation with the surface gravity, with increasing values from main sequence to giant and supergiant stars, and a weak correlation of the residuals with the metallicity. \citet{dtt89} defined and measured the CaT index in a sample of 106 stars and studied the relationships of this index with $\log{g}$ and [Fe/H]. Later, \citet{cen01}, with a new stellar library of 706 stars in the near-IR region, redefined the windows of both line and continuum and introducing a new CaT index de-contaminated from Paschen lines, CaT$^{*}$.  High spectral resolution instruments, now in operation in large aperture telescopes, have opened a new door to re-define some of these indices. Thus, e.g. \citet{rodriguezmerino20} have proposed three new indices in the blue spectral range (3900 –- 4500\,\AA) with low-medium-resolution spectra (R $\approx$ 6000). Recently, MI21 have also computed some indices with the {\sc HR-pyPopStar} model at R $\approx$ 20\,000. 

There are some sources of uncertainty in index measurements, like the continuum fitting algorithm and its associated errors, translated to errors in the corresponding equivalent widths; the wavelength range and the spectral resolving power. This last effect is important as the lower the spectral resolution, the higher contamination with blended lines, and these unresolved observed spectral lines, which might be produced by variations in the physical processes with different dependence on the stellar parameters for each one, could being measured as a single feature. High-resolution echelle spectrographs, with a wide wavelength range, have been used for years to avoid this line-contamination problem, although the way in which these spectra have been used had been to select only a certain wavelength range and/or some given lines to characterise stars of few (or even one) spectral types, as the higher the number of unblended spectral lines, the more complex analysis is needed.

We include in this paper 22 stellar indices measured in the MEGASTAR spectra of our sample to facilitate their use by other groups. In the {\mbox HR-I} set-up, we have measured the classical lines for the  CaT and Balmer lines. In the {\mbox HR-R} window, besides H$\alpha$, we can discriminate and measure a large number of lines not usually measured in previous data obtained with lower spectral resolution. We describe these indices and discuss their dependence on stellar parameters in Appendix~\ref{indices}. These measurements will be also available as data products in our MEGASTAR web page.

\section{Stellar Parameters}
\label{parest}

\subsection{Selecting the best model fit to each spectrum}
\label{chi2}

We have estimated the physical parameters of the MEGASTAR stars subsample by comparing the radial-velocity corrected observed spectra, with the stellar models from MUN05, both resulting from processes described in previous Section~\ref{spectra}, in the range \mbox{$3500 \le T_\mathrm{eff} \le 25\,000$}\,K. We have used a $\chi^{2}$ technique to calculate, for each pair of observed-modelled spectra, the differences at any given wavelength on a certain range using the classical equation:
\begin{equation}
\label{chi}
\chi^{2}=\sum_{i=1}^{nl}\frac{[F_{\mathrm{mod}}(\lambda)-F_{\mathrm{obs}}(\lambda)]^{2}}{\sigma^{2}},
\end{equation}
where $F_{\mathrm{mod}}$ and $F_{\mathrm{obs}}$ are the rectified modelled and observed fluxes, respectively, $nl$ is the number of wavelength bins, and $\sigma$ is the error of the measured flux in the observed spectrum, obtained from the average SNR of each spectrum as described in \citet{stoehr2008}.
\begin{figure*}
\centering
\includegraphics[width=0.46\textwidth,angle=0]{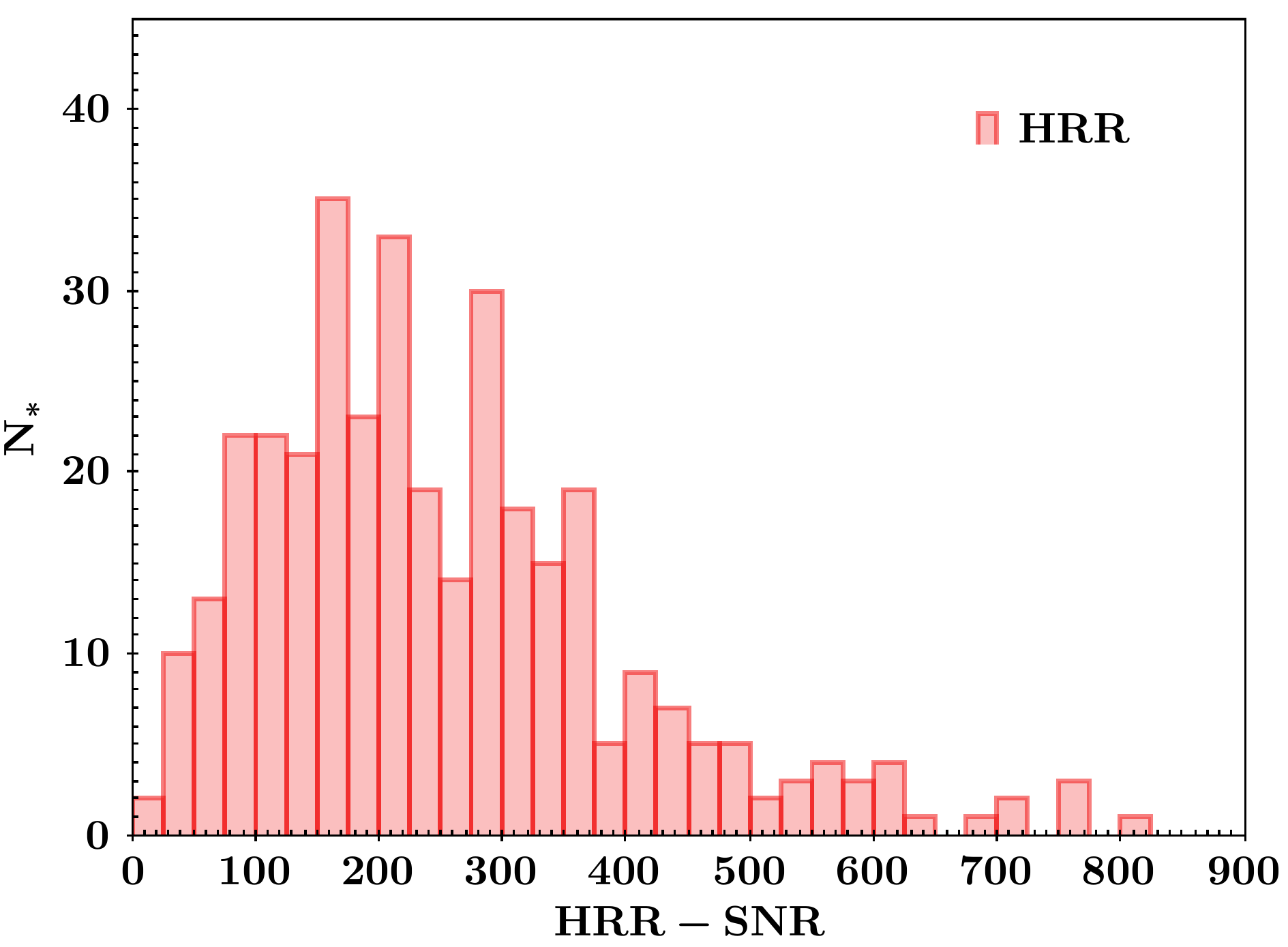}
\includegraphics[width=0.46\textwidth,angle=0]{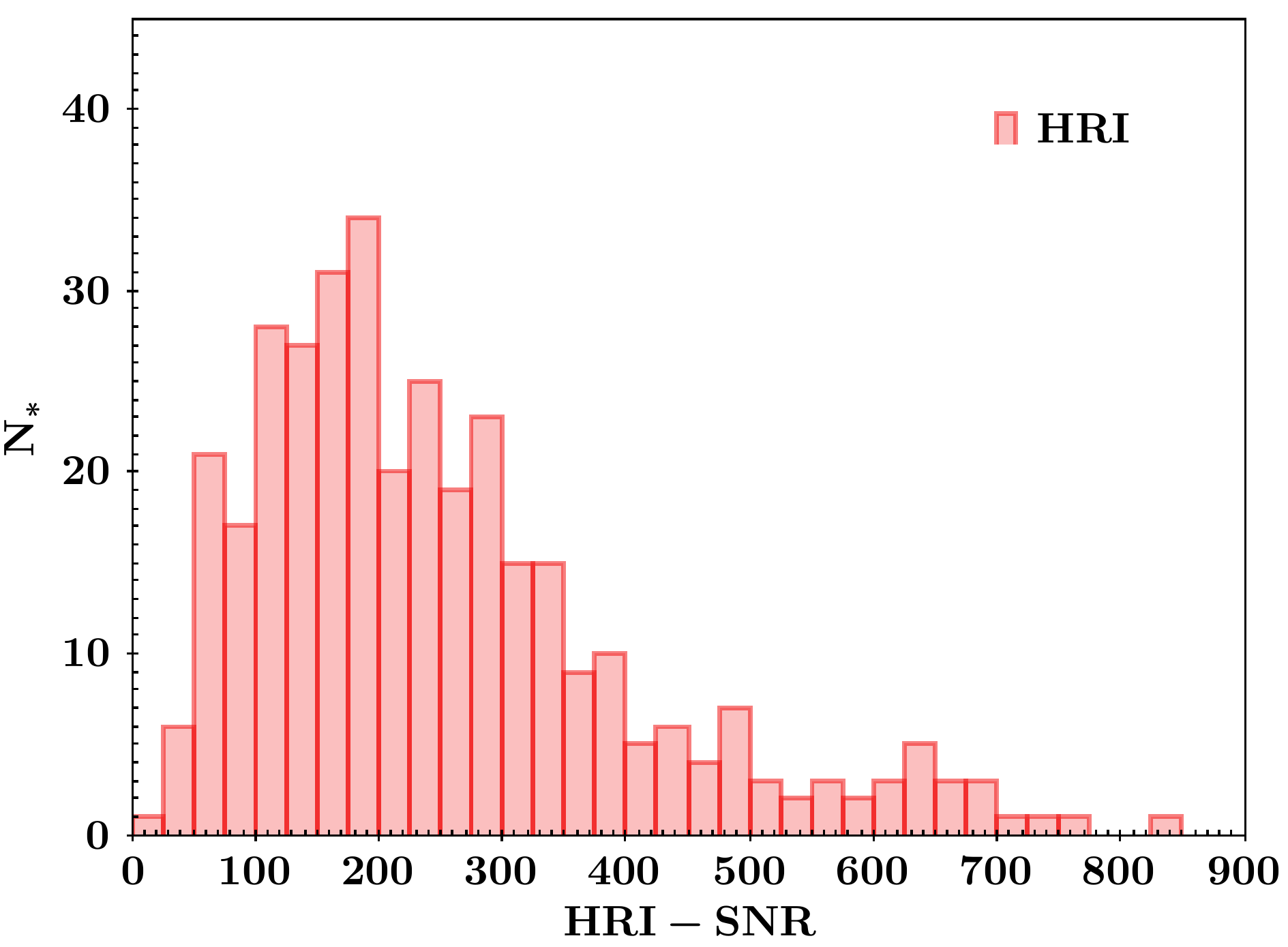}
\caption{Histograms showing the SNR for our subsample of 351 MEGASTAR cool stars spectra in both set-ups: \mbox{HR-R} (left panel) and \mbox{HR-I} (right panel).}
\label{fig:SNR}
\end{figure*}

The averaged value of SNR in our sample $\langle\mathrm{SNR}\rangle$ is $\sim 250$, if measured from the FWHM, or $\sim 150$ when a Gaussian fit is applied, while the maximum (mode) for both \mbox{HR-R} and \mbox{HR-I}  data is $\sim 200$. These values imply an error in the measured normalised flux of $\langle\sigma\rangle\sim 8\%$. The SNR distributions obtained for our subsample in both set-ups are shown in Fig.~\ref{fig:SNR}. 

\begin{figure}
\centering
\includegraphics[width=0.33\textwidth,angle=-90]{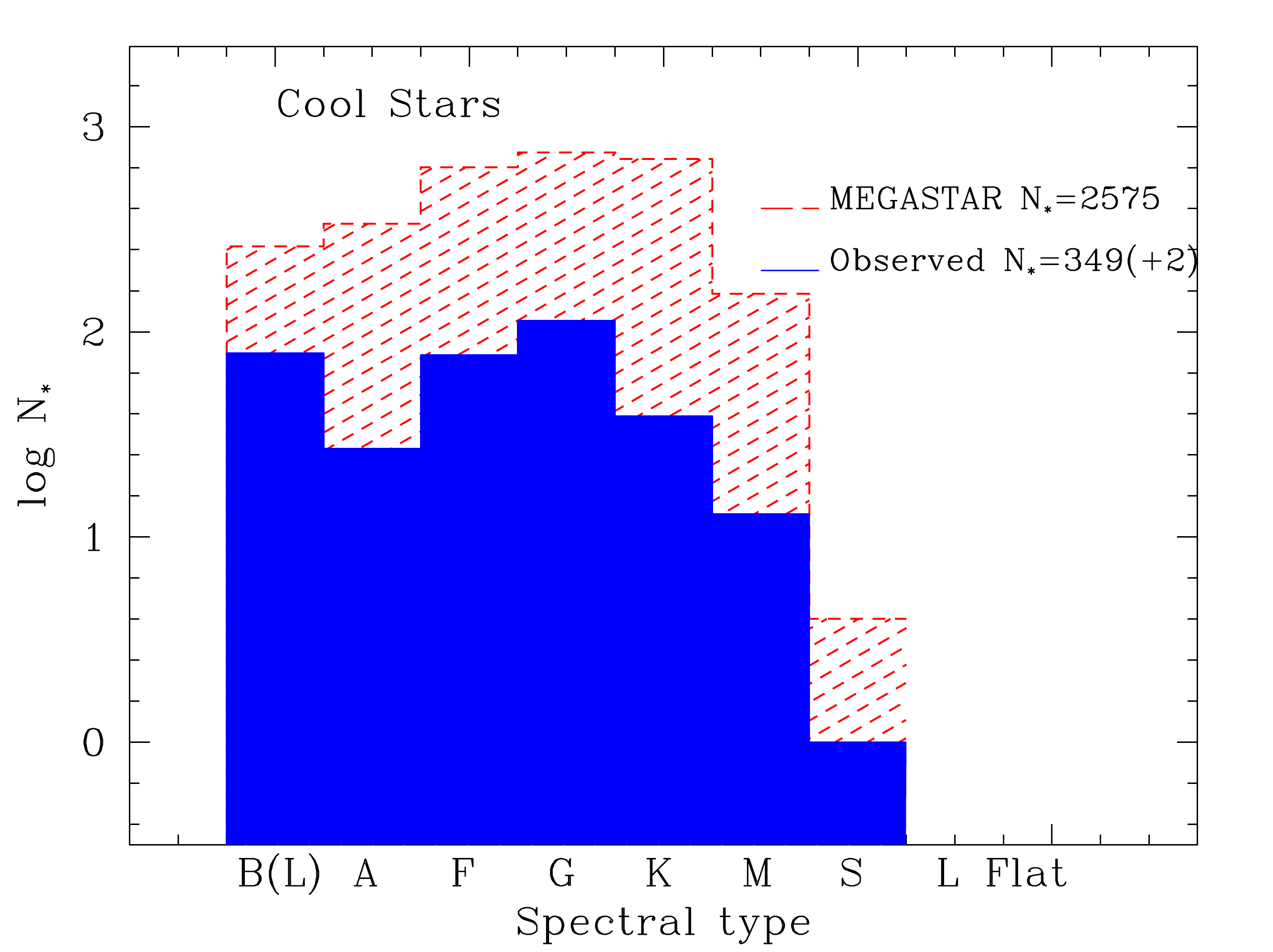}
\caption{Histogram showing the number of stars with spectral types equal or later than B2 in the complete MEGASTAR catalogue (red hashed) and in the subsample of this work (filled blue).}
\label{fig:histo}
\end{figure}

Our stellar sample spans a wide range of spectral types, from late B to S, as shown in Fig.~\ref{fig:histo}, where the histograms of the numbers of stars of the complete MEGASTAR library (excluding  stars of spectral types WR, O and B earlier than --or equal as-- B2) and of the sample employed in this work are depicted. 

Before applying Eq.~(\ref{chi}), we have performed a preliminary selection of models in a given range of $T_{\mathrm{eff}}$ according to the spectral type. Table~\ref{tipos} summarises the applied criteria. We have followed the method described by \citet{arentsen2019}, which led us to divide our sample, following their expected $T_{\mathrm{eff}}$, column 3 of Table~\ref{tipos}, into four groups in column 4: (1) M and cooler stars, which are compared to models with $T_{\mathrm{eff}} < 4550$\,K only; (2) K, G and F spectral types, compared to models with $3000 \le T_{\mathrm{eff}} \le 8000$\,K; (3) A-type stars, compared to models with $7000\,K < T_{\mathrm{eff}}< 15\,000$\,K and (4) late-B stars, compared to models with $9000\,K < T_{\mathrm{eff}}<25\,000$\,K. 

\begin{table}
\caption{Filtering of $T_{\mathrm{eff}}$ models for the stellar parameters determination of our sample. For each spectral type (column 1), column 2 gives the number of stars; column 3 shows the expected $T_{\mathrm{eff}}$ range, while column 4 summarises the $T_{\mathrm{eff}}$ range for the used MUN05 models.}
\centering
\begin{tabular}{crcc}
Spectral & Number & expected $T_{\mathrm{eff}}$& fitting $T_{\mathrm{eff}}$\\
Type    & of stars & range [K] &  range [K]       \\
\hline
S & 1 & $<$ 3000 & $\le 4550$ \\
M & 13 & $< 3700$ & $\le 4550 $ \\  
K & 39 & 3700--5200 & 3000--8000  \\  
G & 113 & 5200--6000 & 3000--8000  \\  
F & 77 & 6000--7500 & 3000--8000  \\
A & 27 & 7500--10000 & 7000--15000 \\
B(L) & 79 & 10000 -- 22500 & 9000-25000 \\
\hline
\end{tabular}
\label{tipos}
\end{table}

\subsection{Testing the technique: application to theoretical models}
\label{checking}

To test the reliability of our method for determining the stellar physical parameters, we have applied it to the MUN05 theoretical spectra used in this work, to check if the same input stellar parameters are recovered as output of our $\chi^{2}$ technique. These models were cropped to the wavelength ranges of the \mbox{HR-R} and \mbox{HR-I} set-ups. 

\begin{figure*}
\centering
\includegraphics[width=0.33\textwidth,angle=0]{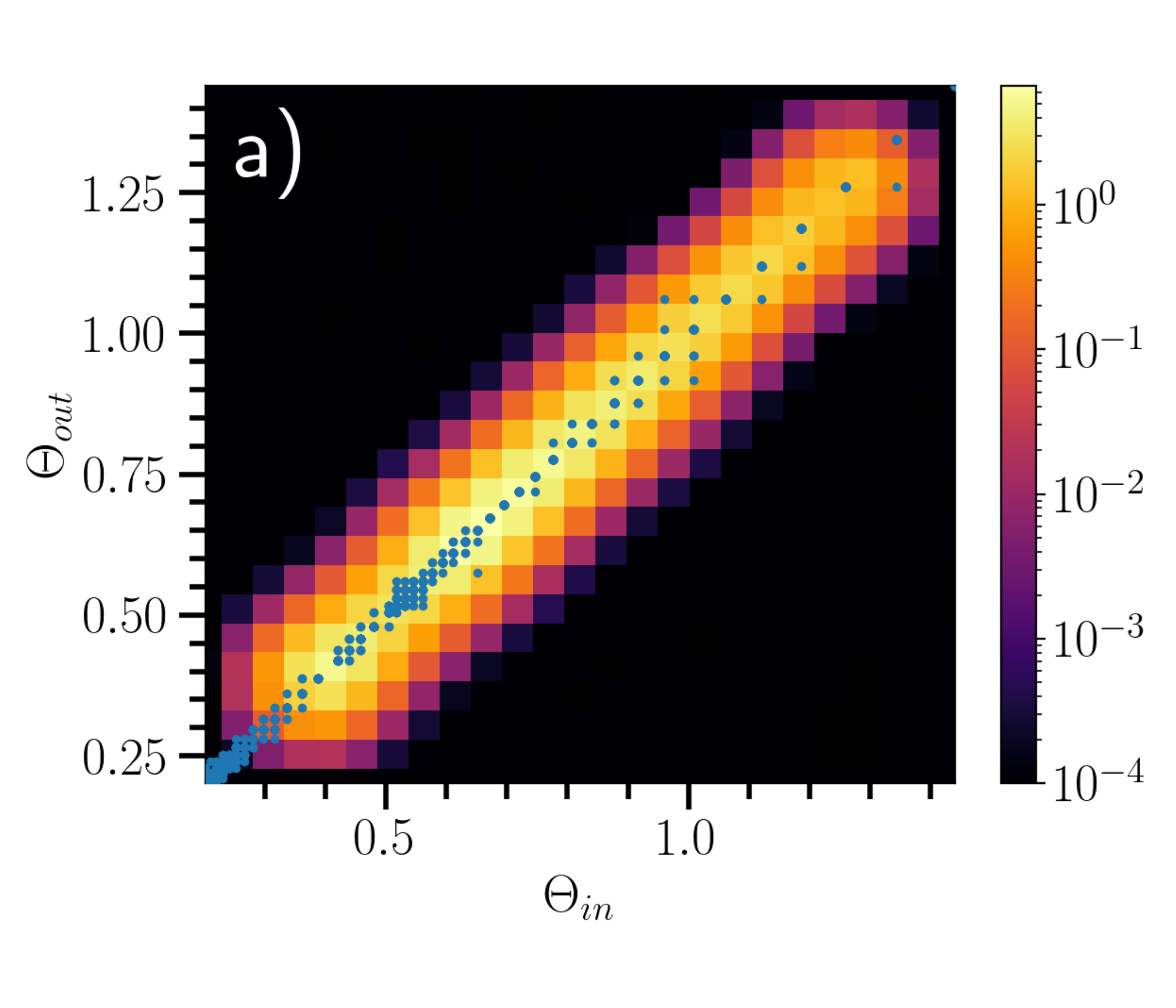}
\includegraphics[width=0.33\textwidth,angle=0]{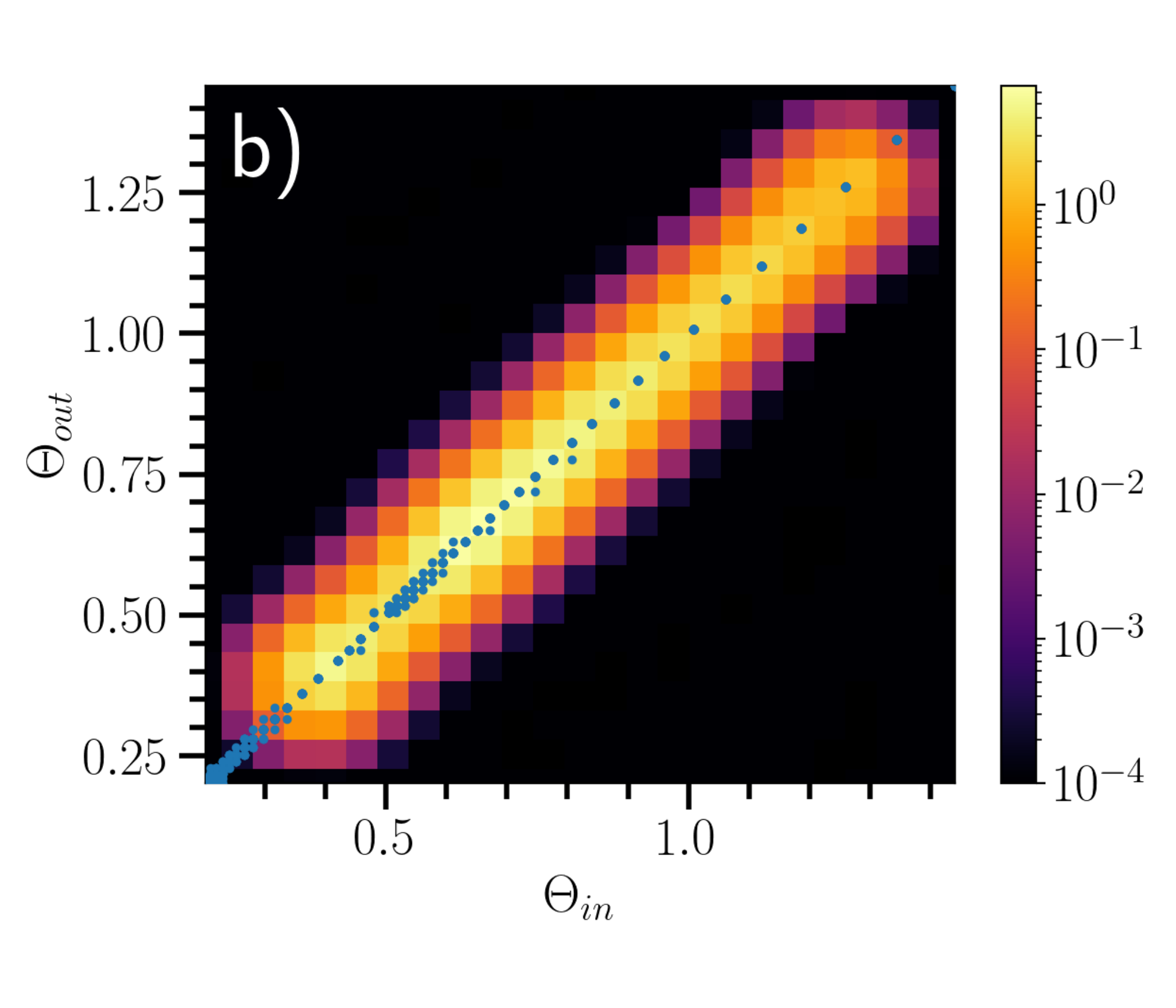}
\includegraphics[width=0.33\textwidth,angle=0]{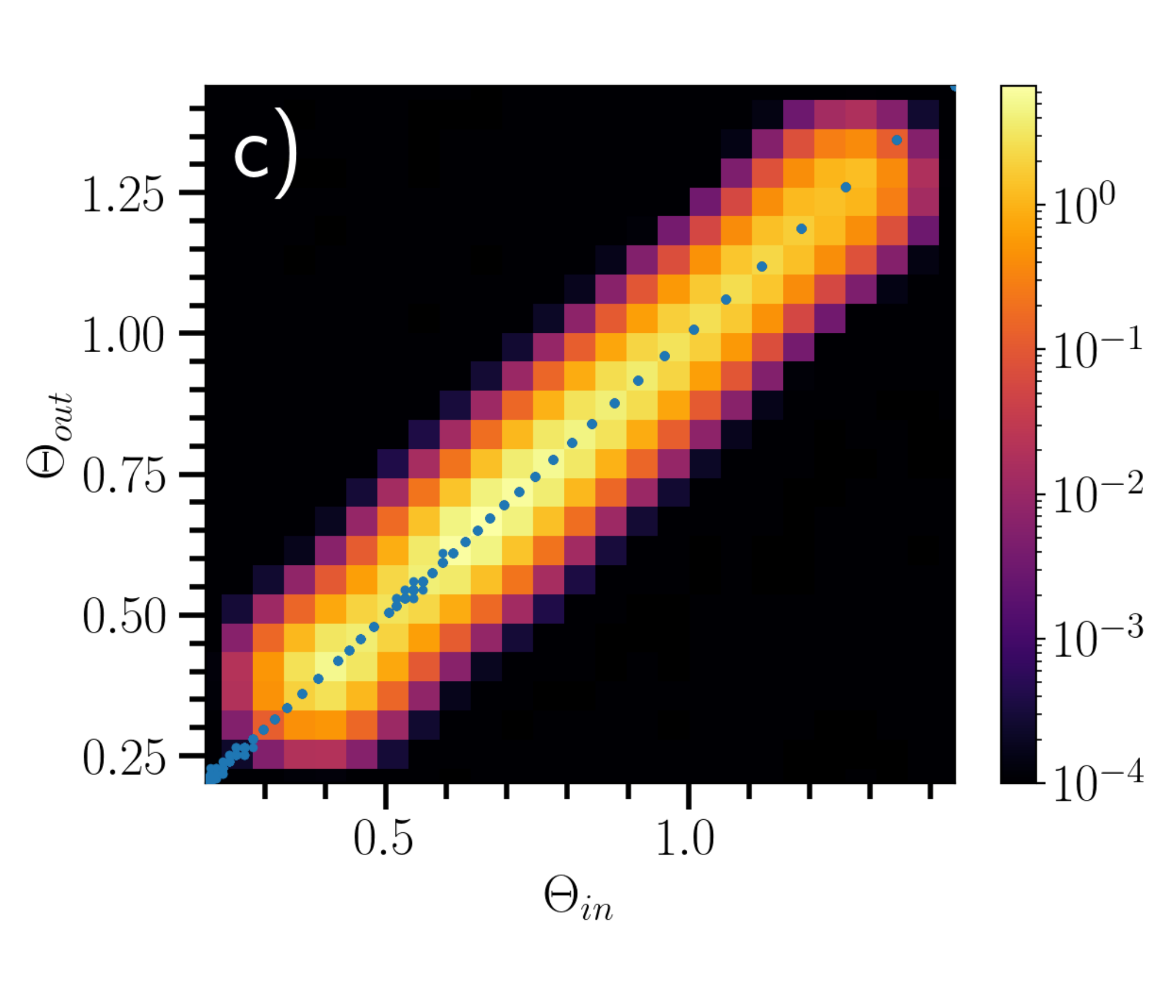}
\includegraphics[width=0.33\textwidth,angle=0]{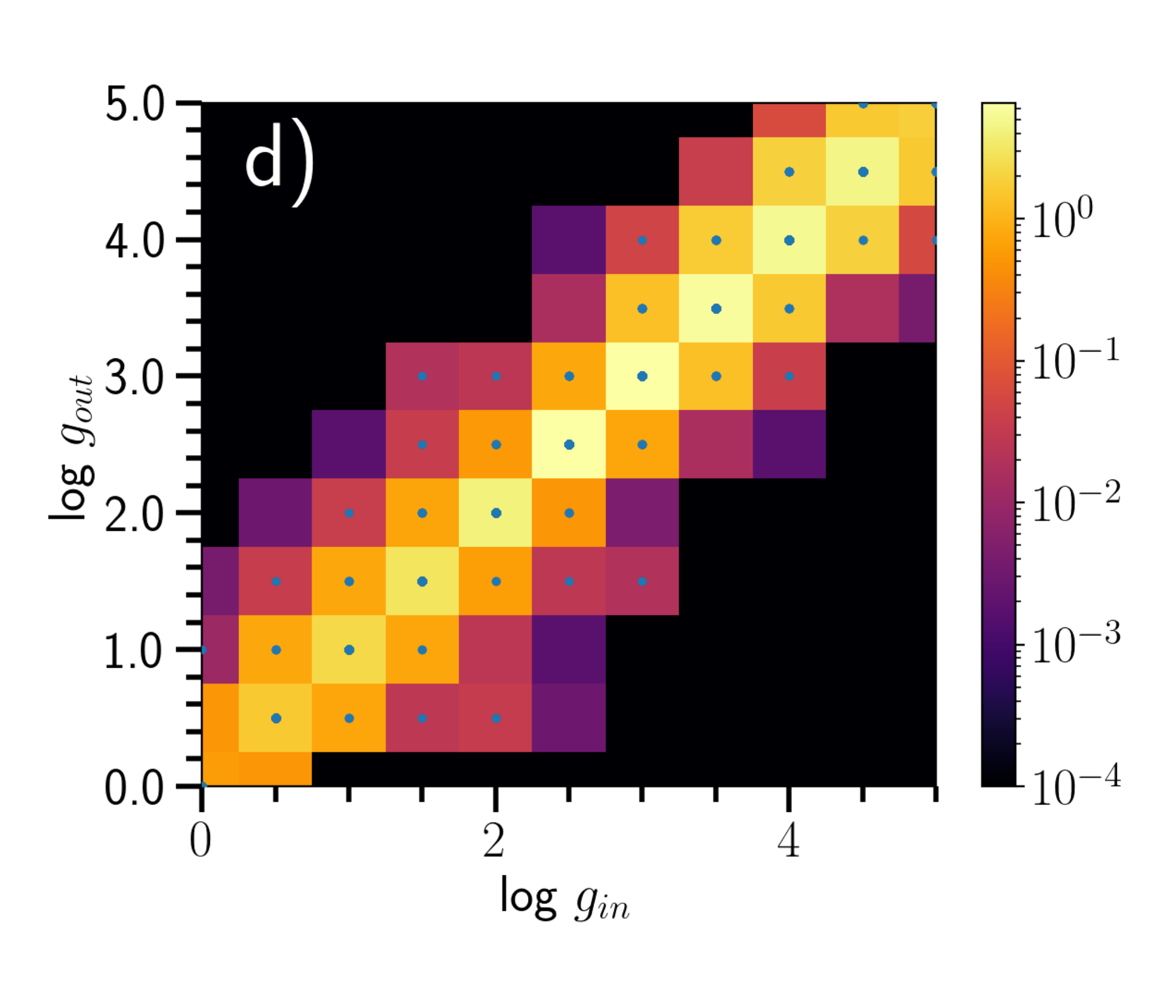} 
\includegraphics[width=0.33\textwidth,angle=0]{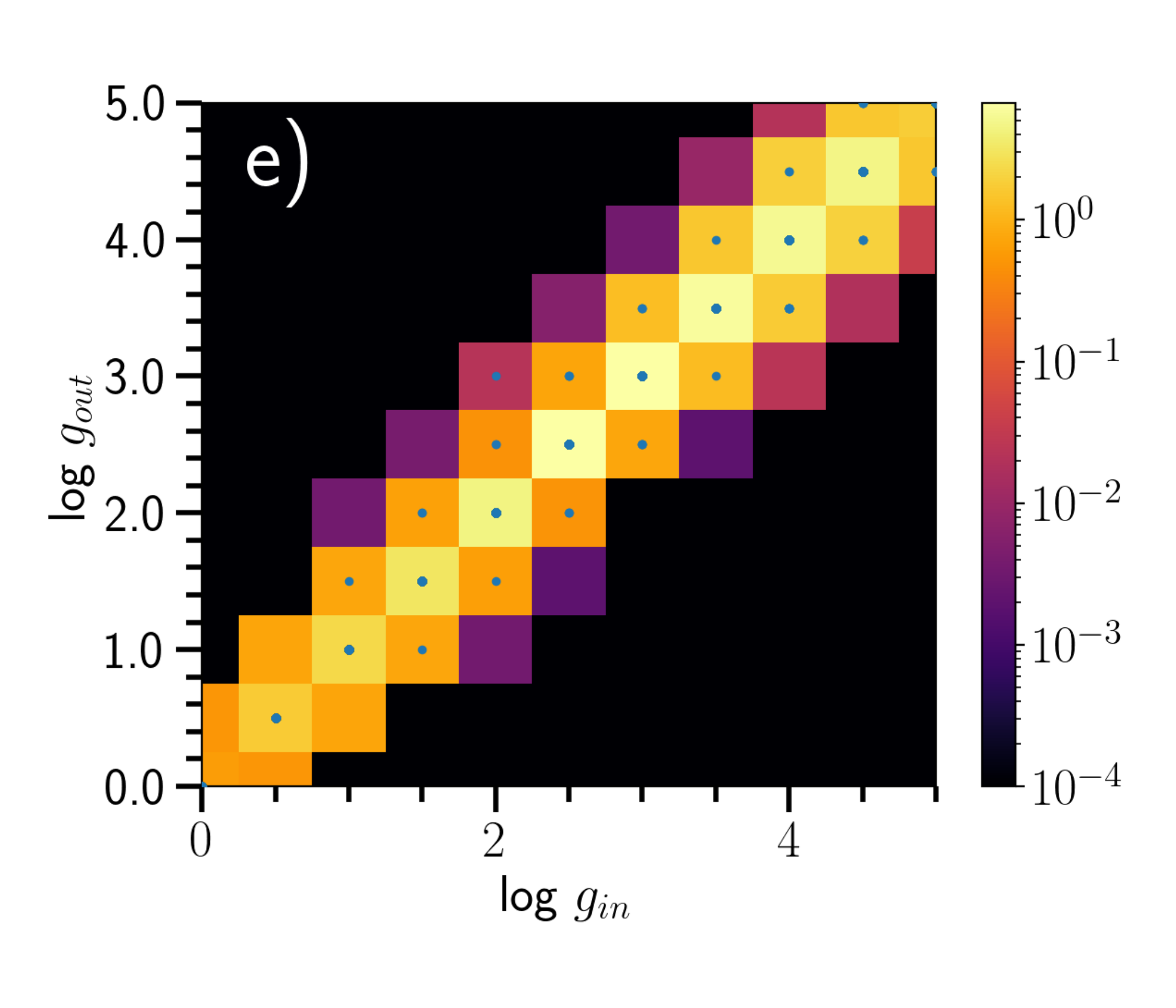} 
\includegraphics[width=0.33\textwidth,angle=0]{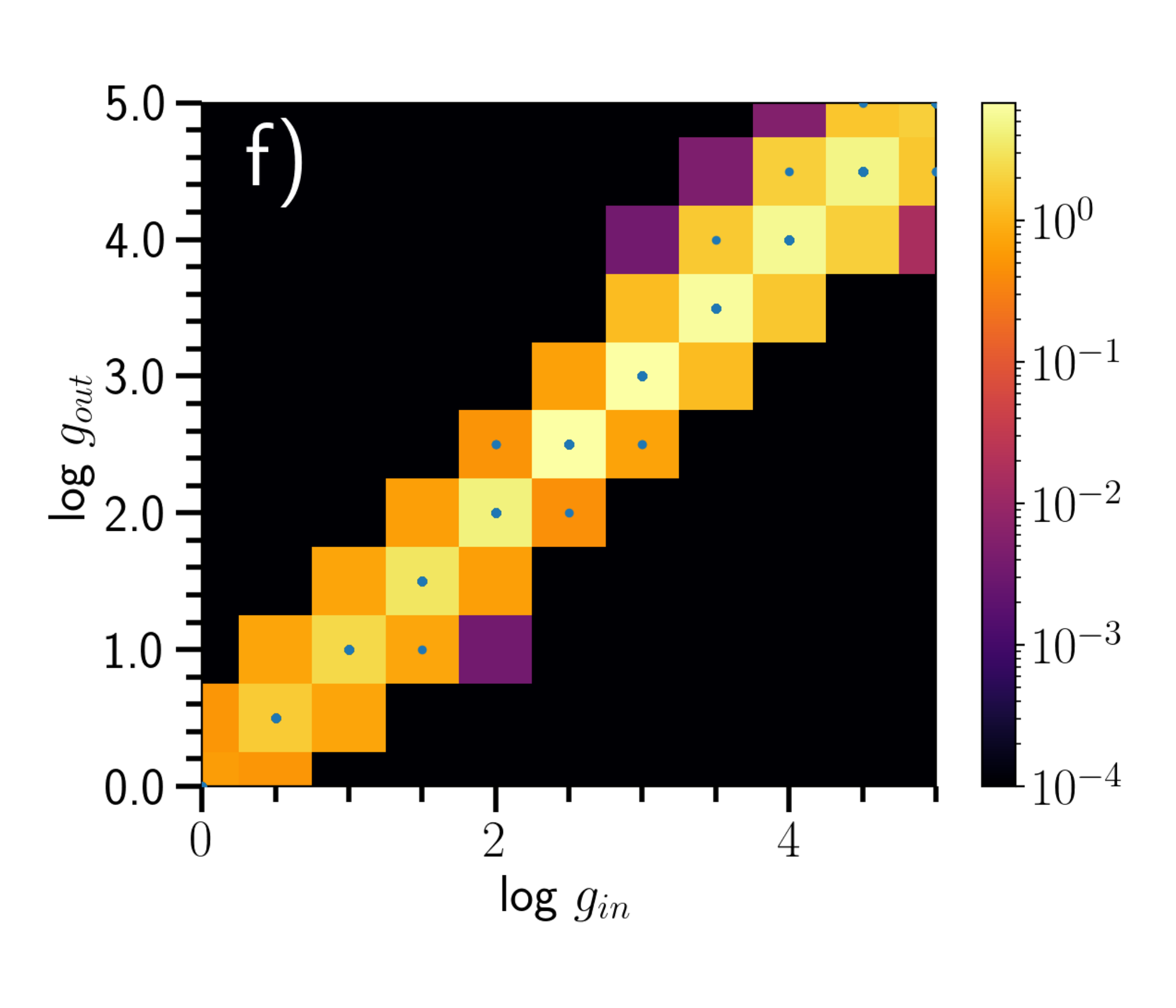}
\includegraphics[width=0.33\textwidth,angle=0]{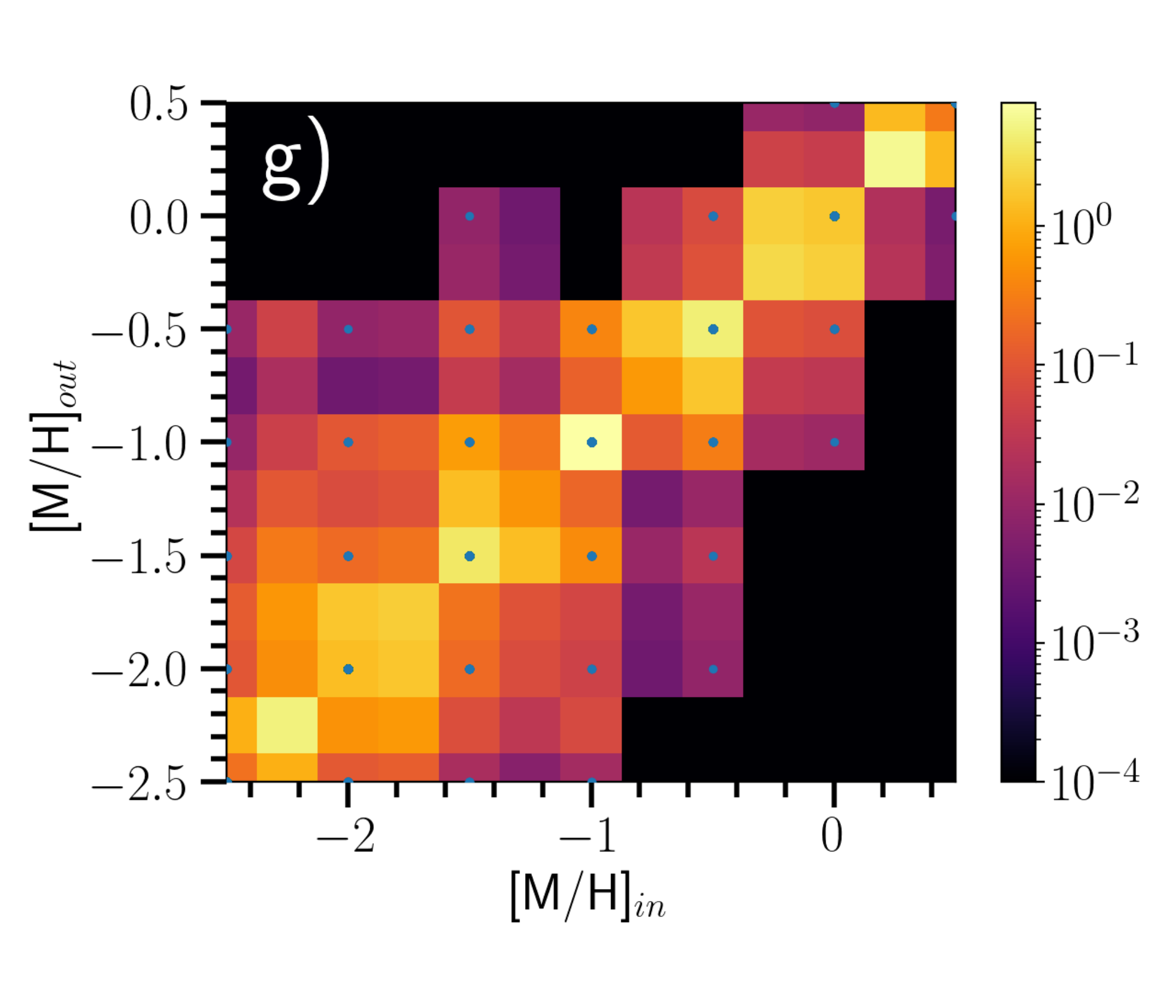}
\includegraphics[width=0.33\textwidth,angle=0]{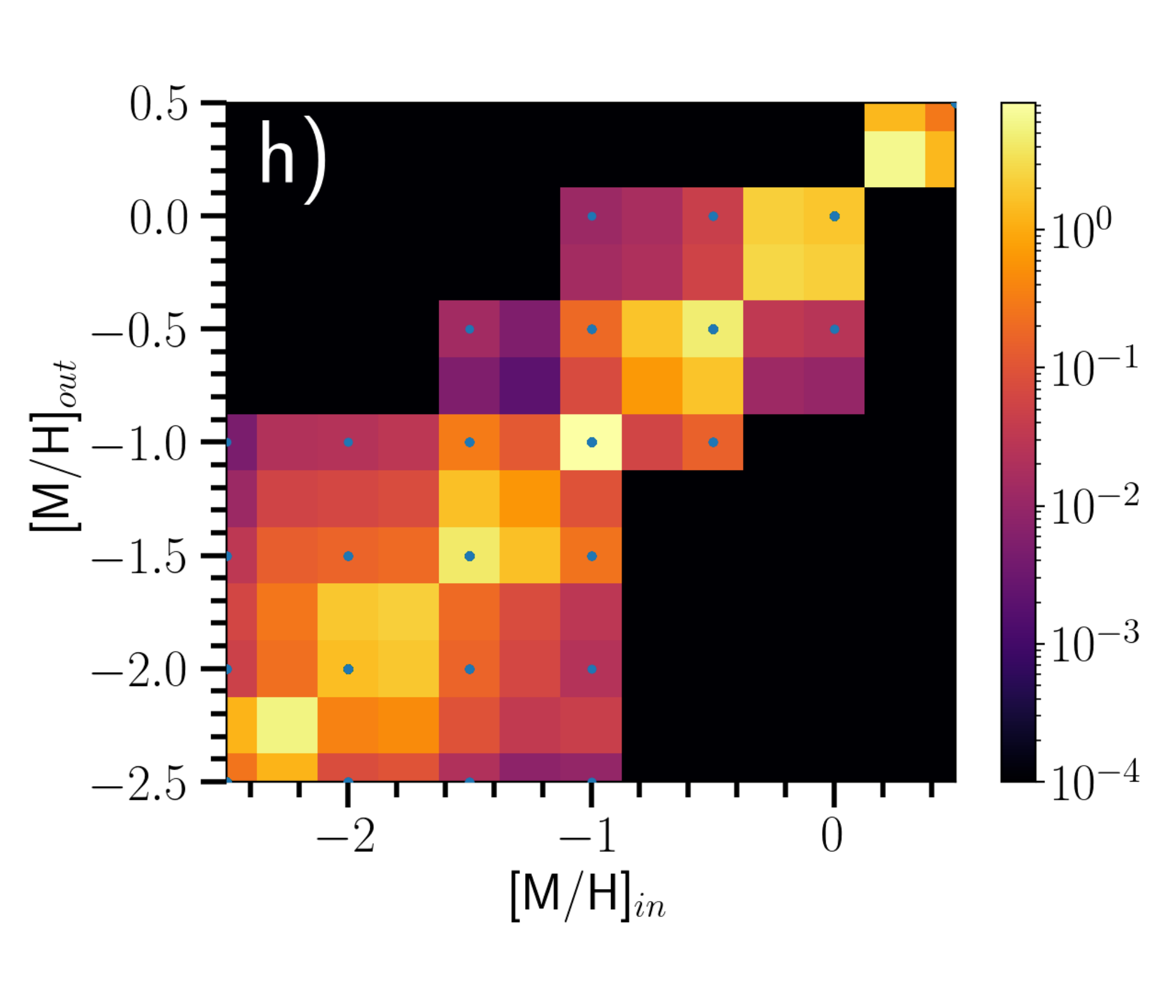}
\includegraphics[width=0.33\textwidth,angle=0]{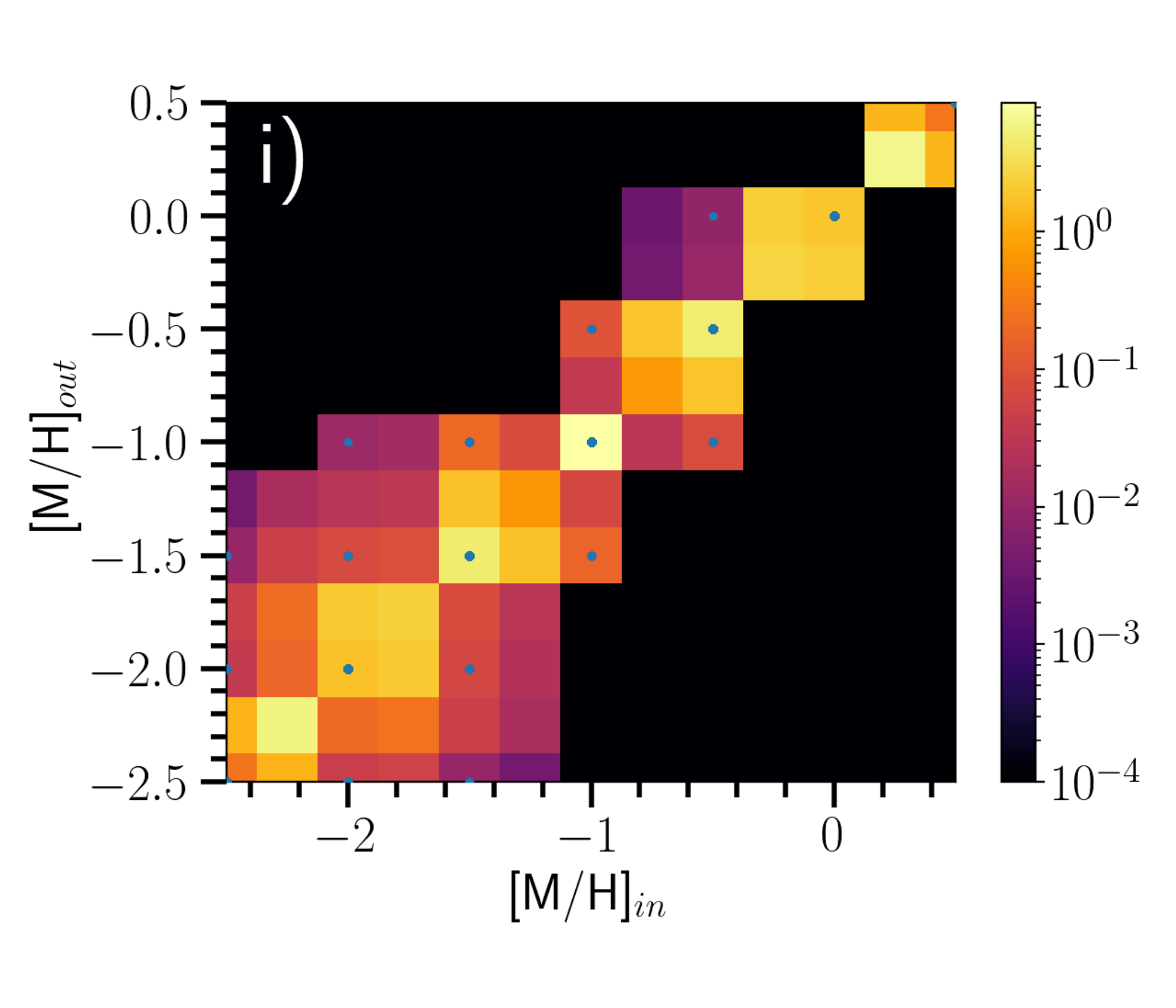}
\caption{Comparison of the stellar parameters obtained with our $\chi^{2}$ technique and the original values from the MUN05 models used in this work. In all panels, the X-axis represents the original parameter of each theoretical spectrum while the Y-axis the resulting value obtained after applying the $\chi^{2}$-technique over the noisy theoretical spectra. The panels include plots for each stellar parameter: top, middle and bottom rows show effective temperature (represented as $\theta$=5040/T$_\mathrm{eff}$), $\log{g}$ and [M/H], respectively. Panels in different columns correspond to $\mathrm{SNR}=50$, 100 and 200, increasing from left to right. The background colour maps represent the relative density of solutions where lighter regions indicate higher density of solutions in logarithmic scale.}
\label{fig:noise}
\end{figure*}

First, we have added Gaussian noise to this grid of theoretical spectra with SNR in the range of our observations by using the Box-Muller Transform \citep{box-muller}, which finds a set of random numbers $X$ following a Gaussian distribution from a pair of random numbers (u,v) uniformly distributed between 0 and ~1:
\begin{eqnarray}
    u &\!\!\! = \!\!\! & \mathrm{rand}()\,;\;\textrm{in the range [0,1)},\\
    v &\!\!\! = \!\!\! & \mathrm{rand}()\,;\; \textrm{other similar random number}\\
    X_{\mathrm{mod,SNR}}(\lambda)
    &\!\!\! = \!\!\! &\sqrt{-2\,\log{(1-u)}}\cdot \cos{(2\,\pi\,v)}.
\end{eqnarray}

\begin{figure*}
\centering
\includegraphics[width=0.33\textwidth,angle=0]{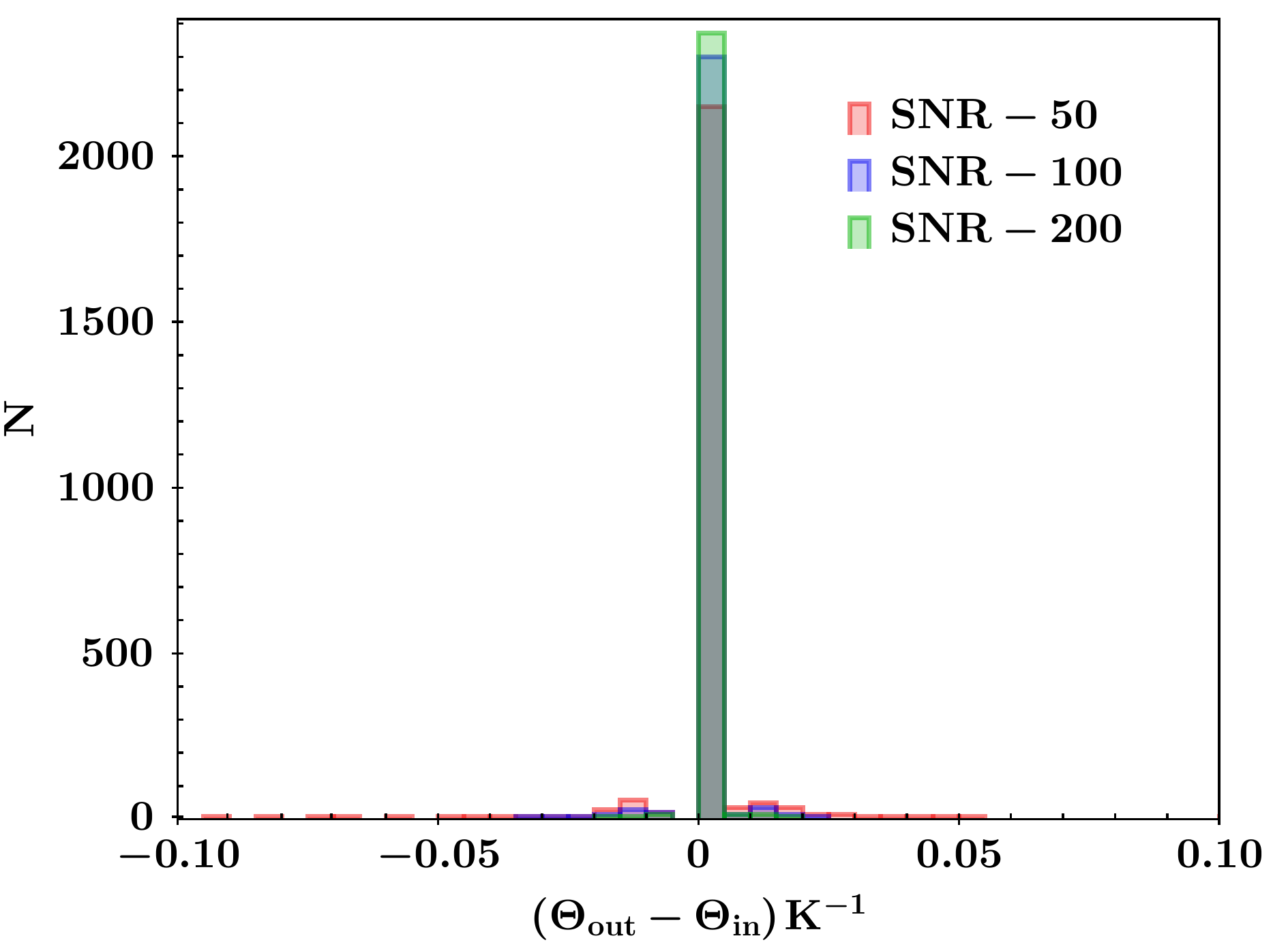}
\includegraphics[width=0.33\textwidth,angle=0]{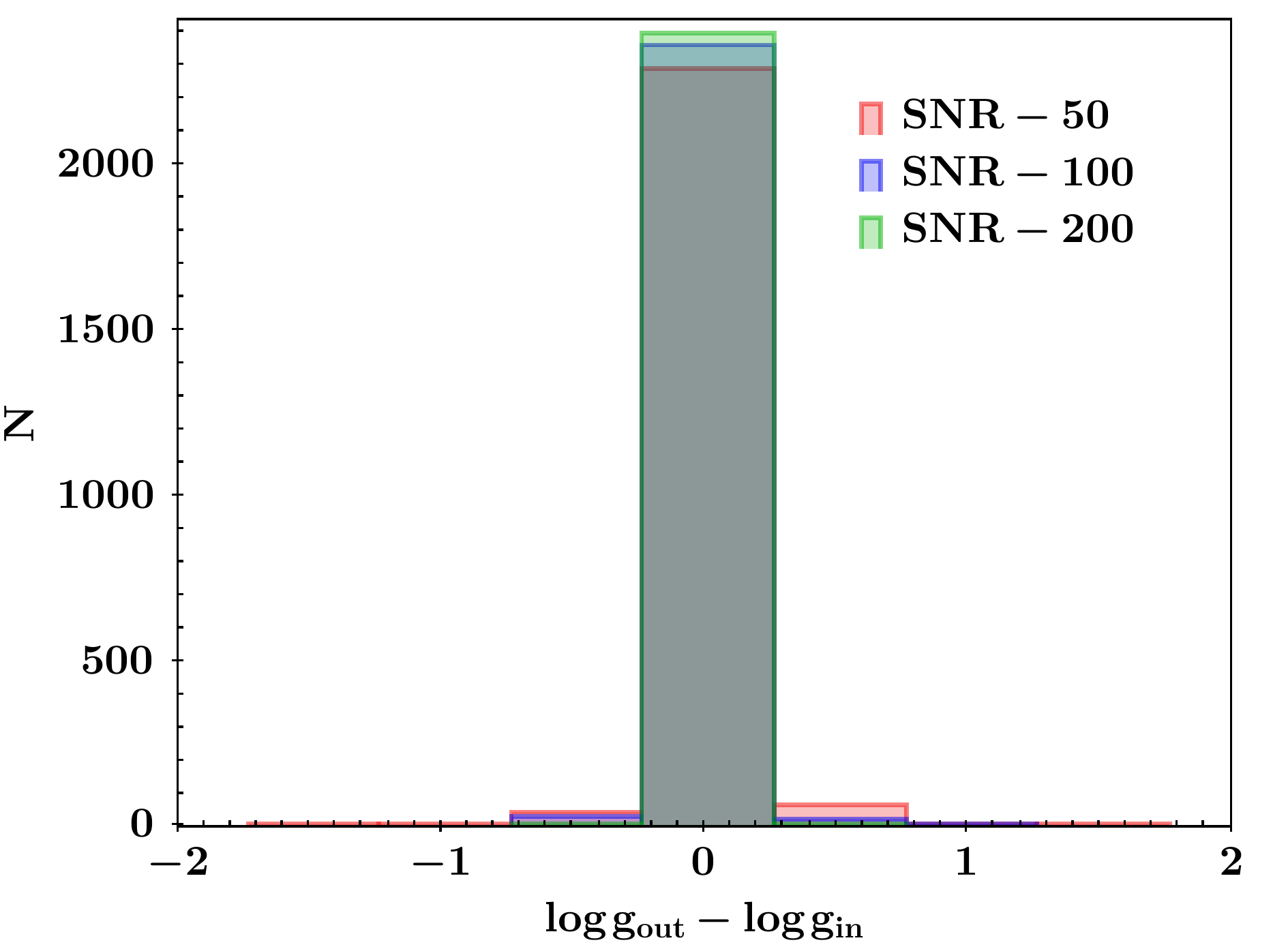}
\includegraphics[width=0.33\textwidth,angle=0]{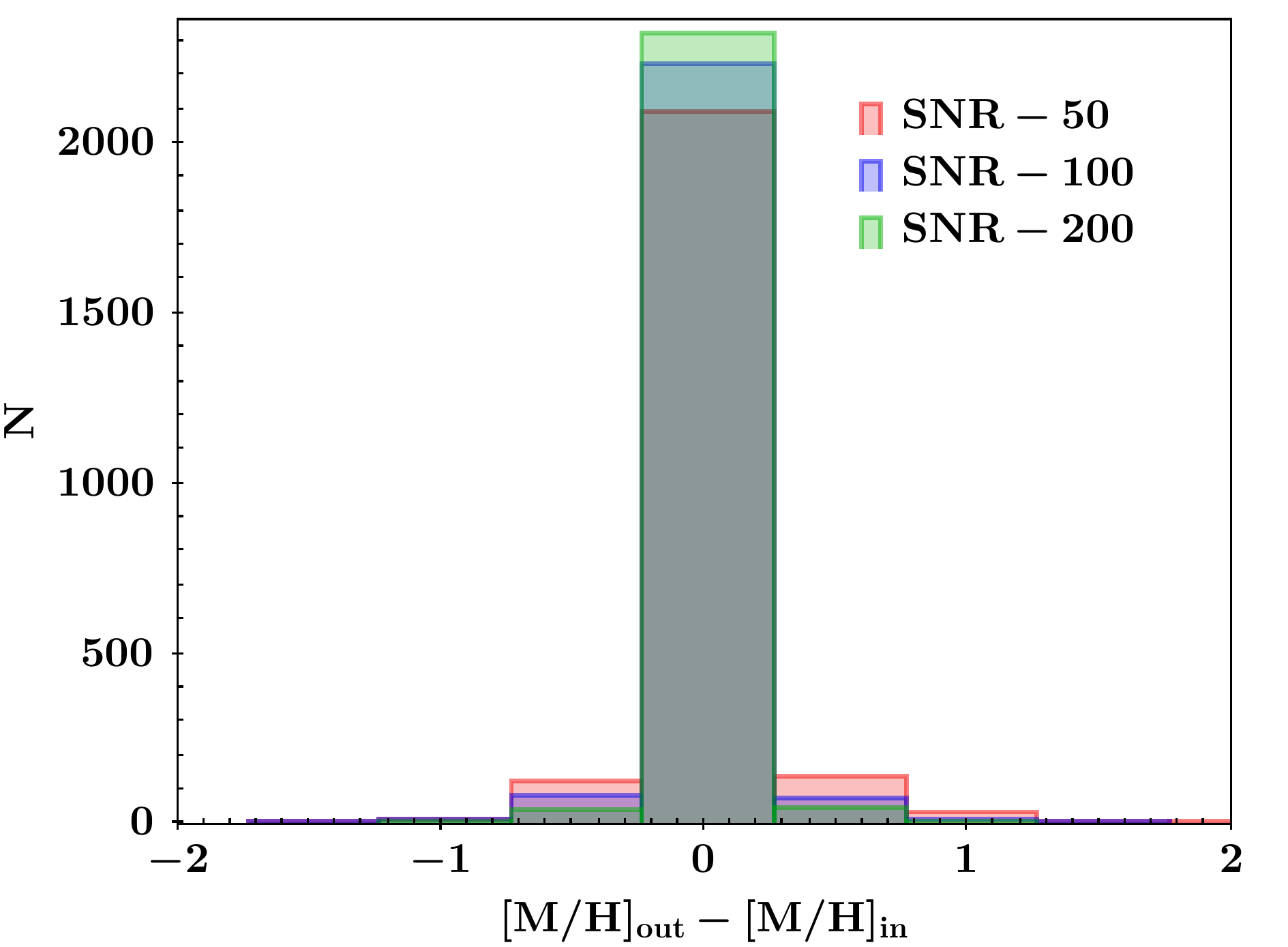}
\caption{Histograms of differences between output and input values obtained when comparing the MUN05 noisy theoretical spectra with the set of original MUN05 models for $\mathrm{SNR}= 50$, 100 and 200, plotted with red, blue and green lines, respectively}.
\label{fig:out_in_histo}
\end{figure*}
This random number $X_{\mathrm{mod,SNR}}$ following a Gaussian curve is multiplied by the selected dispersion and added to the flux $F_{\!\mathrm{mod}}(\lambda)$ as:
\begin{eqnarray}
\sigma_{\mathrm{mod}}(\lambda)&\!\!\!=\!\!\!&\frac{F_{\!\mathrm{mod}}(\lambda)}{\sqrt{\mathrm{SNR}}},\\
F_{\!\mathrm{add,mod}}(\lambda) &\!\!\!=\!\!\!& X_{{\mathrm{mod},SNR}}(\lambda) \times \sigma_{\mathrm{mod}}(\lambda),\\
F_{\!\mathrm{noisy,mod}}(\lambda)&\!\!\!=\!\!\!& F_{\!\mathrm{mod}}(\lambda) + F_{\!\mathrm{add,mod}}(\lambda).
\end{eqnarray}

We have generated noisy spectra \big($\lambda$, $F_{\!\mathrm{noisy}}(\lambda$)\big) with $\mathrm{SNR}$ = 50, 100 and 200 for the 2408 MUN05 models within our limits. Then, we have applied our computation technique to compare each pair noisy-original model for deriving the best fit and, in consequence, the set of estimated stellar parameters: T$_\mathrm{eff}$, $\log{g}$ and [M/H]. 

\begin{figure}
\centering
\includegraphics[width=0.35\textwidth,angle=-90]{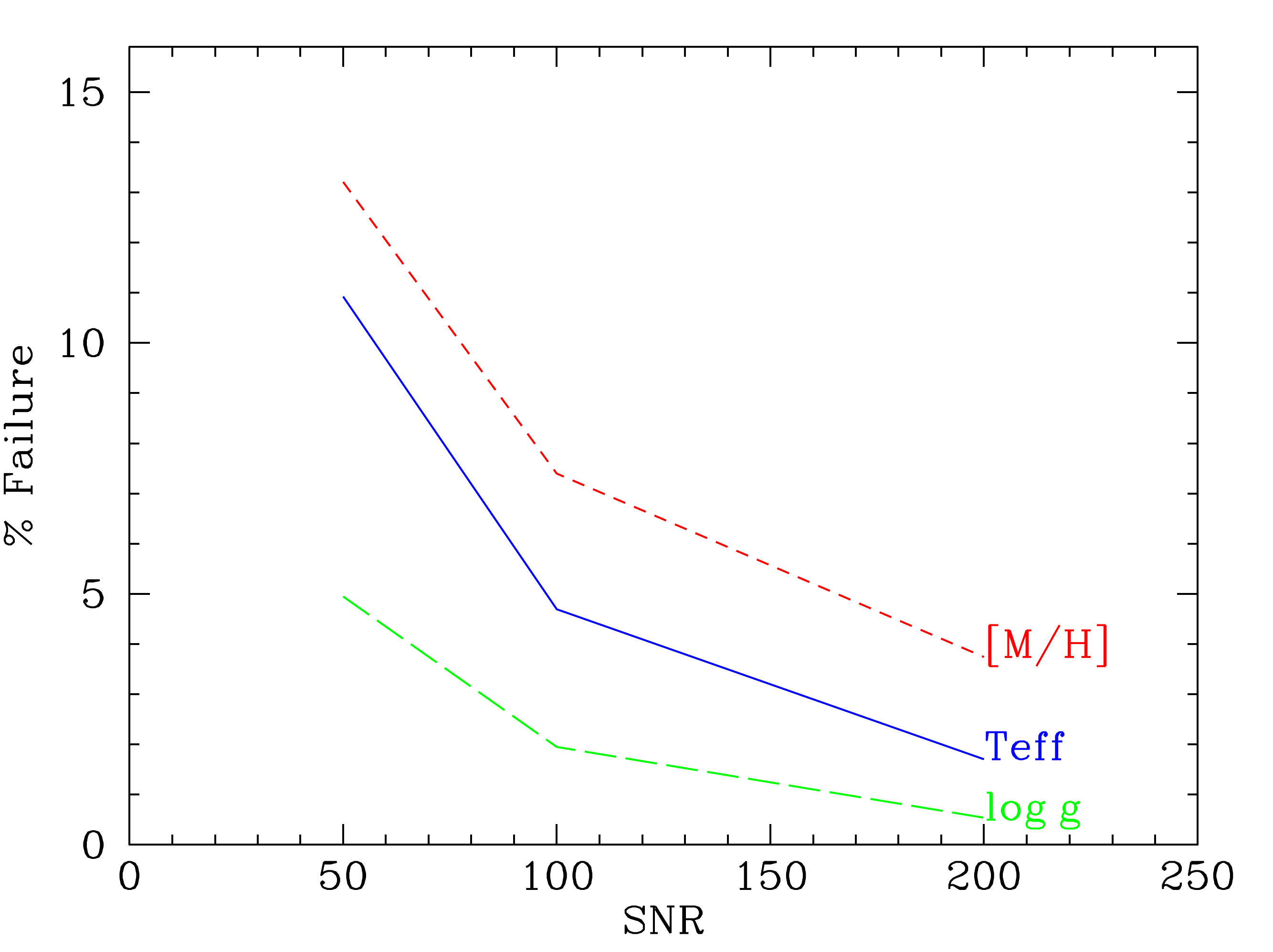}
\caption{Percentage of failures in the estimation of the stellar parameters -- with different colour and lines as labelled in the plot -- as a function of SNR.}
\label{fig:failures_SNR}
\end{figure}

\scriptsize
\begin{table*}
\caption{Stellar Parameters obtained from our best fit to MUN05 models using both \mbox{HR-R} and \mbox{HR-I} set-ups. The complete table for the 351 stars is available on-line (see Section ~\ref{data}). The star name is given in column (1), the stellar parameters: $T_{\rm eff}$ (K), $\log{g}$ and $\rm [M/H]$, as found in the literature, if available, are given in columns (2), (3) and (4), respectively; the spectral type is given in column (5); the radial velocity and its error are shown in columns (6) and (7); the SNR in \mbox{HR-R} and \mbox{HR-I} set-ups are given in columns (8) and (9), respectively; the stellar parameters: $T_{\rm eff}$ (K), $\log{g}$ and $\rm [M/H]$, obtained by the $\chi^{2}$ technique are given in columns (10), (11) and (12), respectively; the $\chi^{2}_{min}$; the associated likelihood $\mathcal{L} $, the number of wavelengths used in each stellar fit and the number of models with similar $\mathcal{L}$ around the value with minimum $\chi^{2}_{min}$ are given in columns (13), (14), (15) and (16), respectively. Finally, the stellar parameters, $T_{\rm eff}$, in K, $\log{g}$ and $\rm [M/H]$, with their corresponding errors, obtained as averaged of those models, are given in columns (17) to (22).}
\resizebox{18cm}{!}{
\centering
\begin{tabular}{lc@{\;\;}c@{\;\;}ccccccc@{\;\;}c@{\;\;}ccccrrllllll@{}}
\multicolumn{1}{c}{Name} & $T_{\mathrm{eff,lit}}$ & $\log{g}_{lit}$&  [M/H]$_{lit}$ & Sp. & \multicolumn{2}{c}{Rvel $\pm \Delta$} & SNR & SNR & T$_{\mathrm{eff,min}}$ & $\log{g}_{min}$ & [M/H]$_{min}$ &  $\chi^{2}_{\mathrm{min,red}}$ &  $\mathcal{L}$ & $N_{\lambda}$ & $N_\mathrm{m}$ & \multicolumn{2}{c}{$\langle T_{\mathrm{eff}} \pm \Delta \rangle$} & \multicolumn{2}{c}{$\langle\log{g} \pm \Delta\rangle$} & \multicolumn{2}{c}{$\langle\mathrm{[M/H]} \pm \Delta\rangle$} \\
& \multicolumn{3}{c}{literature}  & Type & & & \mbox{HR-R} & \mbox{HR-I} & \multicolumn{3}{c}{minimum $\chi^{2}$}  & & & & &\multicolumn{6}{c}{averaged}  \\ 
\multicolumn{1}{c}{(1)} & (2) & (3) & (4) & (5) & \;\;\;\;\;(6) & (7) & (8) & (9) & (10) &
(11) & (12) & (13) & (14) & (15) & (16) & \;\;\;(17) & (18) & (19) & (20) & 
\;\;\;(21) & \!\!\!(22) \\
\hline \hline
BD$-$032525  &  5750 &  3.60 & $-$1.90 &   F3 & \multicolumn{2}{r}{$41.3 \pm 0.8$} & 131.9  & 124.1 & 6250 &  5.0 &    $-$1.5 &  0.0356 &  99.8 & 7721 &  247 & \multicolumn{2}{r}{5750 $\pm$ 375} & \multicolumn{2}{l}{ 3.50 $\pm$ 1.25} & \multicolumn{2}{l}{$-$2.00 $\pm$   0.50} \\
BD$-$122669  &  6955 &  4.00 & $-$1.41 &   A5 & \multicolumn{2}{r}{$48.7 \pm 0.7$} & 103.8   & 89.2 & 7000 &  5.0 &    $-$1.5 &  0.0285 &  99.9 & 7718 &   57 & \multicolumn{2}{r}{ 7250 $\pm$ 250} & \multicolumn{2}{l}{ 4.50 $\pm$ 0.50} & \multicolumn{2}{l}{$-$1.50 $\pm$   0.50} \\
BD+083095    &  5728 &  4.12 & $-$0.36 &  G0V & \multicolumn{2}{r}{$-85.0 \pm 0.3$} & 97.8  & 114.7 &  6000 &  4.0 &    $-$0.5 &  0.0479 &  99.7 & 7703 &  209 & \multicolumn{2}{r}{ 5500 $\pm$ 375} & \multicolumn{2}{l}{ 3.00 $\pm$ 1.25} & \multicolumn{2}{l}{$-$1.00 $\pm$   0.50} \\
BD+092190    &  6316 &  4.56 & $-$2.93 &   A0 & \multicolumn{2}{r}{$256.3 \pm 0.1$} & 124.0  & 128.3 & 7000 &  5.0 &    $-$1.5 &  0.0654 &  99.6 & 7622 &   31 & \multicolumn{2}{r}{ 7000 $\pm$ 125} & \multicolumn{2}{l}{ 4.50 $\pm$ 0.50} & \multicolumn{2}{l}{$-$1.50 $\pm$   0.50} \\
BD+203603    &  6121 &  4.32 & $-$2.09 &   F0 & \multicolumn{2}{r}{$-268.2 \pm 0.7$} & 208.5 &  199.7 & 6250 &  4.5 &    $-$1.5 &  0.0409 &  99.8 & 7623 &  158 & \multicolumn{2}{r}{ 6000 $\pm$ 375} & \multicolumn{2}{l}{ 3.50 $\pm$ 1.00} & \multicolumn{2}{l}{$-$2.00 $\pm$   0.50} \\
BD+262606    &  ...   &   ... &  ...   &  A5V & \multicolumn{2}{r}{$4.6 \pm 0.1$} & 173.9  & 173.5 & 7000 &  5.0 &    $-$1.5 &  0.0989 &  99.2 & 7737 &   16 & \multicolumn{2}{r}{ 7000 $\pm$ 125} & \multicolumn{2}{l}{ 5.00 $\pm$ 0.25} & \multicolumn{2}{l}{$-$1.50 $\pm$   0.50} \\
HD\,017081  &    13320 &  3.64 &  0.03 &  B7IV & \multicolumn{2}{r}{$7.5 \pm 0.3$} & 413.2  & 428.4 & 13000 &  4.0 & +0.0 & 0.1050 &  99.1 & 7737  & 40 & \multicolumn{2}{r}{14000 $\pm$ 500} &  \multicolumn{2}{l}{4.00 $\pm$ 0.25} &  \multicolumn{2}{l}{$-$1.00 $\pm$  0.75} \\
\hline
\end{tabular}
\label{results-chi2}
}
\end{table*}
\normalsize

We show in Fig.~\ref{fig:noise} the resulting stellar parameters for each noisy theoretical spectrum.  The panels include plots for each stellar parameter, where the X-axis represents the original parameter of each theoretical spectrum, and the Y-axis the resulting value obtained after performing the fit over the noisy theoretical spectra.  Fig.~\ref{fig:out_in_histo} shows the difference \mbox{$\Delta={\rm Parameter}_{\rm Output}-{\rm Parameter}_{\rm Input}$} between the output and the input parameters obtained when the $\chi^{2}$ technique is applied to MUN05 models. The width of the histogram is narrower for $\mathrm{SNR} = 200$ compared to distributions obtained for $\mathrm{SNR} = 100$ or 50, as expected. Most best-fitting results yield differences between the input and the output spectra equal to zero. The results are accurate enough for the three cases, meaning that the method provides reliable results when applied to observations with $\mathrm{SNR} \geq 50$.

We consider a {\sl failure} each case in which the input set of physical parameters of a MUN05 model is not fully recovered as output from our technique, that is, when the identity is not obtained in Fig.~\ref{fig:noise}.
From a total of 2408 models, and for $\mathrm{SNR}$ values of 50, 100, and 200, we obtain 263, 113 and 41 failures for $T_{\mathrm{eff}}$; 119, 47, and 13 failures for $\log{g}$ and 318, 178, and 90 failures for [M/H], showing that our method is more accurate to predict effective temperature and surface gravity than to estimate the metallicity. We summarise these results in Fig.~\ref{fig:failures_SNR}. For $\mathrm{SNR}$ values of 100 and 200, all stellar parameters are recovered with a failure percentage below 8\%. For spectra with $\mathrm{SNR}$ = 50, the percentage of failures when predicting $T_{\mathrm{eff}}$, $\log{g}$ and [M/H] are $\sim$ 11\%, 5\% and 13\%, respectively.  

\subsection{Estimates of stellar parameters and comparison with the literature.}
\label{results}

We have then applied the $\chi^{2}_{\mathrm{min}}$ technique to our observed MEGASTAR sample. As explained in Paper~I, when this technique is used, a likelihood or confidence level, $\mathcal{L}$ is obtained for a $\chi^{2}$ distribution. The minimum value $\chi^{2}_{\mathrm{min}}$ corresponds to the most likely model, or to the maximum value of $\mathcal{L}$, obtaining therefore the most likely stellar parameters associated to that model. Thus, we assign to each observed star the physical stellar parameters corresponding to the model that gives the minimum value of $\chi^{2}$. Columns 1 to 15 of Table~\ref{results-chi2} summarise these results (see caption for explanation of columns). 

However, when analysing the $\chi^{2}$ values, obtained from the fitting of every model of the theoretical grid to a given observed star, we usually find several models with rather similar likelihood that would still provide a good fit to our data. Following the method described in Paper~I, we select models with likelihood similar to that for $\chi^{2}_{\rm min}$, by choosing those within a region $R_{\nu,\alpha}$ \citep{avni76}, such as:
\begin{equation}
    \chi^{2}-\chi^{2}_{min}\le \Delta(\nu,\alpha),
\end{equation}
where $\nu$ is the number of free parameters (3 in our case), and $\alpha$ is the significance level. We have considered all models with $\alpha=\,0.01$, implying differences between the $\chi^{2}$ values of:
\begin{equation}
\chi^{2}-\chi^{2}_{\rm min} \le \Delta(3,0.01)=0.115,
\label{condition}
\end{equation}
which means we are selecting $N_\mathrm{m}$ models with a likelihood of 0.99 to be around the one with the minimum value of $\chi^{2}$, still yielding very good fits. From all these models, we calculate the averaged-best stellar parameters, $\langle T_{\mathrm{eff}} \rangle$, $\langle \log{g} \rangle$ and $\langle [M/H] \rangle$, as well as their dispersion. All these values are given in Table~\ref{results-chi2}.

\begin{figure*}
\centering
\includegraphics[width=0.24\textwidth,angle=-90]{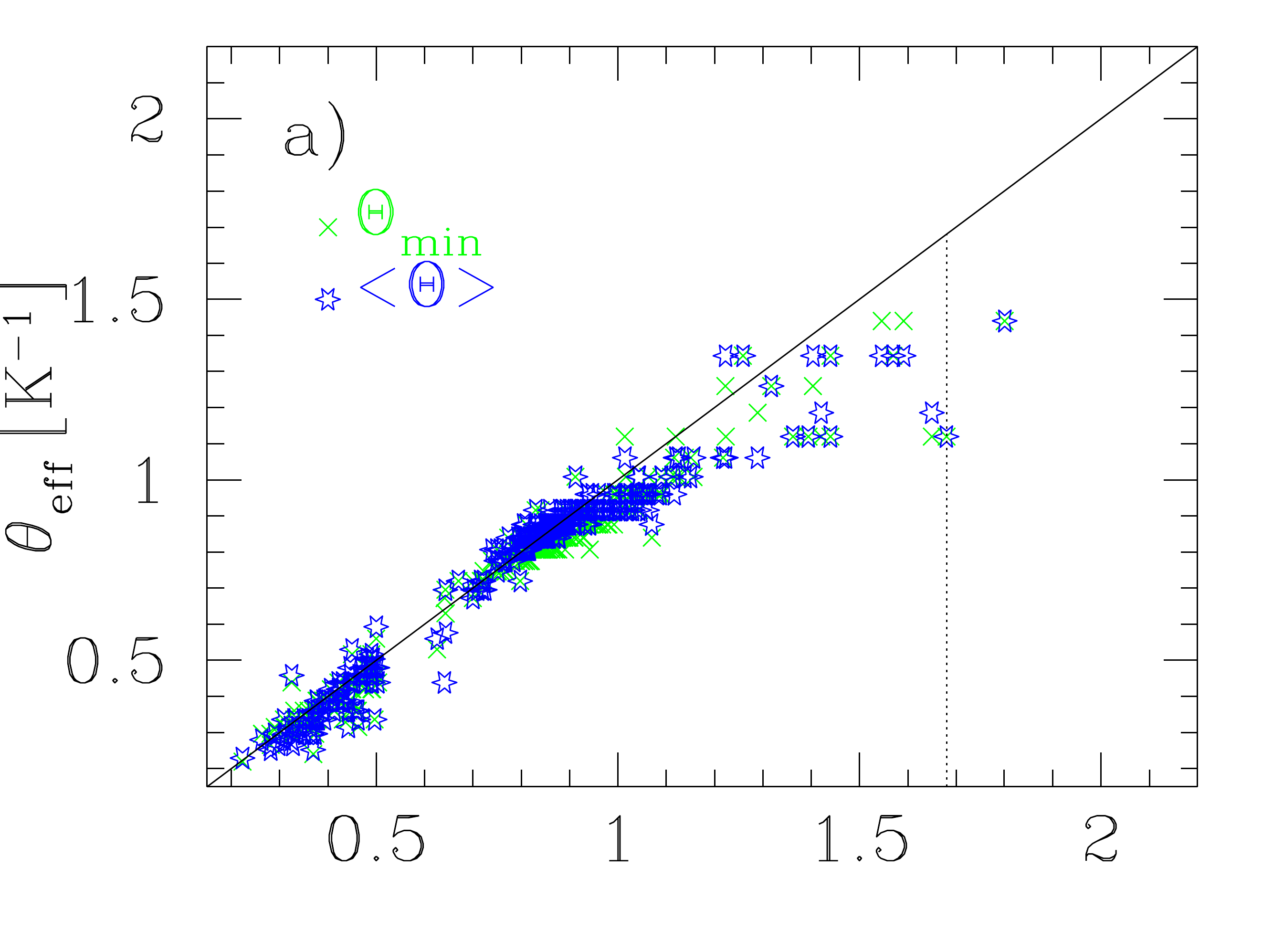}
\includegraphics[width=0.24\textwidth,angle=-90]{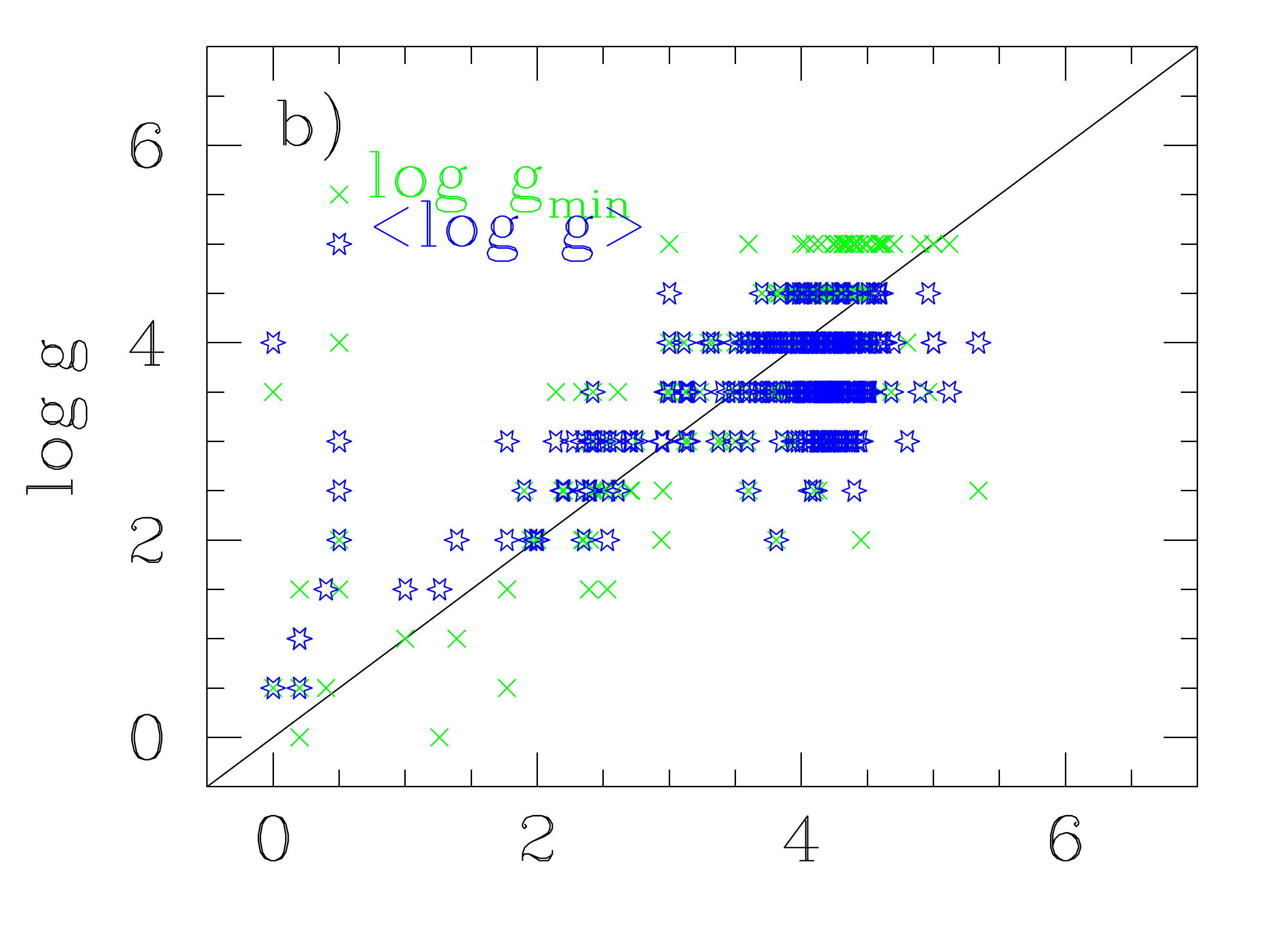}
\includegraphics[width=0.24\textwidth,angle=-90]{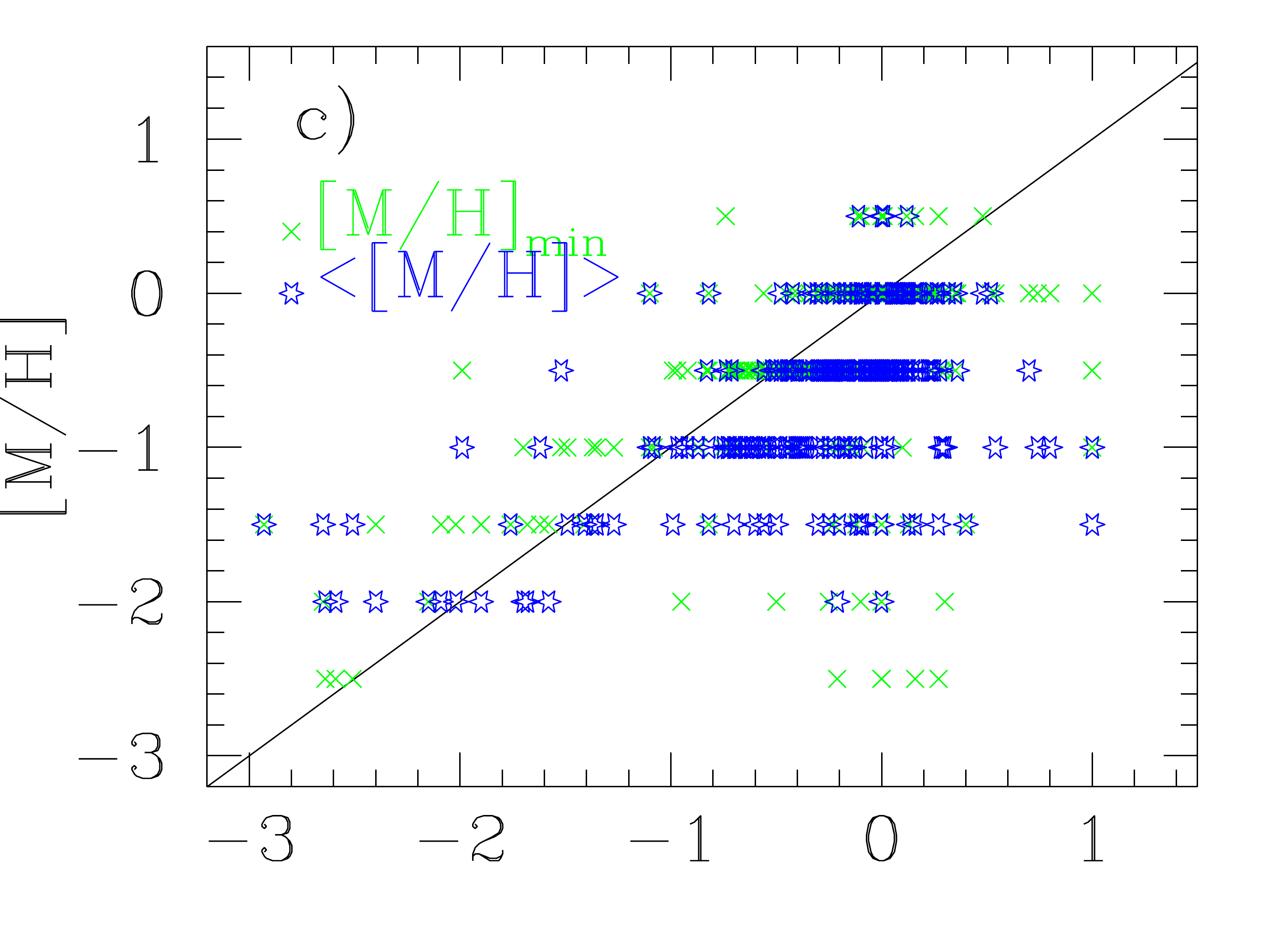}
\includegraphics[width=0.24\textwidth,angle=-90]{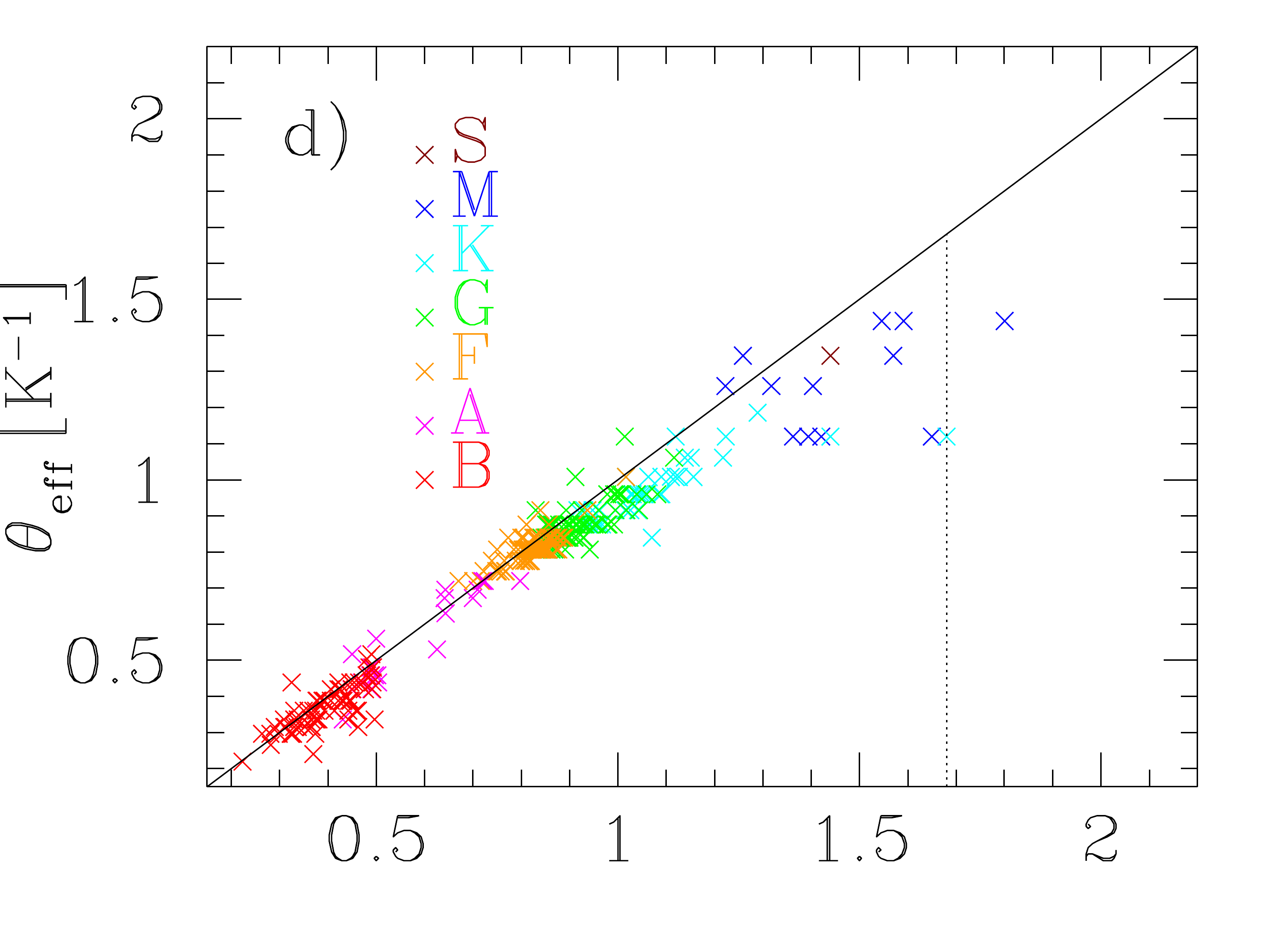}
\includegraphics[width=0.24\textwidth,angle=-90]{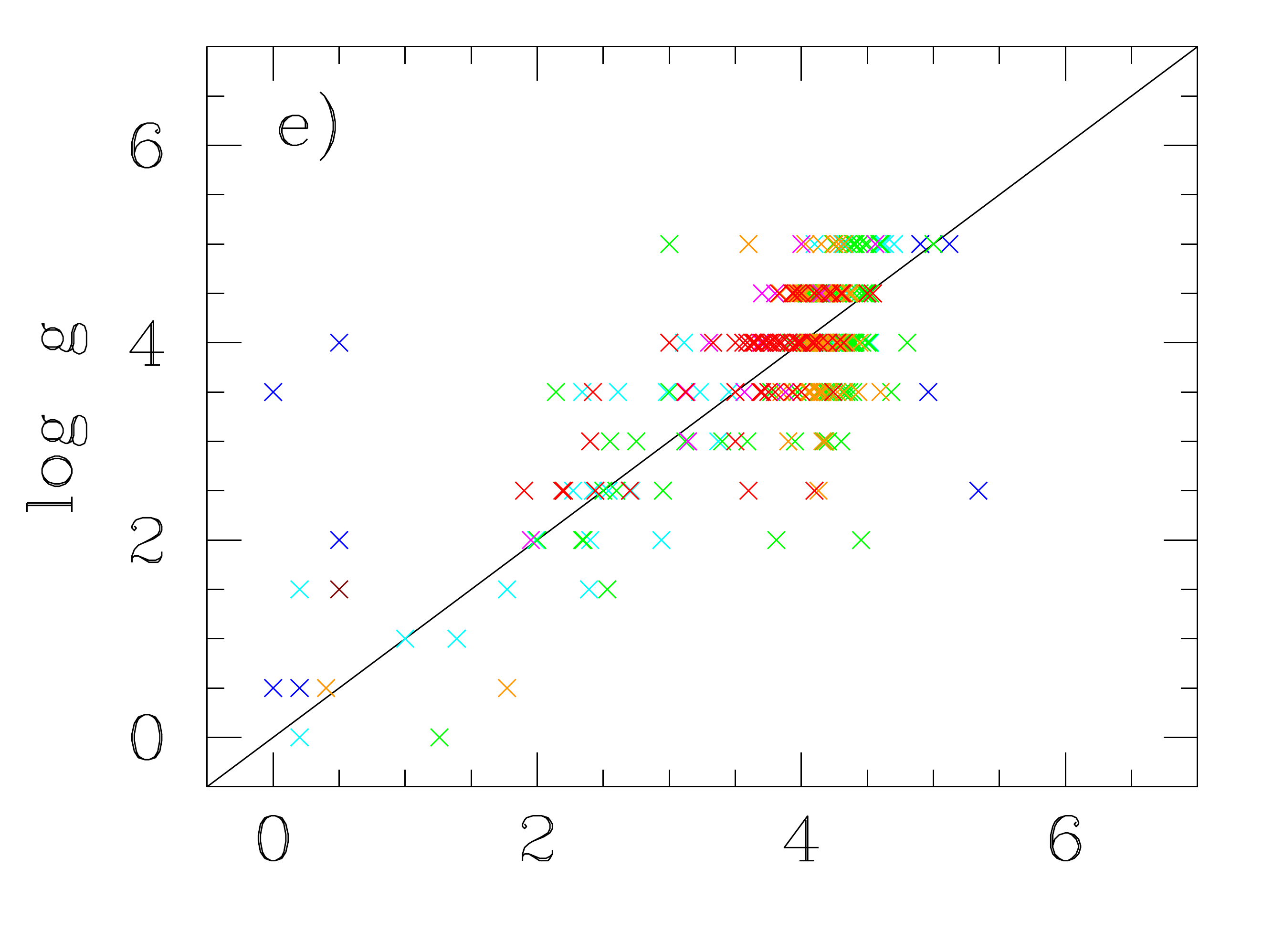}
\includegraphics[width=0.24\textwidth,angle=-90]{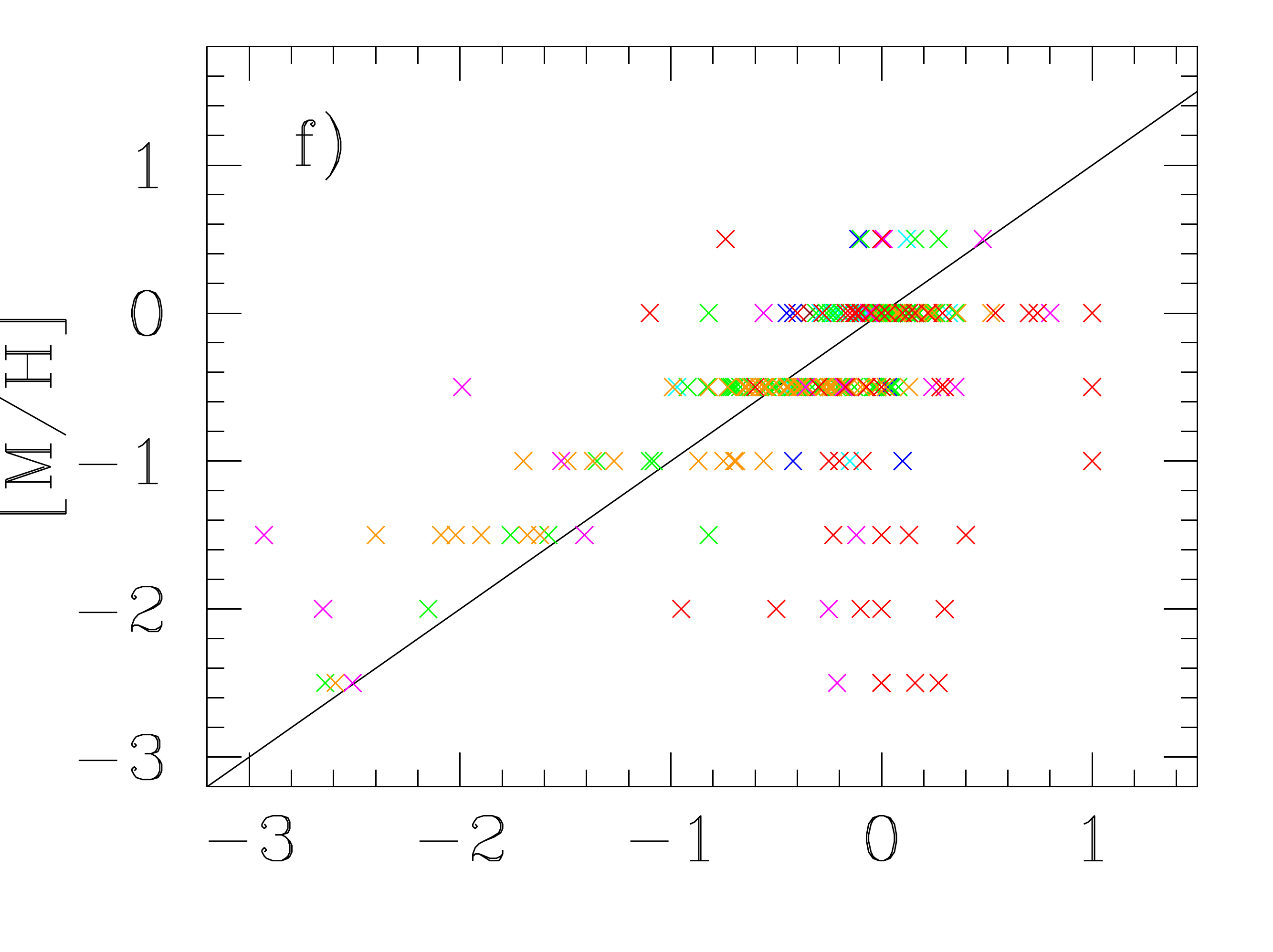}
\includegraphics[width=0.24\textwidth,angle=-90]{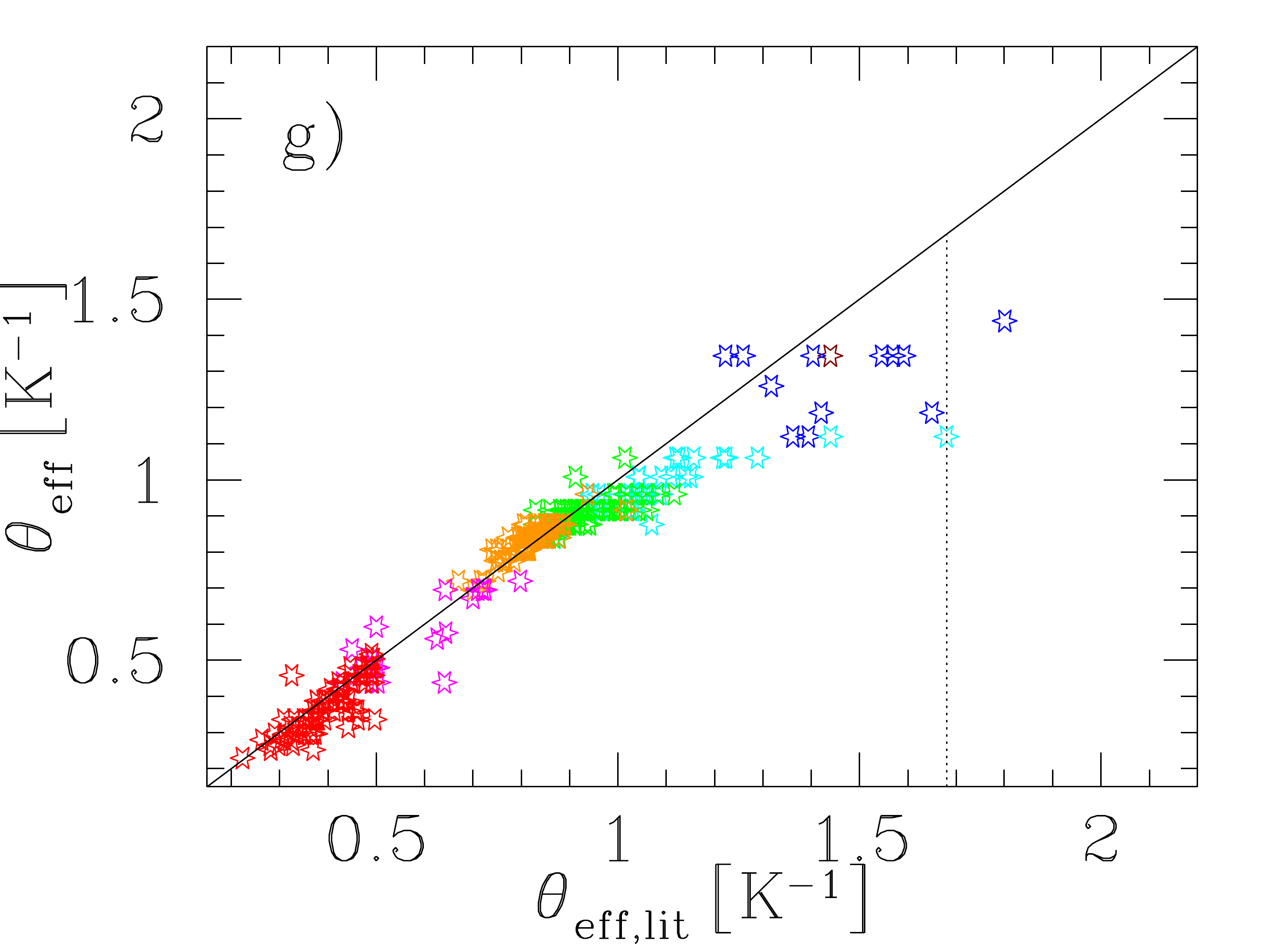}
\includegraphics[width=0.24\textwidth,angle=-90]{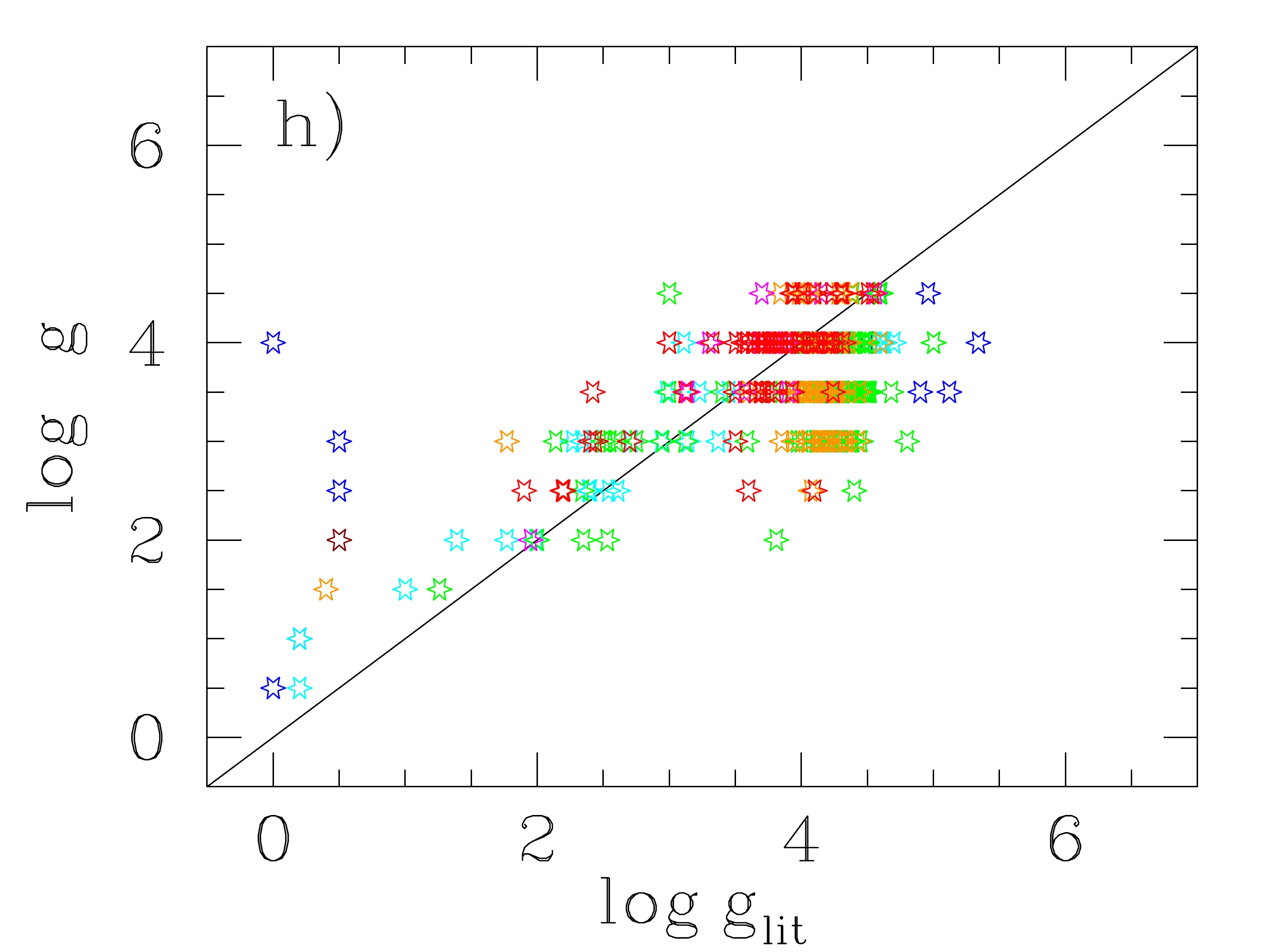}
\includegraphics[width=0.24\textwidth,angle=-90]{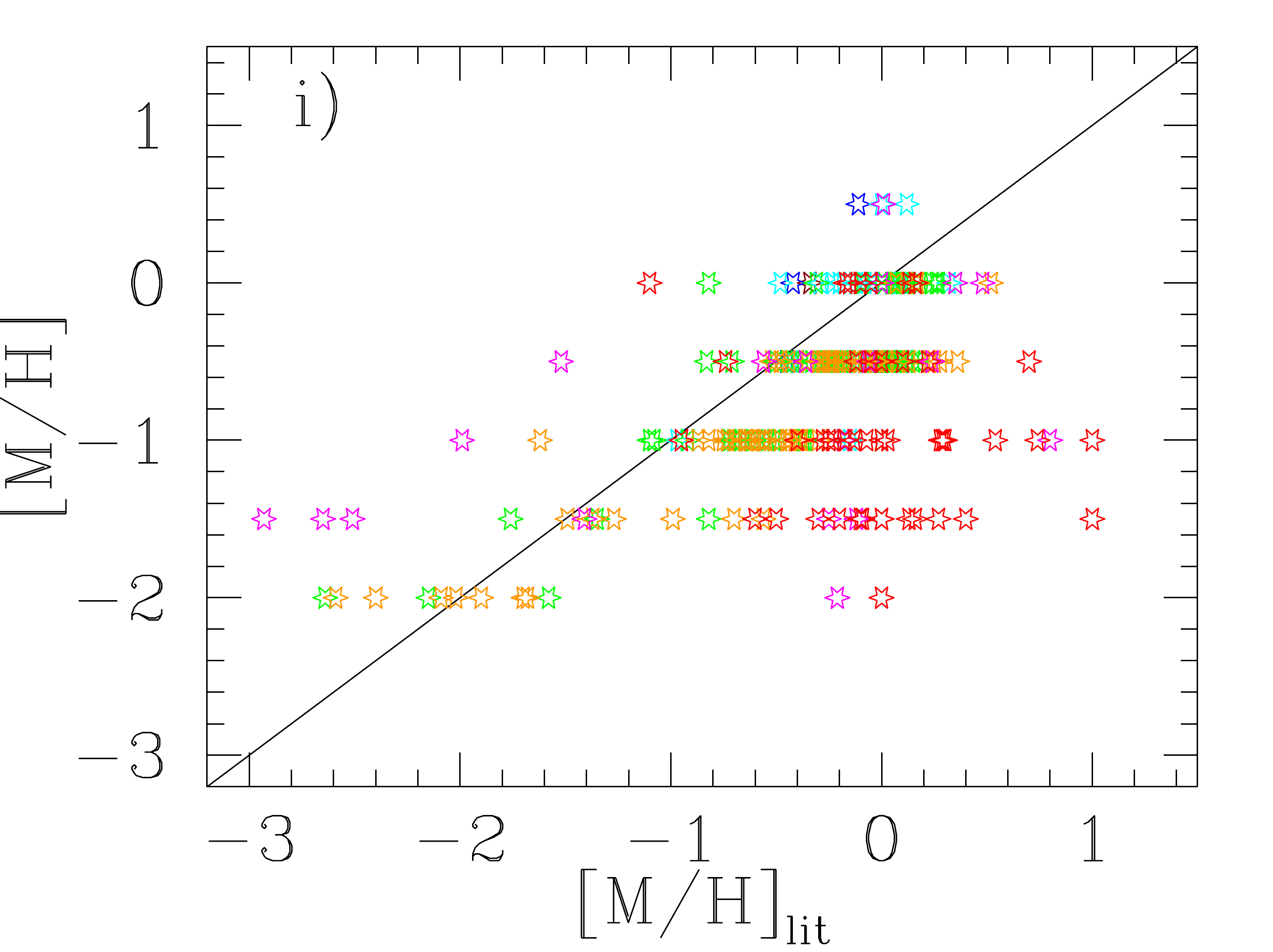}
\includegraphics[width=0.24\textwidth,angle=-90]{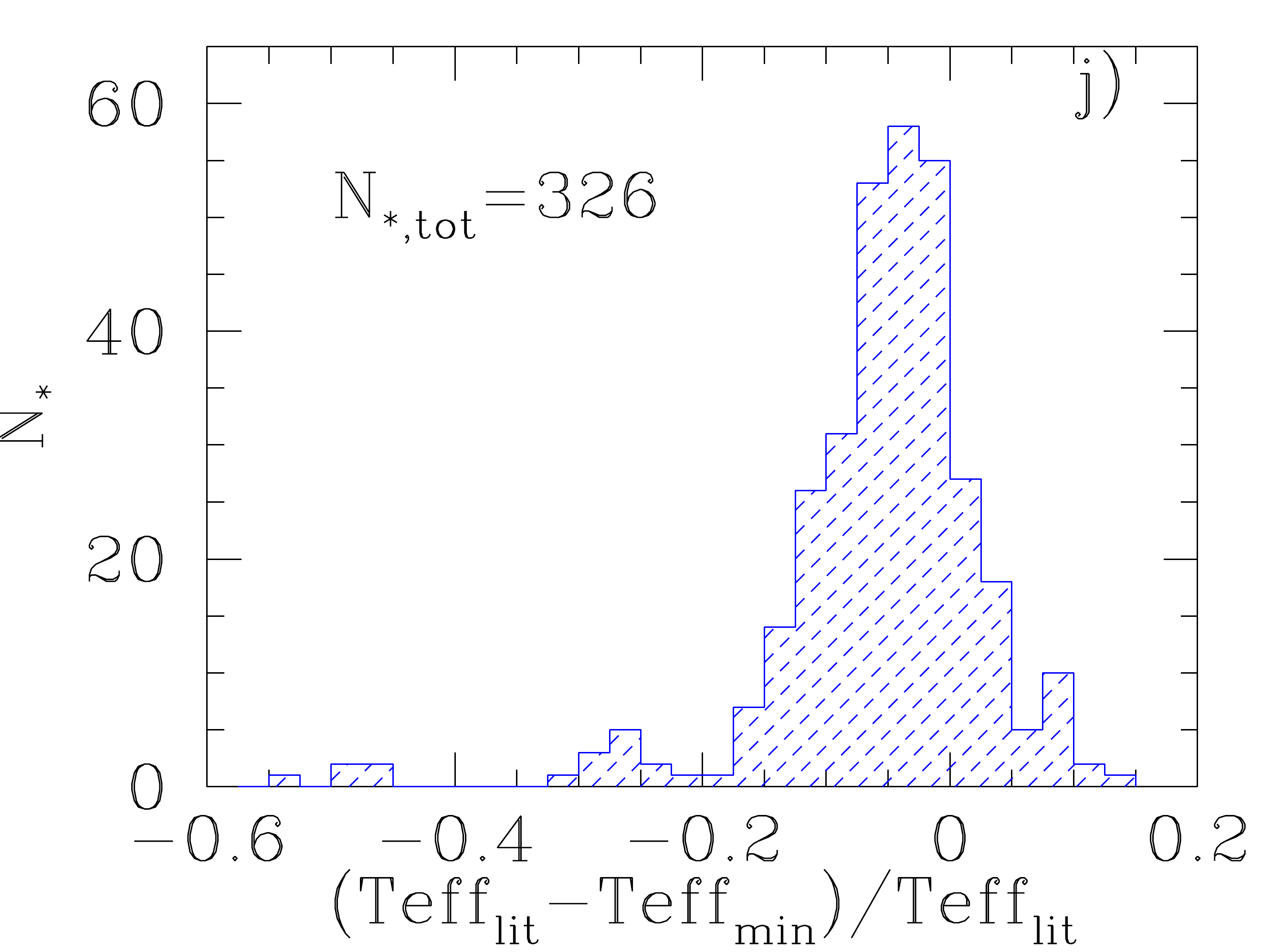}
\includegraphics[width=0.24\textwidth,angle=-90]{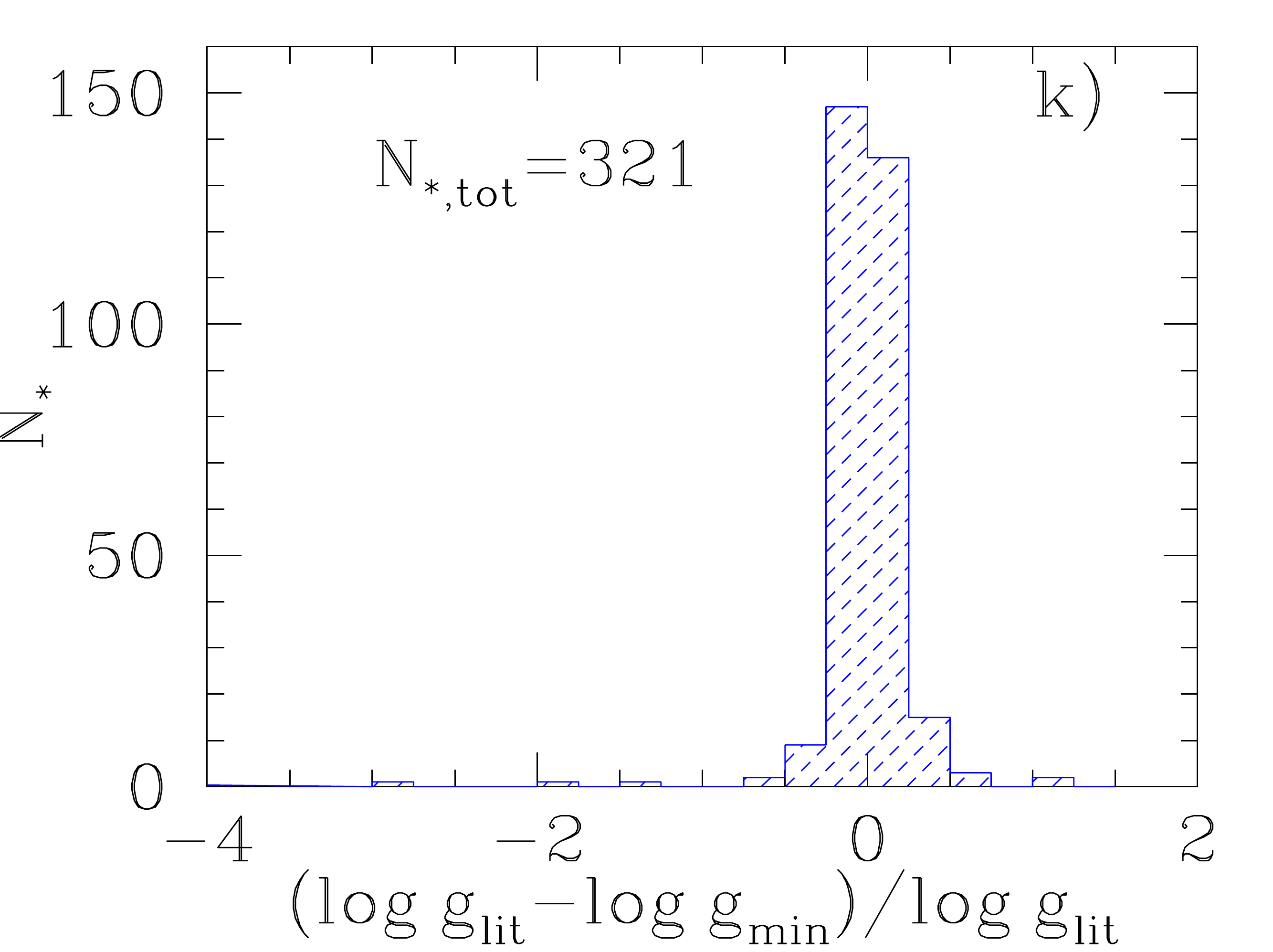}
\includegraphics[width=0.24\textwidth,angle=-90]{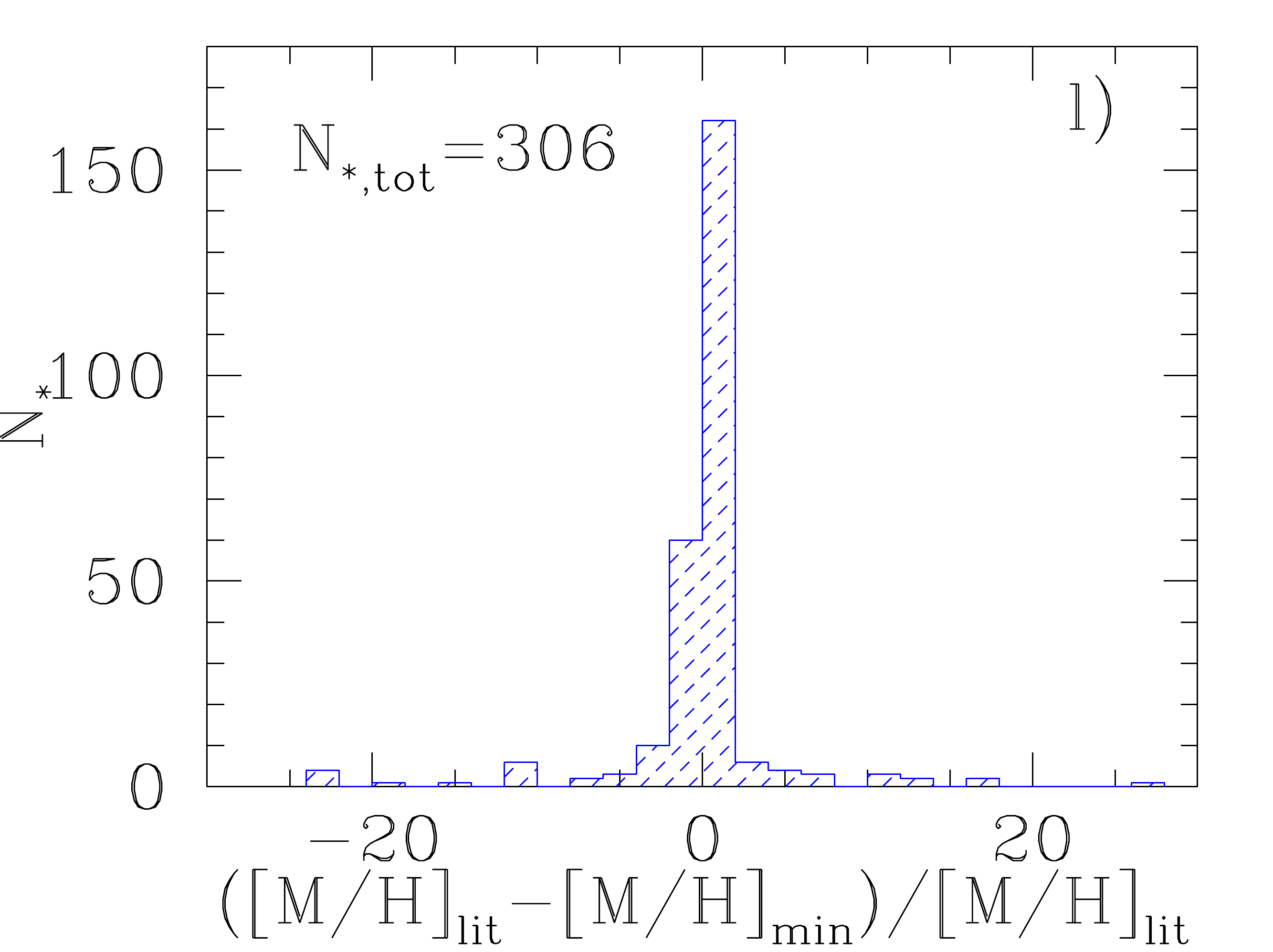}
\caption{Comparison between the resulting stellar parameters (Y-axis) and the ones obtained from the literature for each individual star (X-axis). The top panels show our results obtained with the absolute minimum $\chi^{2}$ (green crosses) and the best-averaged values around it (blue stars) compared to the literature values. The second and third panels separate both determinations (minimum $\chi^{2}$ and average in second and third rows, respectively) where our results are drawn using different symbols to identify the spectral types, as labelled in panel d).  The bottom panels show histograms of relative differences between the literature values and the assigned stellar parameters with the $\chi^{2}$ technique for stars with available values (the number of these stars is given in each plot).}
\label{fig:parest}
\end{figure*}

Fig.~\ref{fig:parest} shows the derived stellar parameters compared to those obtained from the literature (see the database presented in Paper~I and II for the corresponding references). The vertical panels represent, from left to right, $T_{\mathrm{eff}}$, displayed as $\theta=5040/T_\mathrm{eff}$, $\log{g}$ and [M/H], respectively. The diagonal line represents the identity locus. In the top panels, we show the parameters obtained with the two methods described in Section~\ref{chi2}, as a function of the corresponding reference value from the literature. The green crosses are the results directly obtained from the minimum $\chi^{2}$ model, while the blue stars correspond to the averaged values when selecting models around that minimum $\chi^{2}$. We find a good agreement in the panel corresponding to $\theta$, being only the coolest stars (the highest $\theta$ values) the ones that deviate from the identity line. This was expected as the MUN05 grid does not include any spectra for stars cooler than 3500\,K. 

\begin{figure*}
\centering
\includegraphics[width=0.42\textwidth]{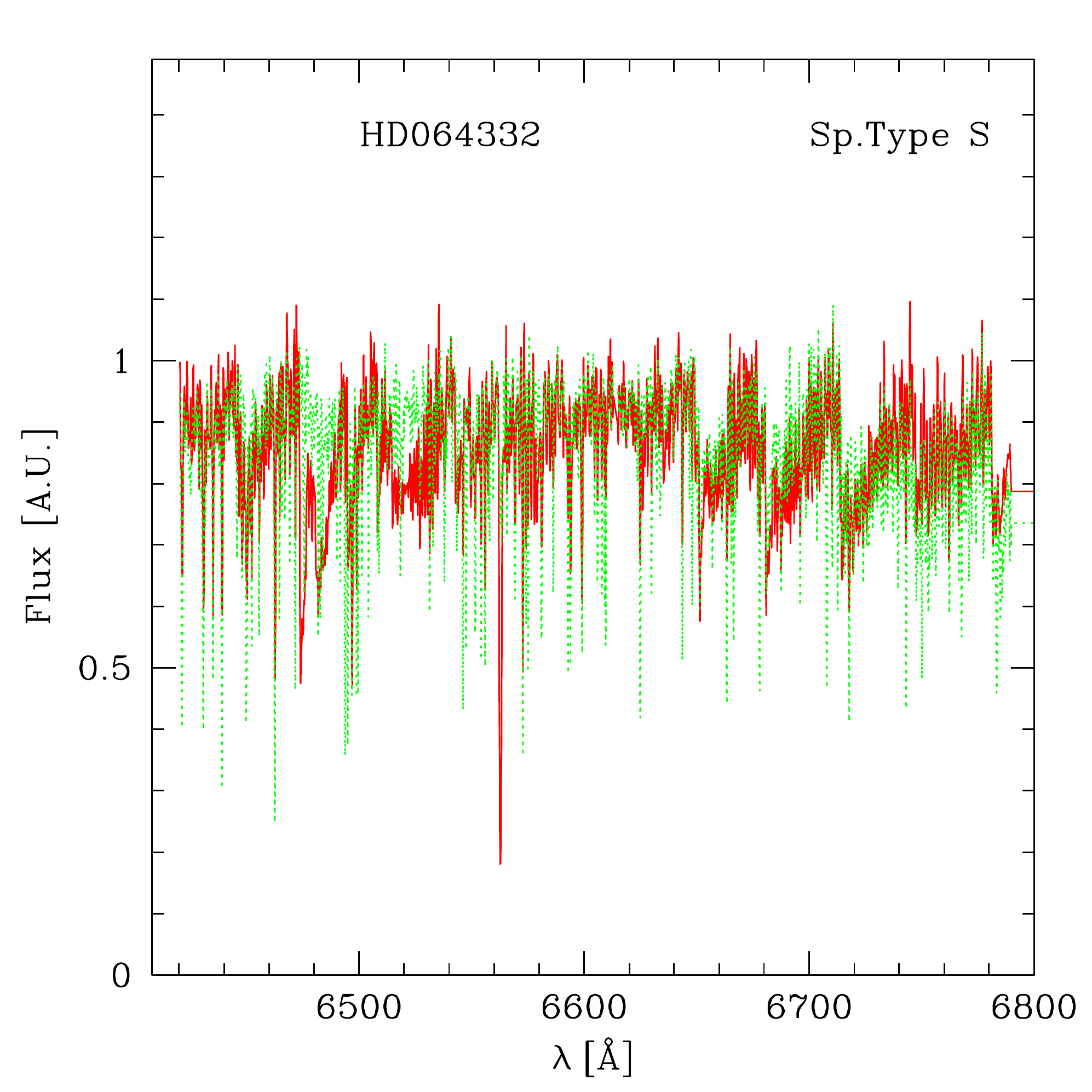}
\includegraphics[width=0.42\textwidth]{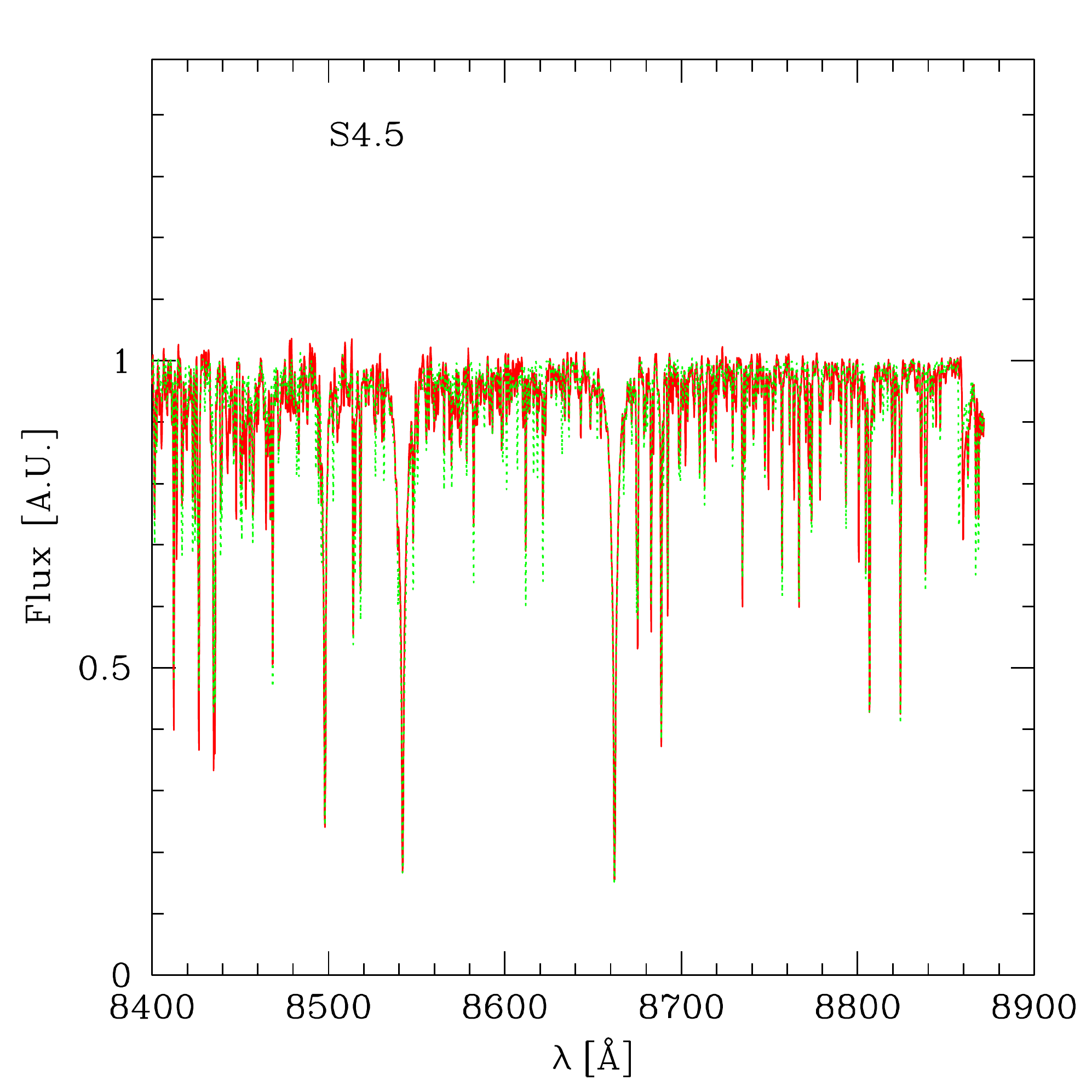}
\caption{Some examples of our sample of stellar spectra (red colour) overplotted with their corresponding best-fitting model (green colour) for spectral types S, M, K, G, F, A, and (late) B, as labelled.}
\label{examples}
\end{figure*}

Regarding $\log{g}$ and [M/H] estimates, the larger dispersion is partly due to the shorter number of possible values of $\log{g}$ and [M/H] in the model grid. MUN05 models have $T_\mathrm{eff}$ values in the range \mbox{3500 -- 50\,000\,K} with steps as small as 250\,K up to 10\,000\,K, increasing to 1000\,K for $T_\mathrm{eff}$ from 11\,000\,K to $\sim 25\,000$\,K, while the steps for both $\log{g}$ and [M/H] are 0.5\,dex. In other words, the mapping of $\log{g}$ and [M/H] in MUN05 models has less resolution than the one for $T_\mathrm{eff}$, which limits the accuracy in the stellar parameter determination. Another reason for the discrepancies is that the values from the literature come from different sources, mostly from spectra with lower resolution and wider wavelength range than in MEGASTAR.
 
In the second and third rows of Fig.~\ref{fig:parest}, we represent our results colour-coded by spectral type, as shown in panel d). In the second row panels, we display the minimum-$\chi^{2}$ results, while in the third row panels, we show the equivalent averaged-best values. There are no large differences between the minimum-$\chi^{2}$ and the averaged results derived from our method, which implies that we can just compute only the minimum -$\chi^{2}$ values for the whole MEGASTAR catalogue when available. Nevertheless, the computed error in the average values provides an indication of the precision of the method, as shown in columns 17 to 22 of Table \ref{results-chi2}.  We confirm that, as expected, M stars show the largest discrepancies with respect to the published values, especially in $T_\mathrm{eff}$ and $\log{g}$, while A and B-type stars exhibit the poorest fits in metallicity.  

We observe that there are few stars catalogued with solar abundance in the literature for which we obtain low values of [M/H]. Most of these cases ($\sim 12$) are B spectral-type stars, which are on the galactic disk and would have young ages, and therefore a solar abundance as found \citep{takeda_honda2016} is expected. The rotation velocity of these stars might be the cause of this wrong assignment.  This is an important point that deserves a particular study. We will carefully analyse this question in our next paper by Berlanas et al. (in preparation) on the study of the hot stars of MEGASTAR DR1 subsample.

To better visualise the differences between the stellar parameters from the literature, {\it lit}, and from MEGASTAR data and our minimum-$\chi^{2}$ technique, {\it min}, we represent in the bottom panels of the same Fig.~\ref{fig:parest} (as in Fig.\ref{fig:out_in_histo}), the distributions for these differences as histograms, showing the differences in $T_\mathrm{eff}$, $\log{g}$ and [M/H] at the left, middle and right panels. Most of the differences are smaller than a 10\%, finding only few values out of this range.

\subsection{Best-fitting models and observed spectra from M to late-B spectral types}
\label{cool}

In Fig.~\ref{examples} we show illustrative examples of the observed spectra and the best-fitting models derived in this work. We display the \mbox{HR-R} spectra at the left and the \mbox{HR-I} spectra at the right, for six different spectral types: S, M, K, G, F, A, and late-B (from top to bottom panels). In each panel, the observed spectrum is represented by a red line, while the best-fitting model is drawn as a green line. When the colour appears orange, it  means that the fit is quite good since both lines are overlapped. We have selected five stars of each spectral type,  except for S type for which we have only one star in DR1. The plots for the entire sample are given in  Appendix B (Supplementary Material), where we also include the averaged spectra as an additional blue-coloured line.

\begin{figure*}
\setcounter{figure}{9}
\centering
\includegraphics[width=0.42\textwidth]{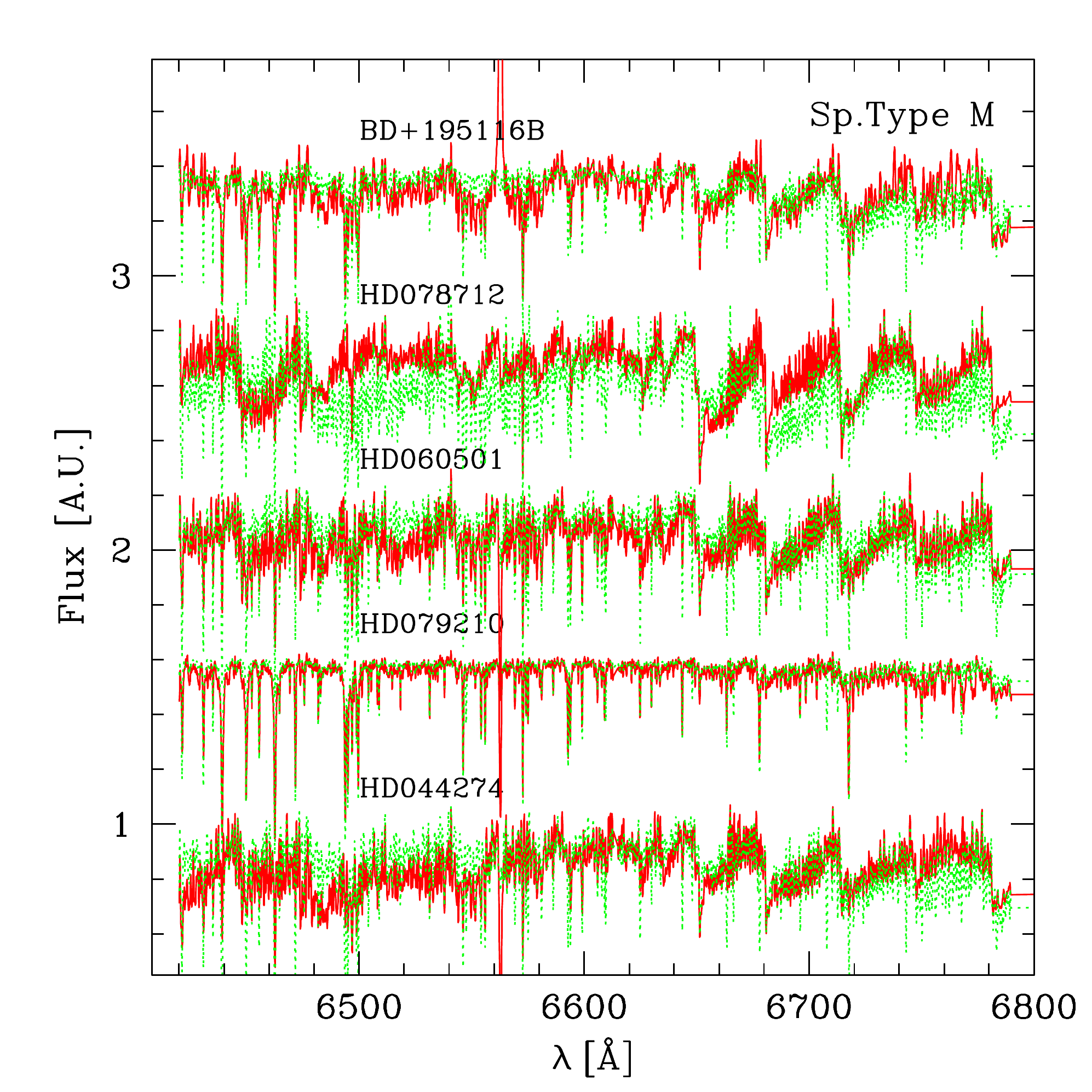}
\includegraphics[width=0.42\textwidth]{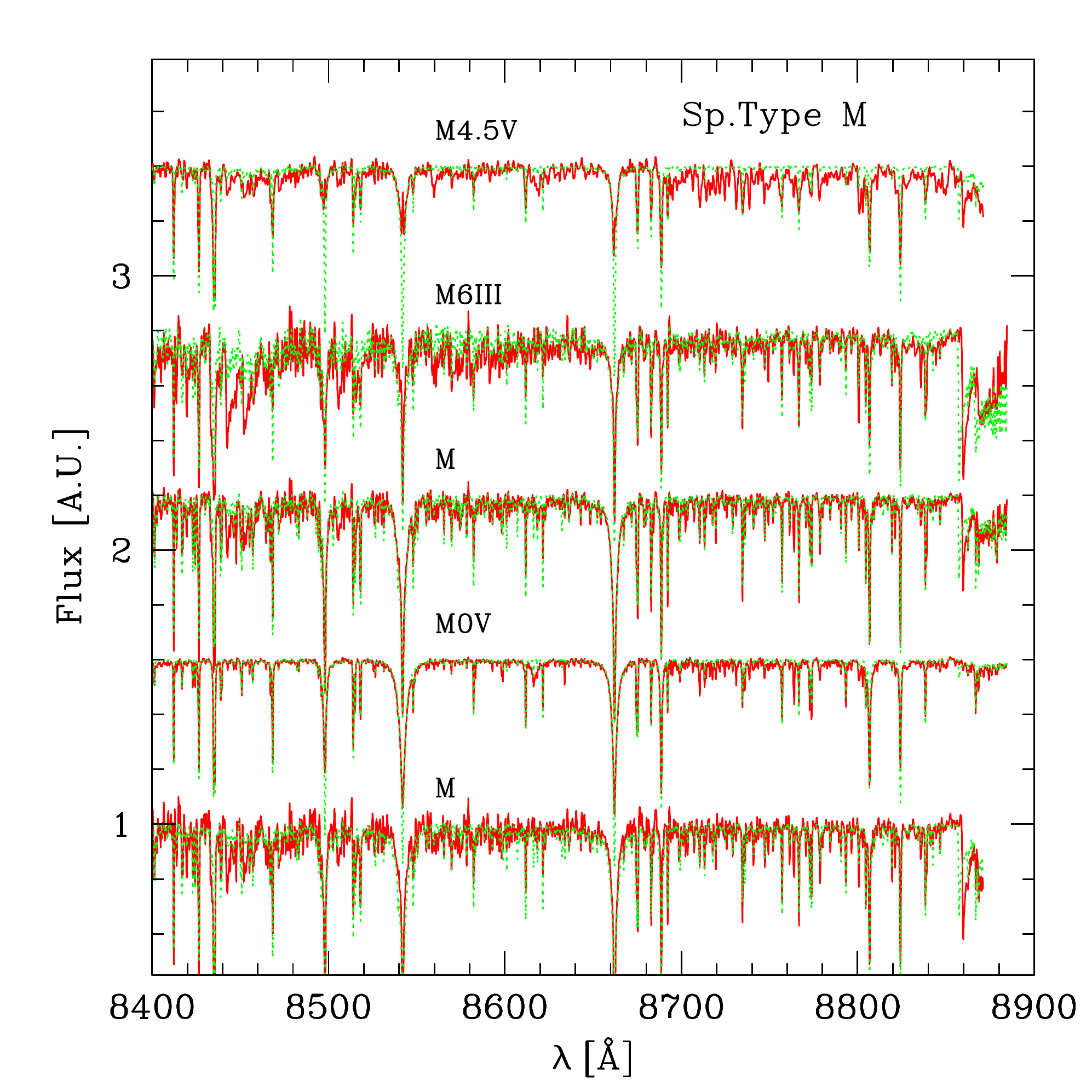}
\includegraphics[width=0.42\textwidth]{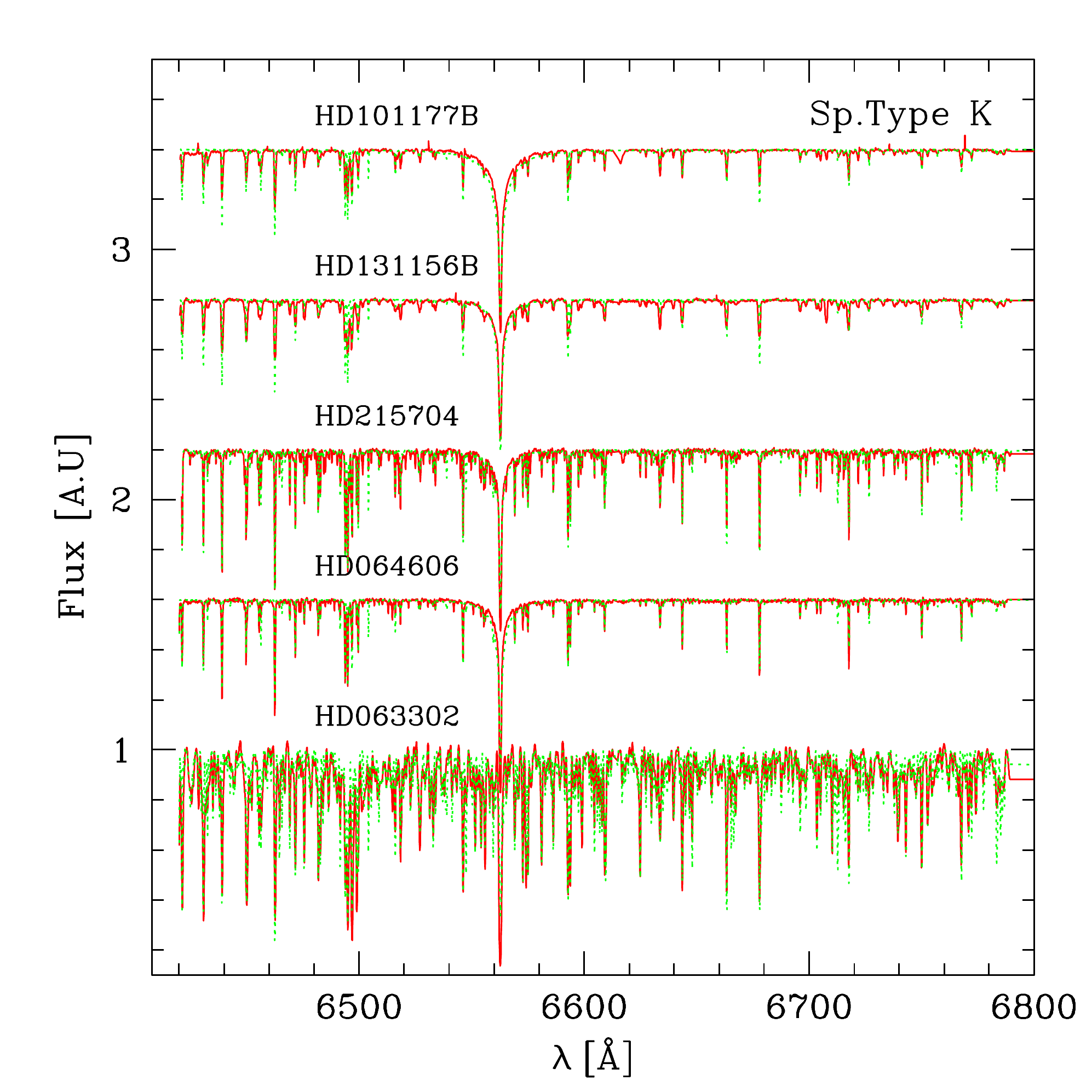}
\includegraphics[width=0.42\textwidth]{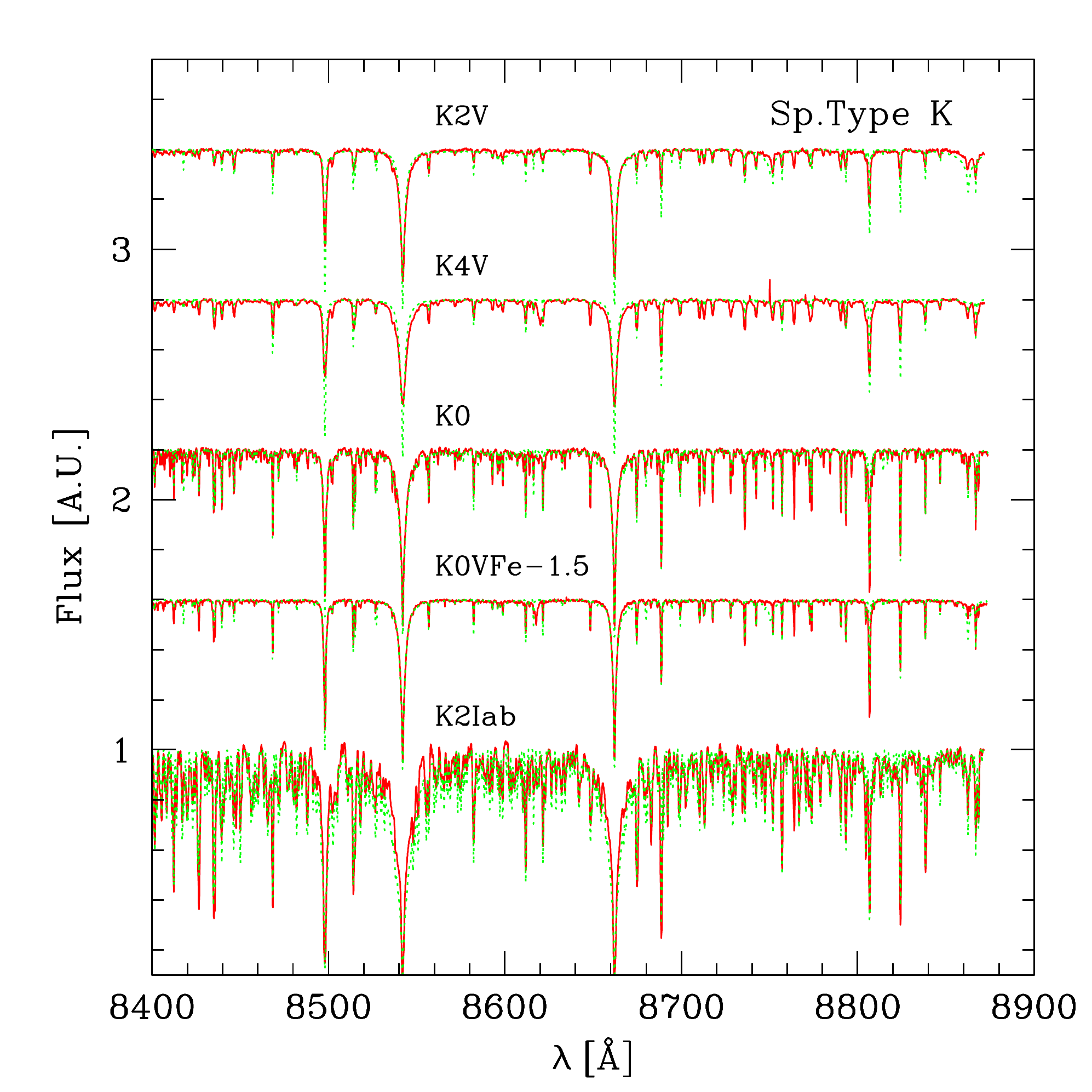}
\includegraphics[width=0.42\textwidth]{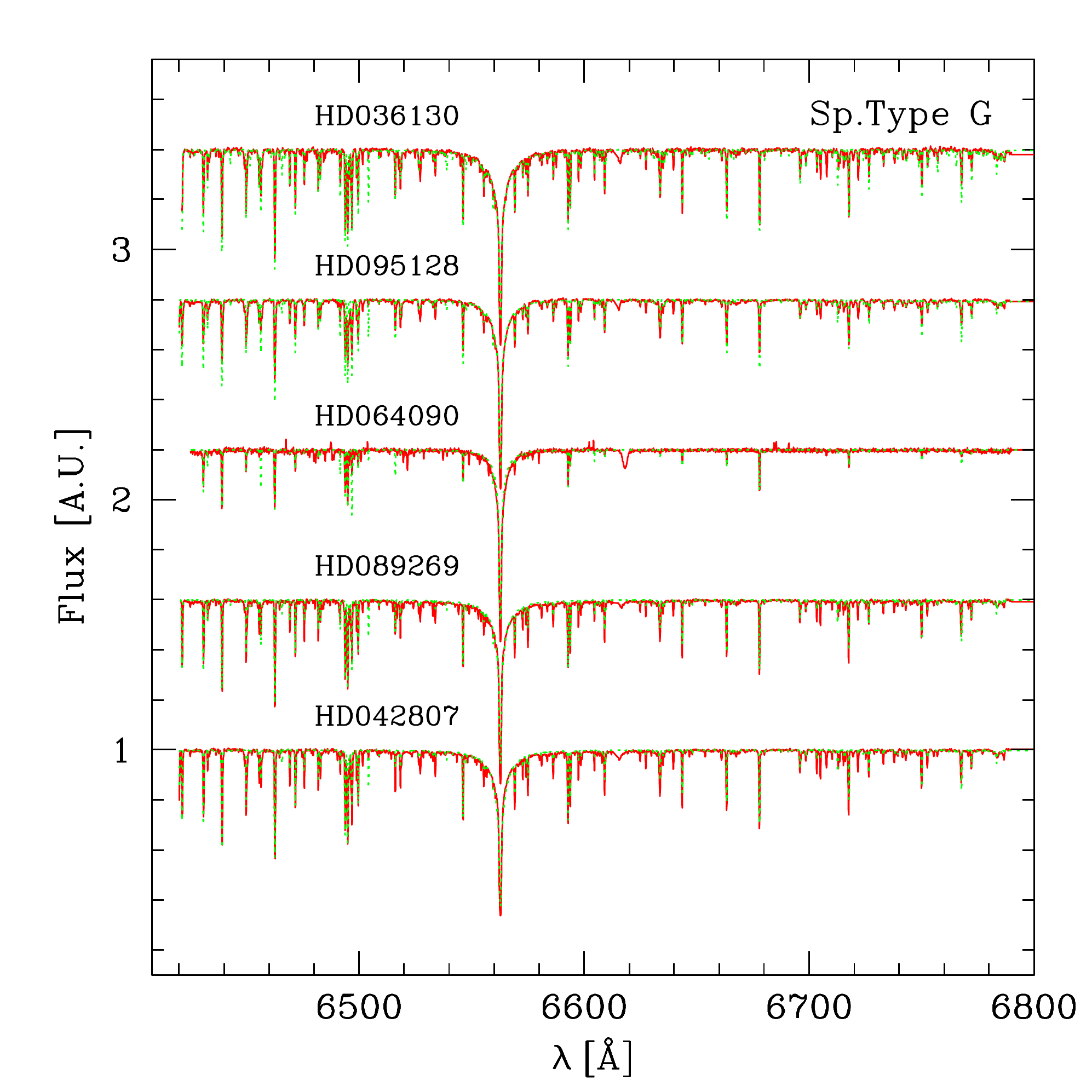}
\includegraphics[width=0.42\textwidth]{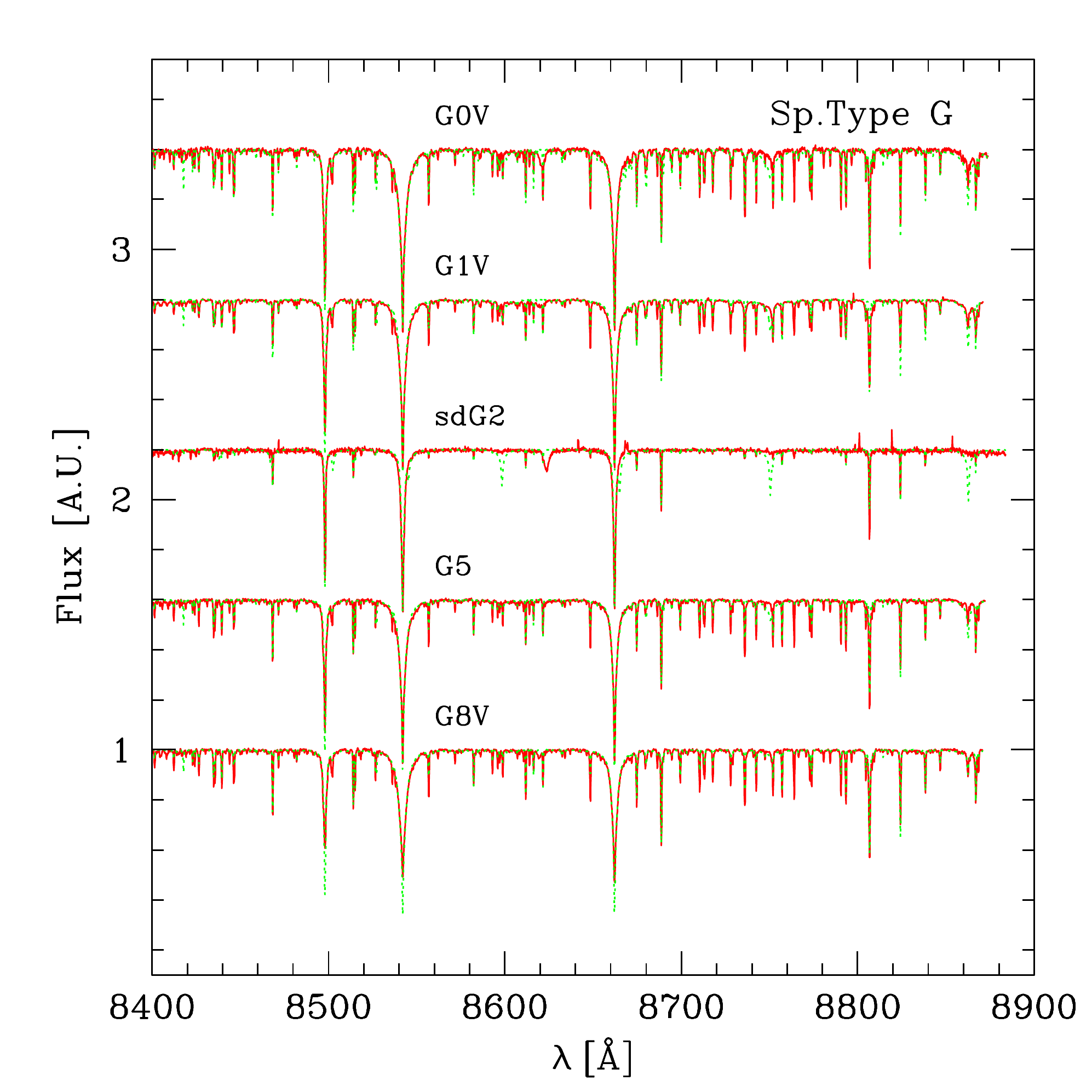}
\caption{Cont. Some examples of our sample of stellar spectra (red colour) overplotted with their corresponding best-fitting model (green colour) for spectral types S, M, K, G, F, A, and (late) B, as labelled.}
\end{figure*}
\begin{figure*}
\setcounter{figure}{9}
\centering
\includegraphics[width=0.42\textwidth]{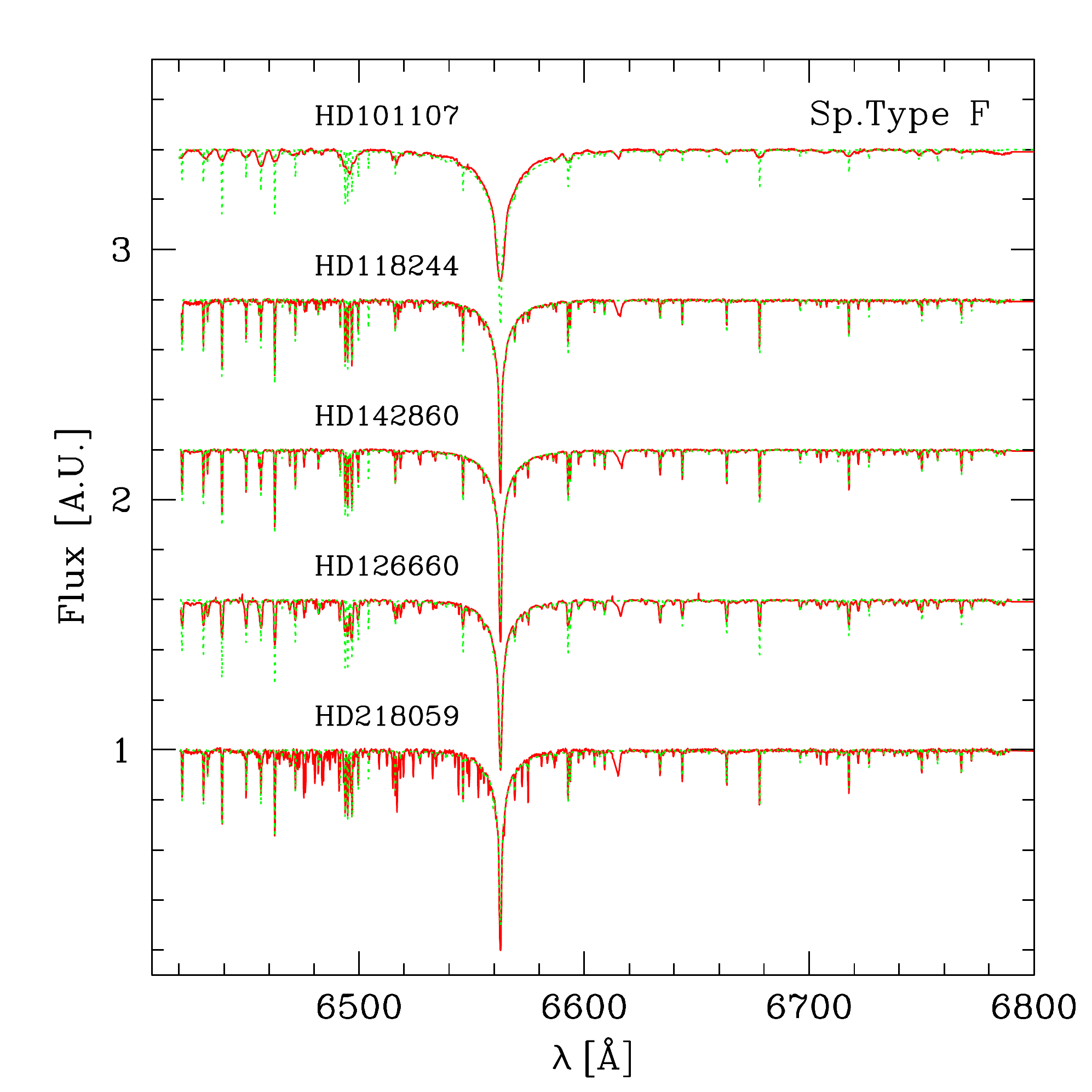}
\includegraphics[width=0.42\textwidth]{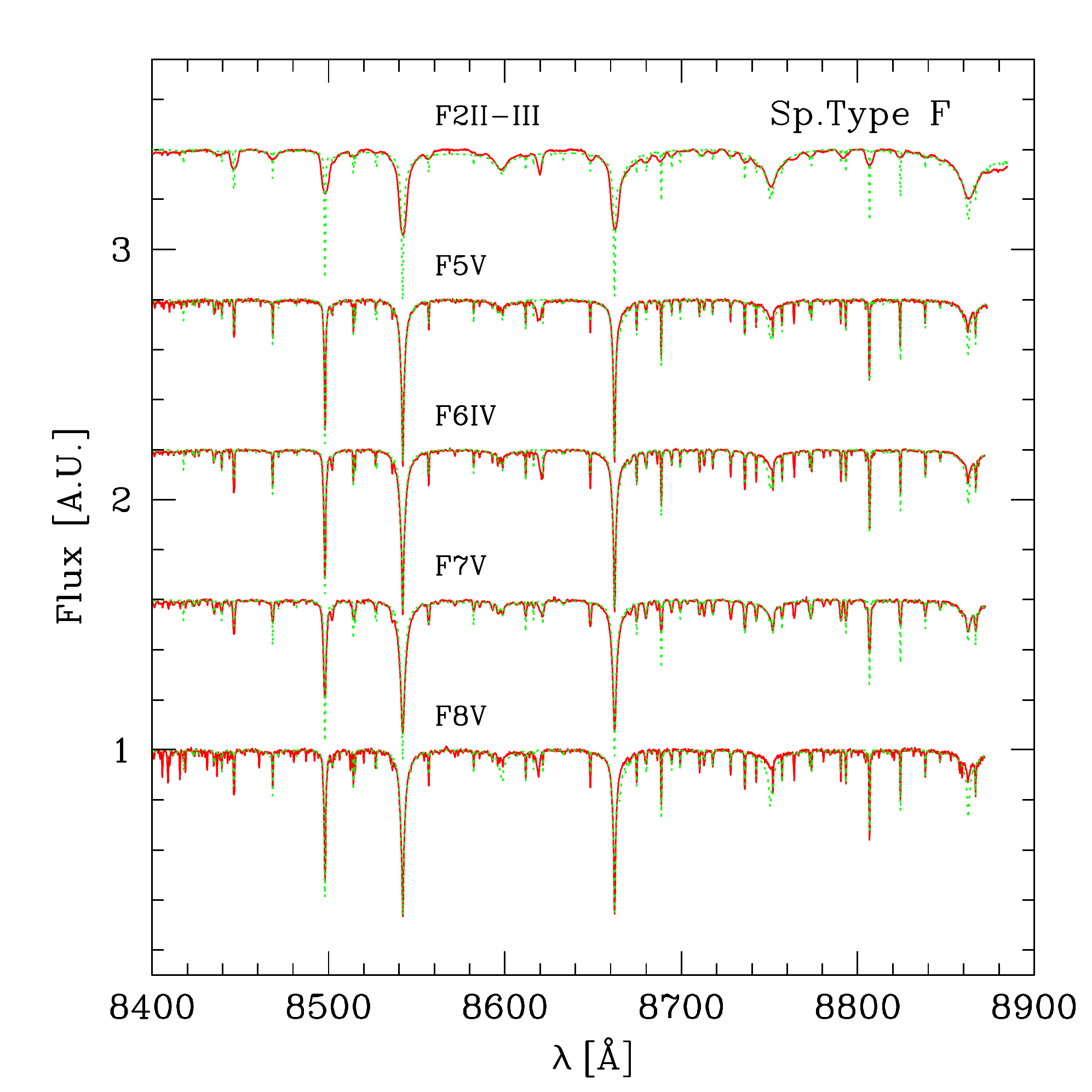}
\includegraphics[width=0.42\textwidth]{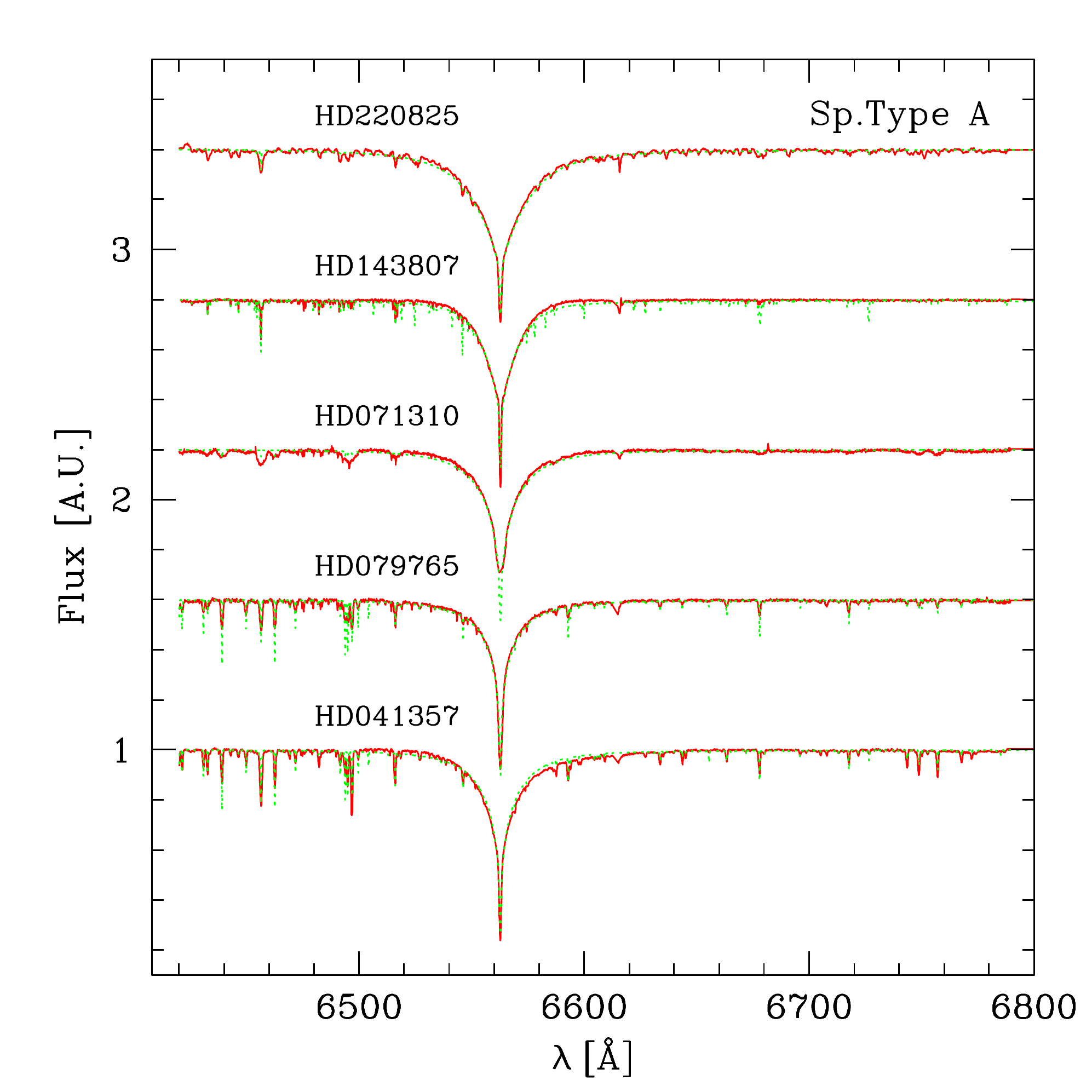}
\includegraphics[width=0.42\textwidth]{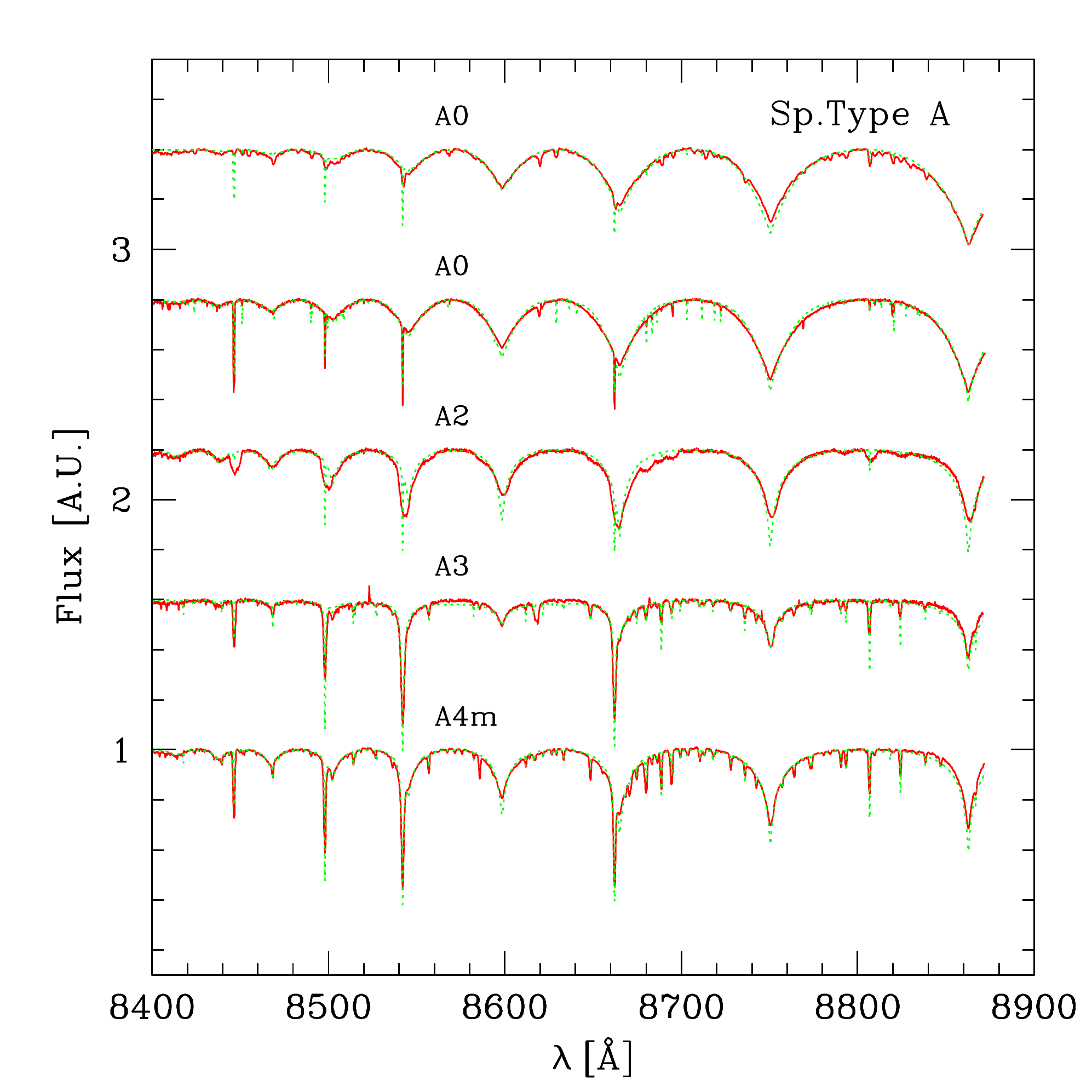}
\includegraphics[width=0.42\textwidth]{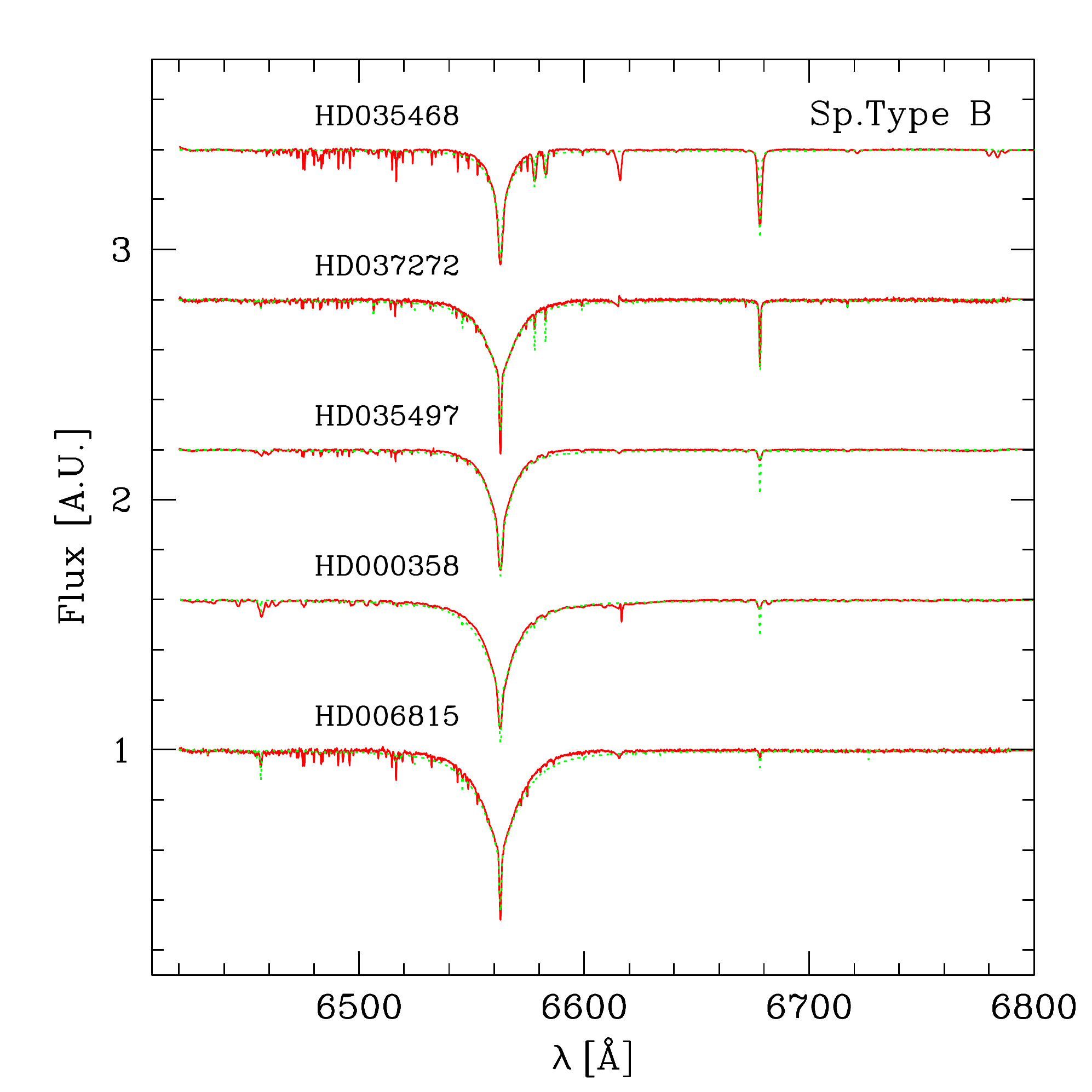}
\includegraphics[width=0.42\textwidth]{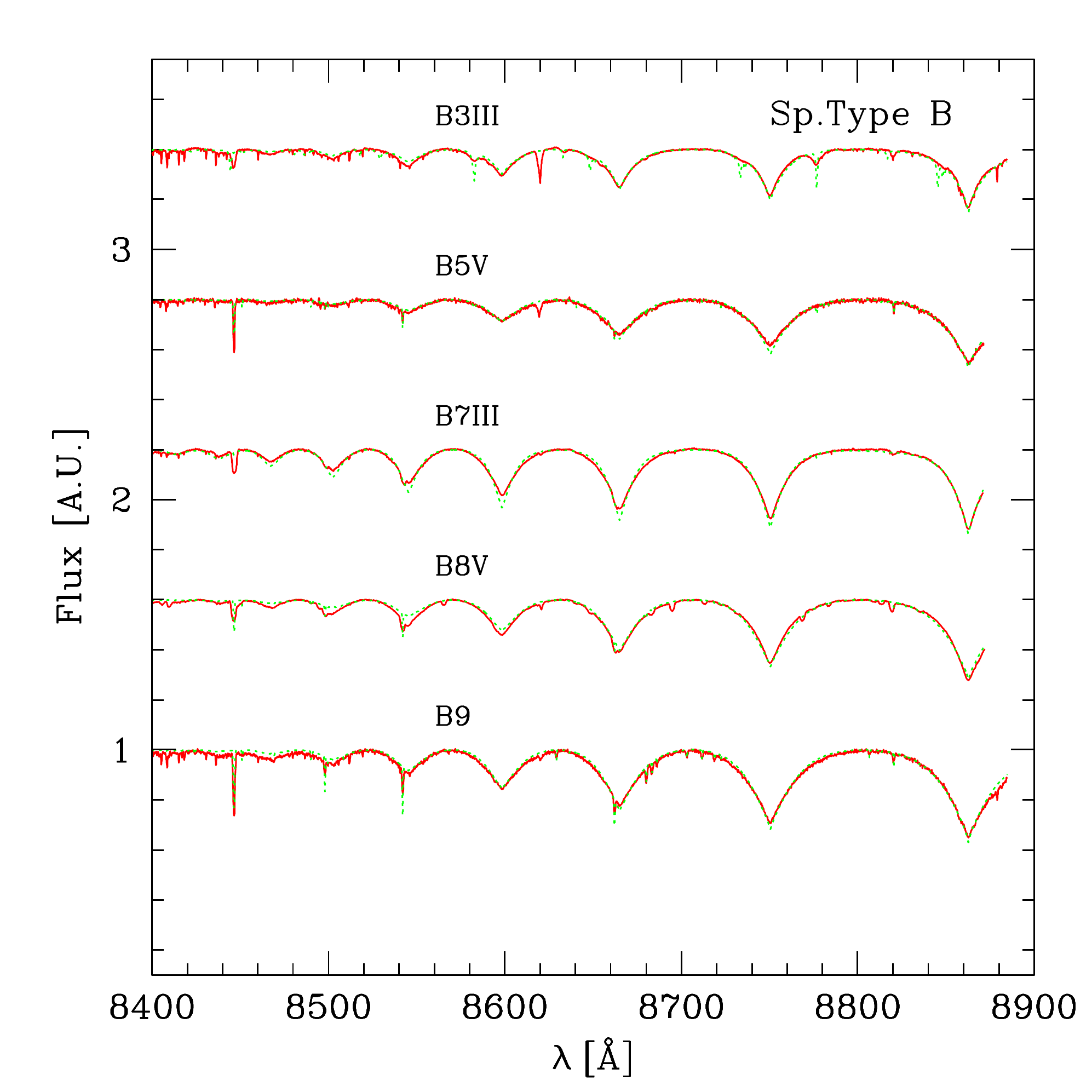}
\caption{Cont. Some examples of our sample of stellar spectra  (red colour) overplotted with their corresponding best-fitting model (green colour) for spectral types S, M, K, G, F, A, and (late) B, as labelled.}
\end{figure*}

From the spectra and their best fitting models, we emphasize:
\begin{itemize}
\item[-]The fits for stars cooler than M have strong limitations as MUN05 grid does not cover temperatures lower than 3\,500\,K. This effect is particularly evident at the blue-end of HR-R spectra. 
\item[-] For stars hotter than K, we find a clear sequence in metallicity with stronger or deeper lines in the spectra. We expect that the methodology presented in this paper will allow us to estimate the metallicity of MEGASTAR stars with a high level of accuracy.
\item[-]It is possible to discriminate peculiar stars. For example, LHS10 (see its spectra in Appendix B, Supplementary Material, page 37), is a binary star, which does not have the same spectral features as any other M-type star, preventing a good fitting with any MUN05 models. 
\end{itemize}

Finally, we note that once the spectra are rectified and stellar parameters are obtained, it is possible to tackle the task of developing an
evolutionary synthesis code. In Appendix C (Supplementary Material), we show a first prototype of a new {\sc MegaPopStar} evolutionary synthesis code, with the heritage of {\sc HR-pyPopStar} code, but using MEGASTAR library as the input spectral atlas. We have used a set of 134 stars of solar metallicity from the sample presented in this paper, using their rectified and radial velocity corrected spectra, as well as their derived $T_{\rm eff}$ and $\log{g}$ with the method described in Section ~\ref{parest}, to synthesize SSP SEDs in the MEGARA \mbox{HR-R} and \mbox{HR-I} ranges. 

We have used the \citet{Kroupa2002} Initial Mass Function (IMF) and the Padova's isochrones at $Z_{\sun}$. For each age we have computed the corresponding SED. To do this, we assign to each point of the isochrone the spectrum corresponding to the closest star, in the HR diagram, by computing the geometrical distance in the $\log{g}$ -- $T_{\rm eff}$ plane. Then, we weight each MEGASTAR spectrum according to its contribution to the luminosity of the isochrone. We have modelled SEDs for intermediate ages between 10\,Myr and 10\,Gyr. We cannot compute SEDs for younger ages as they would require O, B or WR stars, not present in the sample of this paper. 

In Appendix C (Supplementary Material), we compare the {\sc MegaPopStar} SEDs for $\log{age}$ = 7.00, 7.60, 8.00, 8.60, 9.00 and 10.00 to the ones obtained for the same ages with the theoretical synthesis models computed with {\sc HR-pyPopStar} by using MUN05 stellar spectra \citep[][]{millan-irigoyen2023}, describing the differences between both codes and discussing the main issues found as a consequence of the low number (at this moment) of MEGASTAR spectra. 

\section{Summary and Conclusions}
\label{conclusions}

We have performed a careful analysis of 349/351 spectra/stars with spectral types later than B2 from the MEGASTAR library DR1 \citep{carrasco2021}, in order to estimate their stellar parameters: effective temperature, $T_{\rm eff}$, surface gravity $\log{g}$, and metallicity, [M/H]. 

This task has required, for each \mbox{HR-R} and \mbox{HR-I} MEGASTAR spectrum of our sample, the measurement of the radial velocity and the rectification if the spectrum, what we have done with the tool {\sc bounfit} \citep{cardiel09} for all the 702 spectra. We have applied the same procedure to rectify the selected spectra of MUN05 stellar model atlas.  

We have compared the complete (\mbox{HR-R} and \mbox{HR-I}) rectified and radial velocity corrected spectra of each MEGASTAR star to the MUN05 models and, using a $\chi^{2}$-technique, we have selected the model that best reproduce the observed spectrum, assigning the stellar parameters ($T_{\rm eff}$, $\log{g}$, and [M/H]) of this model to the observed star. 

As the MEGASTAR library spectral range is very short in comparison with MUN05 spectra, we have previously tested the goodness of the stellar parameter assignment method by applying it to the rectified theoretical spectra of MUN05 atlas after adding them some noise (SNR $=$ 50, 100, and 200), with the aim of verifying that our procedure returns the same stellar parameters than the original values of the input MUN05 spectra. We find an excellent agreement in more than 90\% of the cases, with the most accurate results being logically obtained for the highest values of SNR. We have also discussed the results for each specific stellar parameter. 

The derived stellar parameters have been plotted versus the equivalent ones from the literature whenever available. We confirm a very good fit for $T_{\rm eff}$, while a certain dispersion around the identity line is obtained for $\log{g}$ and [M/H], which is somehow expected since MUN05 models are only provided at constant steps of $\pm$ 0.5\,dex in these two parameters. Also, we report under-solar metallicity assignments for few B spectral type stars with solar or super-solar metallicity in the literature, interpreting our results as a consequence of the rotation velocity of these stars, which we have not taken into account here. This point will be analyzed in a forthcoming paper by Berlanas et al. (in preparation), focused on the study of the hot stars spectra of MEGASTAR DR1, using appropriate models including rotation velocity and mass loss.

Summarising, in this work we have:
\begin{itemize}
\item[(i)] rectified the observed spectra of our stars sample.
\item[(ii)] estimated their topocentric radial stellar velocities.
\item[(iii)] estimated the stellar parameters, T$_{\rm eff}$, $\log{g}$, and [M/H], by comparing each observed spectrum with the theoretical models using the $\chi^{2}$-technique explained in Section 3.
\item[(iv)]validated our method by checking that the homogeneous set of derived physical parameters have sufficient precision to be used in an evolutionary synthesis code
\item[(v)] and, by using the rectified and radial velocity corrected spectra, we have also measured 22 EW-type indices in both \mbox{HR-R} and \mbox{HR-I} spectral intervals, and discussed their dependence on the stellar parameters in Appendix~\ref{indices}.
\end{itemize}

The most important conclusion is that we can use the spectra obtained in the  narrow \mbox{HR-R} and \mbox{HR-I} wavelength intervals to estimate the stellar parameter with good accuracy.  

The complete atlas of rectified spectra for the MEGASTAR sample presented in this work is given in the Appendix~B (Supplementary Material), along with the MUN05 best-fitting models corresponding the minimum-$\chi^{2}$ and to the averaged spectra. The page number where the spectra for each star can be found is given in Table~B1 of that appendix. 

The stellar parameters of MEGASTAR atlas are key for being used in any SSP evolutionary synthesis code. These parameters allow to assign each observed spectrum to a point in the isochrone, and then the synthetic SED may be computed by combining and weighting the spectra of the atlas. In the near future, we will include all MEGASTAR spectra in our {\sc MegaPopStar} code, based on the {\sc HR-pyPopStar} evolutionary synthesis model \citep{Millan2021,millan-irigoyen2023}, to produce SSP models. For now, in Appendix C (Supplementary Material) we present a first prototype of this new {\sc MegaPopStar} evolutionary synthesis model, using a set of 134 solar metallicity stars from the MEGASTAR sample studied in this piece of work as input spectral atlas. We have taken their rectified and velocity corrected spectra as well as the corresponding stellar atmospheric parameters obtained in Section~\ref{results} to synthesise SSPs SEDs in the MEGARA \mbox{HR-R} and \mbox{HR-I} intervals. These first computed SEDs are only valid for $Z_{\sun}$, and ages older than 10\,Myr, when only stars with spectral types later than B2 are present. 

MEGASTAR is a work in progress and we are preparing a second release with almost 1000 stars (some of them covering a parameter space missed in DR1) that, once studied in detail and their stellar parameters derived, will be used as part of the input atlas for {\sc MegaPopStar} evolutionary synthesis models.

\section{Acknowledgements}
This work is part of the grants I+D+i RTI2018-096188-B-I00 and PID2019-107408GB-C41, which have been funded by Ministerio de Ciencia e Innovación and Agencia Estatal de Investigaci{\'o}n  (MCIN/AEI/10.13039/501100011033). It has been also partially funded by FRACTAL, INAOE and CIEMAT.  S.R.B. thanks the financial support by MCIN/AEI/10.13039/501100011033 (contract FJC 2020-045785-I) and NextGeneration EU/PRTR and MIU (UNI/551/2021) through a {\sl Margarita Salas} grant.

This work is based on observations made with the Gran Telescopio Canarias (GTC), installed in the Spanish Observatorio del Roque de los Muchachos, in the island of La Palma. This work is based on data obtained with MEGARA instrument, funded by European Regional Development Funds (ERDF), through {\it Programa Operativo Canarias FEDER 2014-2020}. The authors thank the support given by Dr$.$ Antonio Cabrera and Dr$.$ Daniel Reverte, GTC Operations Group staff, during the preparation and execution of the observations at the GTC. 

This research made use of Astropy \citep{Astropy18}, a community-developed core Python package for Astronomy. This research has made use of the SIMBAD database and the VizieR catalogue access tool, CDS, Strasbourg, France \mbox{(DOI: 10.26093/cds/vizier)}. The original description of the VizieR service was published in \mbox{A\&AS 143, 23}. This work has made use of data from the European Space Agency (ESA) mission {\it Gaia} (\url{https://www.cosmos.esa.int/gaia}), processed by the {\it Gaia} Data Processing and Analysis Consortium (DPAC, \url{https://www.cosmos.esa.int/web/gaia/dpac/consortium}). Funding for the DPAC has been provided by national institutions, in particular the institutions participating in the {\it Gaia} Multilateral Agreement.

We are very grateful to the reviewer whose comments and suggestions have helped to improve the manuscript.

\section{Data Availability}
\label{data}

\begin{itemize}

\item[I] The rectified spectra, with the radial velocities, the measured equivalent widths of a set of spectral lines and the stellar parameters will be publicly available in the web page of the MEGASTAR stellar library \footnote{\url{https://www.fractal-es.com/megaragtc-stellarlibrary/private/home}, access with username {\it{public}} and password {\it{Q50ybAZm}.}}.

\item[II] The description of the 22 columns of  Table 2, available only online, is the following:
    \begin{enumerate}
    \item[1] Star name.
    \item[2] Effective temperature, $T_{\mathrm{eff}}$, in K units, as given in the literature.
    \item[3] Surface gravity, $\log{g}$, as given in the literature.
    \item[4] Metallicity, [M/H], as given in the literature.
    \item[5] Spectral type.
    \item[6] Radial velocity, in km\,s$^{-1}$ units.
    \item[7] Error of in the radial velocity determination, in km\,s$^{-1}$ units.
    \item[8] Signal-to-noise ratio, SNR, in HR-R spectrum
    \item[9] Signal-to-noise ratio, SNR, in HR-I spectrum.
    \item[10] Effective temperature obtained from the minimum $\chi^{2}, $T$_{\rm eff,min}$, in K units.
    \item[11] Surface gravity obtained from the minimum $\chi^{2}$, $\log{g}_{min}$.
    \item[12] Metallicity obtained from the minimum $\chi^{2}$, $\mathrm [M/H]_{min}$.
    \item[13] The minimum reduced $\chi^{2}$.
    \item[14] The maximum likelihood $\mathcal{L} $ corresponding the minimum $\chi^{2}$
    \item[15] The number of wavelength used in the fit model-data.
    \item[16] The number of models around the minimum value model with probability within a 0.05 of difference with $\mathcal{L} $ .
    \item[17] Effective temperature, as averaged of the valid models, <T$_{\rm eff}$>, in K units.
    \item[18] Associated errors to values in column 17.
    \item[19] Surface gravity, , as averaged of the valid models, <$\log{g}$>.
    \item[20] Associated  errors to values in column 19.
    \item[21] Metallicity, , as averaged of the valid models, <[M/H]>. 
    \item[22] Associated errors to values in column 21 .
    \end{enumerate}
    
\item[III] Columns description of Table A3. See example in Appendix~\ref{indices}. It has a total of 45 columns that represent:
    \begin{enumerate}    
    \item[1] The name of the star.
    \item[2] The equivalent width for \ion{Ca}{i}\,6439\,\AA.
    \item[3] Error associated to values in column 2.
    \item[4] The equivalent width for \ion{Ca}{i}\,6439w\,\AA.
    \item[5] Error associated to values in column 4.
    \item[6] The equivalent width for \ion{Fe}{i}\,6463\,\AA.
    \item[7] Error associated to values in column 6.
    \item[8] The equivalent width for \ion{Fe}{i}\,6463w\,\AA.
    \item[9] Error associated to values in column 8.    
    \item[10] The equivalent width for \ion{Ca}{i}\,6494\,\AA.
    \item[11] Error associated to values in column 10.     
    \item[12] The equivalent width for \ion{Fe}{i}\,6495\,\AA.
    \item[13] Error associated to values in column 12.  
    \item[14] The equivalent width for \ion{Ca}{i}\,6494+\ion{Fe}{i}\,6495\,\AA.
    \item[15] Error associated to values in column 14.      
    \item[16] The equivalent width for \ion{Fe}{i}\,6593\,\AA.
    \item[17] Error associated to values in column 16.  
    \item[18] The equivalent width for \ion{Fe}{i}\,6594\,\AA.
    \item[19] Error associated to values in column 18.      
    \item[20] The equivalent width for \ion{Fe}{i}\,6593+6594\,\AA.
    \item[21] Error associated to values in column 20.
    \item[22] The equivalent width for \ion{Fe}{i}\,6717\,\AA.
    \item[23] Error associated to values in column 22.  
    \item[24] The equivalent width for H$\alpha_{GON}$\,\AA.
    \item[25] Error associated to values in column 24.  
    \item[26] The equivalent width for CaT1$_{CEN}$\,\AA.
    \item[27] Error associated to values in column 26.      
    \item[28] The equivalent width for CaT2$_{CEN}$\,\AA.
    \item[29] Error associated to values in column 28.  
    \item[30] The equivalent width for CaT3$_{CEN}$\,\AA.
    \item[31] Error associated to values in column 30.    
    \item[32] The equivalent width for \ion{Mg}{i}\,\AA.
    \item[33] Error associated to values in column 32.   
    \item[34] The equivalent width for Pa1$_{CEN}$\,\AA.
    \item[35] Error associated to values in column 34.      
    \item[36] The equivalent width for Pa2$_{CEN}$\,\AA.
    \item[37] Error associated to values in column 36.  
    \item[38] The equivalent width for Pa3$_{CEN}$\,\AA.
    \item[39] Error associated to values in column 38.   
    \item[40] The equivalent width for CaT1$_{MEG}$\,\AA.
    \item[41] Error associated to values in column 40.      
    \item[42] The equivalent width for CaT2$_{MEG}$\,\AA.
    \item[43] Error associated to values in column 42.  
    \item[44] The equivalent width for CaT3$_{MEG}$\,\AA.   
    \item[45] Error associated to values in column 44. 
    \end{enumerate}
    
\item[IV]   The complete atlas of rectified spectra for the MEGASTAR subsample presented in this work is given in the Appendix B (Supplementary Material), along with the MUN05 best-fitting models corresponding to the minimum-$\chi^{2}$ and to the averaged values. The page number where the spectra for each star can be found is given in Table~B1 of that appendix. 

\item[V] Preliminary SEDs for SSPs of ages older than 10\,Myr, as described in Appendix~C (Supplementary Material), will be public in the web page of FRACTAL\footnote{\url{https://www.fractal-es.com/PopStar/}, free access}.

\item[VI] Both Appendices B and C are included as Supplementary Material.
\end{itemize}

\appendix

\section{Spectral lines and measurements of stellar indices}
\label{indices}

\begin{figure*}
\centering
\includegraphics[width=0.22\textwidth,angle=-90]{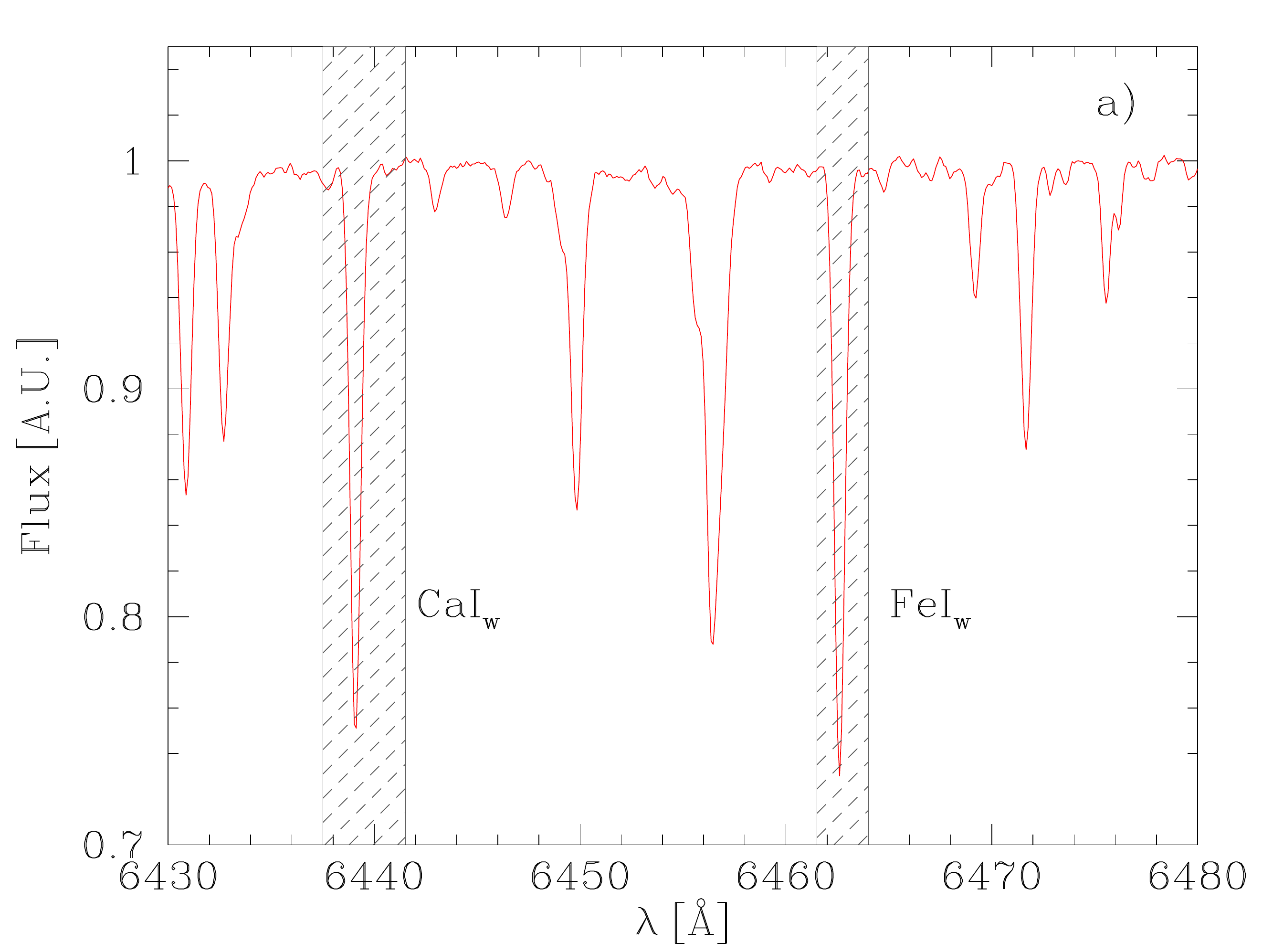}
\includegraphics[width=0.22\textwidth,angle=-90]{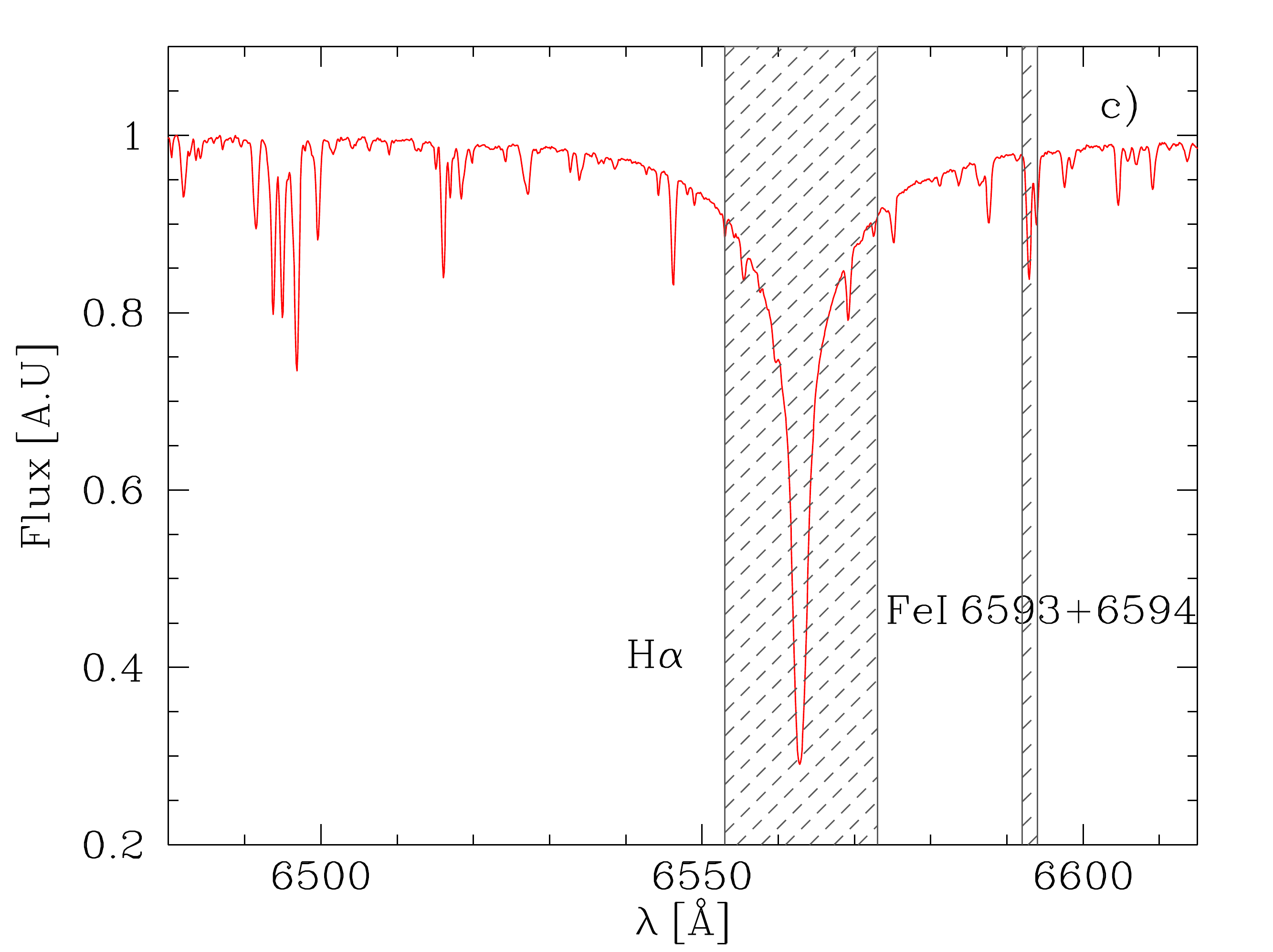}
\includegraphics[width=0.22\textwidth,angle=-90]{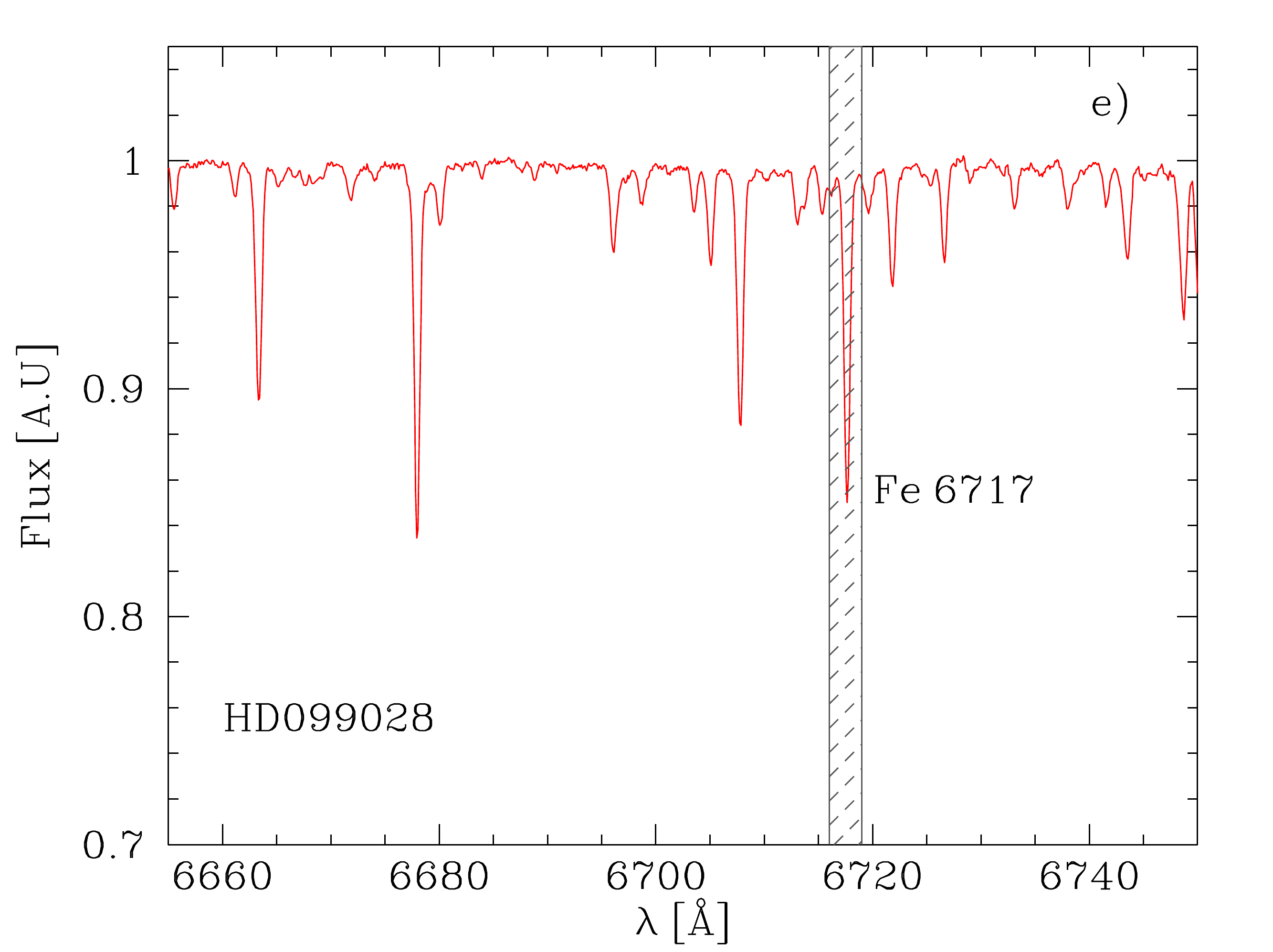}
\includegraphics[width=0.22\textwidth,angle=-90]{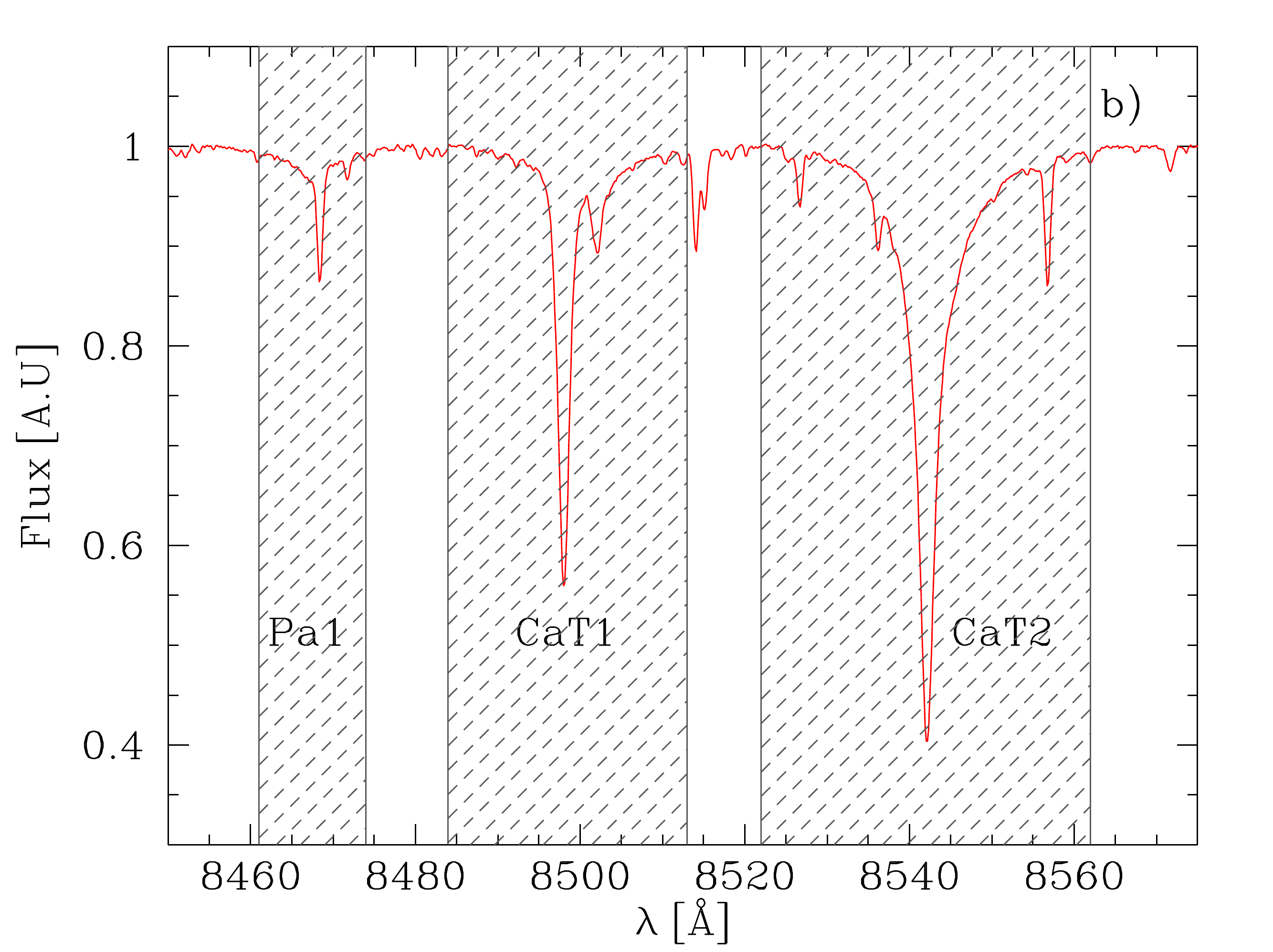}
\includegraphics[width=0.22\textwidth,angle=-90]{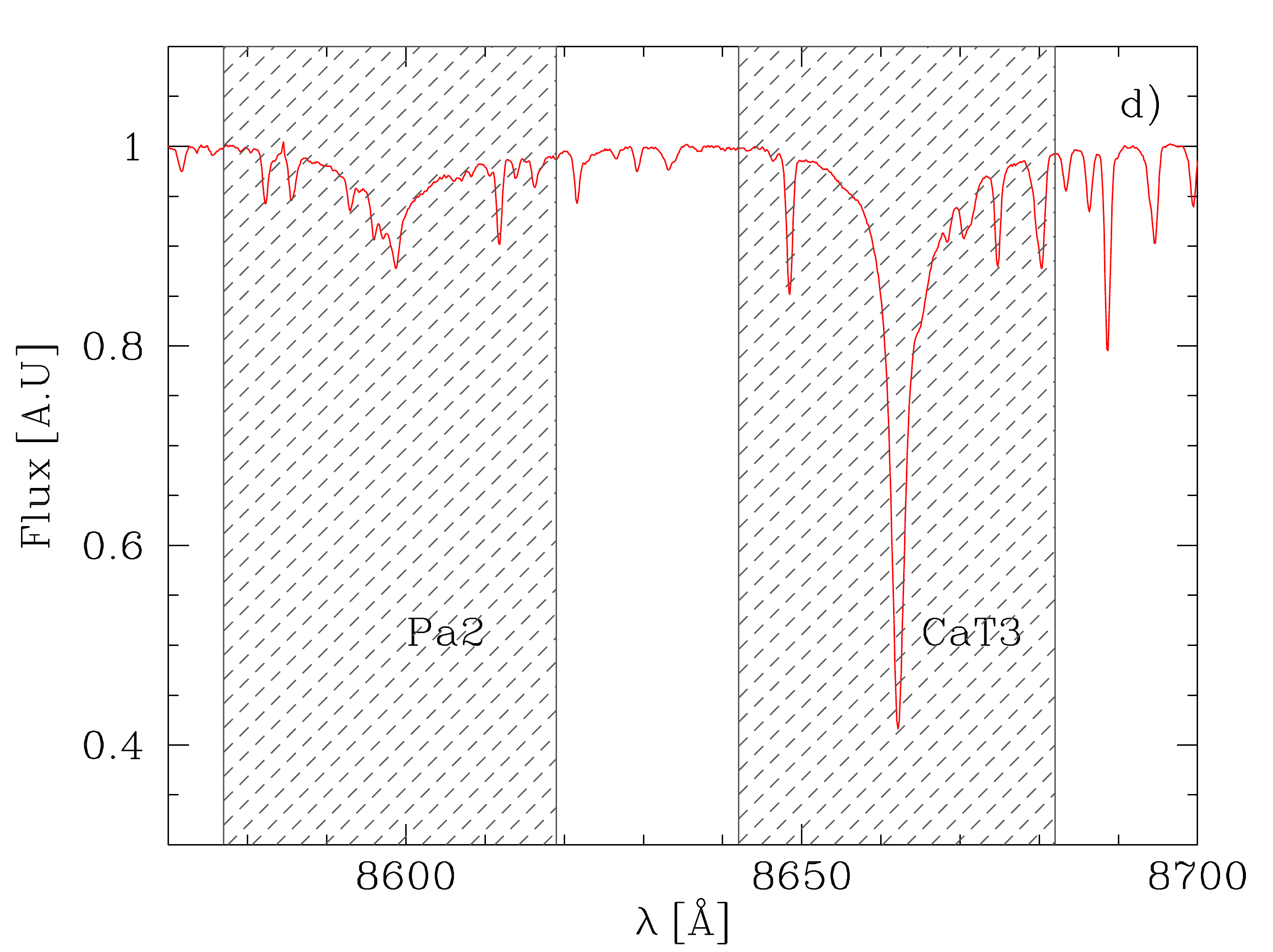}
\includegraphics[width=0.22\textwidth,angle=-90]{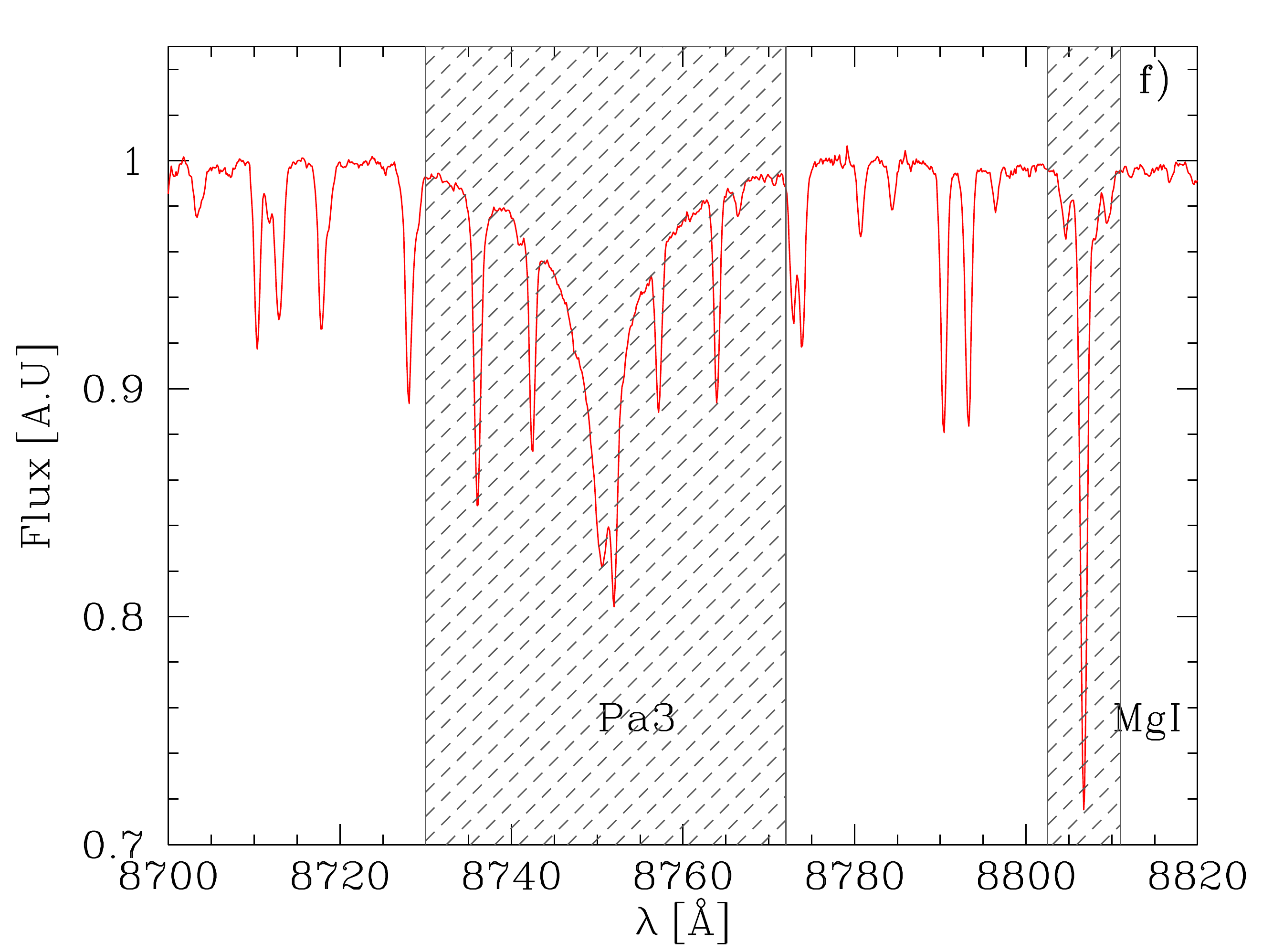}
\caption{Lines windows definitions  for HR-R (top) and HR-I (bottom) for some lines and indices measured in this work. The spectra correspond to HD099028. }
\label{fig:indices1}
\end{figure*}

We carried out measurements of 22 indices to facilitate their use by other groups. There are many strong spectral lines, as we already identified in Paper~I. The list of some of these lines is shown in Table~\ref{Table:lines}.  In the case of the \mbox{HR-R} set-up, except for the H$\alpha$ line, there are relatively few indices-related works in this wavelength range. The study of these spectral lines could be relevant for finding physical parameters in individual stars or for defining new indices to contribute to the understanding of stellar populations in clusters and galaxies. At present, we have measured in MEGASTAR  \mbox{HR-R} set-up, the H$\alpha$ index as proposed by \citet[hereinafter GON05]{Gonzalez+2005}, plus some other strong lines of \ion{Ca}{i} and \ion{Fe}{i} as defined in Table~\ref{Table:definition}. The lines and indices of  the \mbox{ HR-I} set-up measured in this  sample are also described in Table~\ref{Table:definition}. Fig.~\ref{fig:indices1} shows the detail of these definitions for \mbox{HR-R} and  \mbox{HR-I} for the different lines and indices we have measured. The spectra correspond to the star HD~099028. 

\begin{figure*}
\centering
\includegraphics[width=0.3\textwidth,angle=-90]{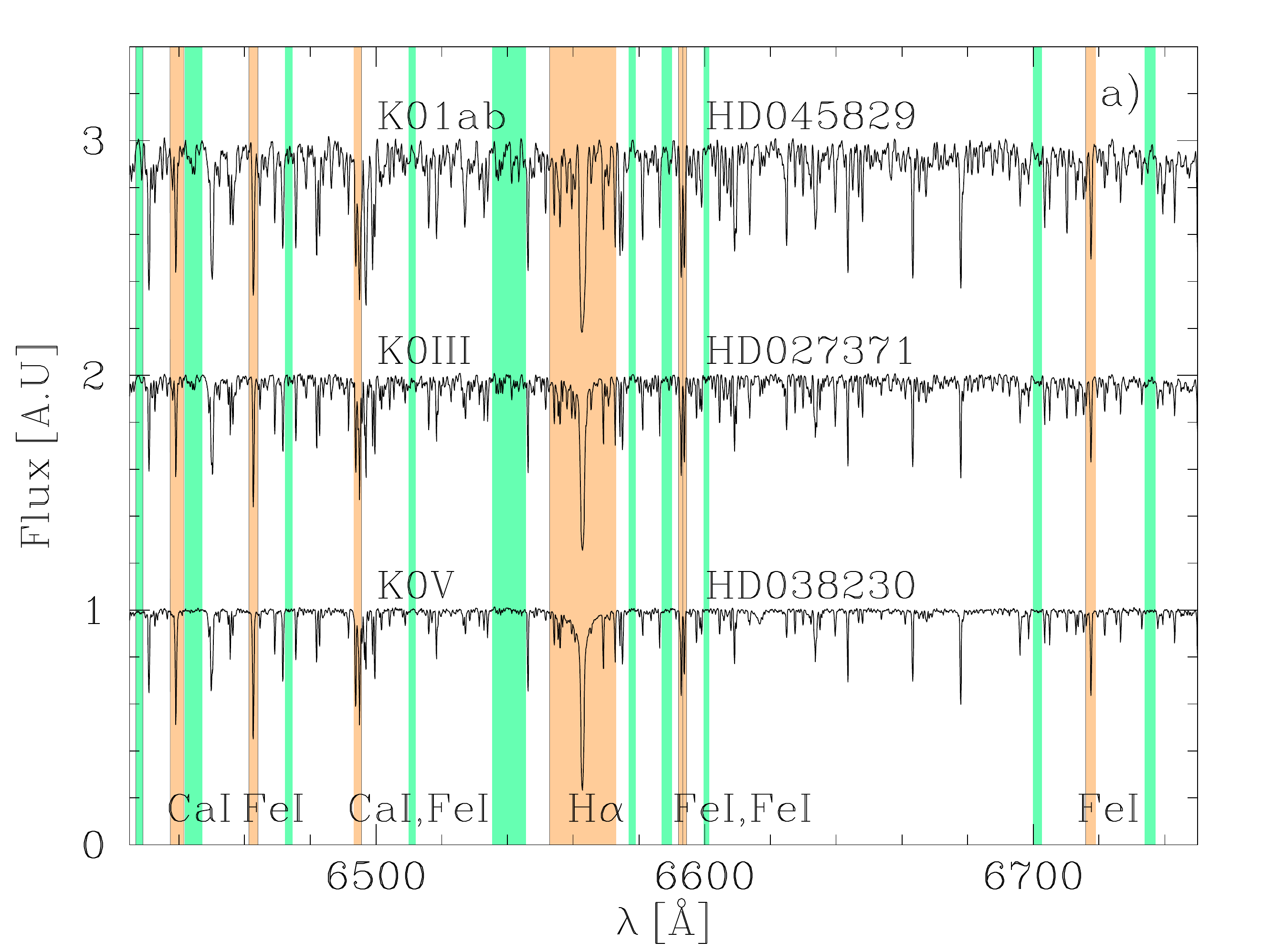}
\includegraphics[width=0.3\textwidth,angle=-90]{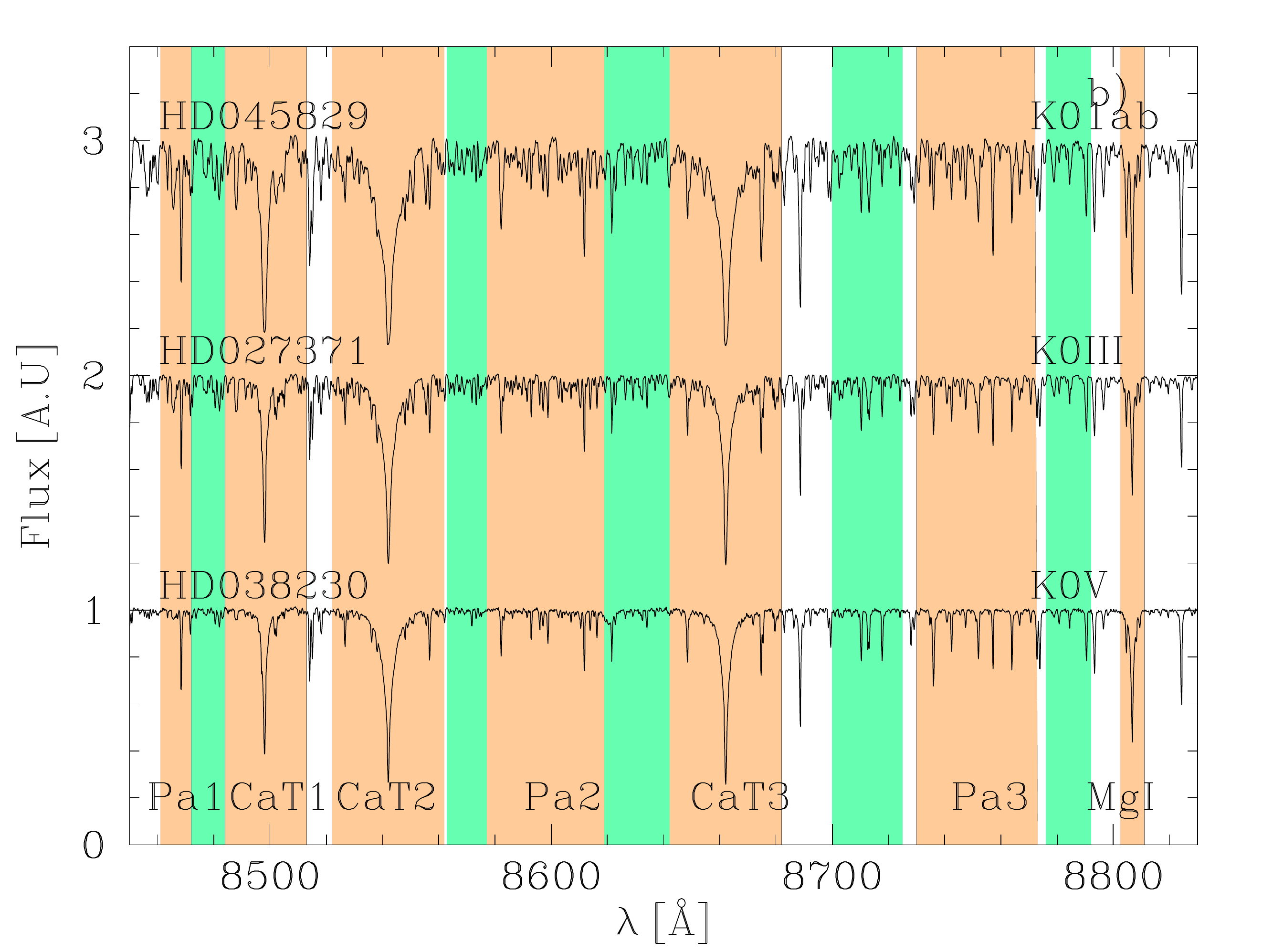}
\includegraphics[width=0.3\textwidth,angle=-90]{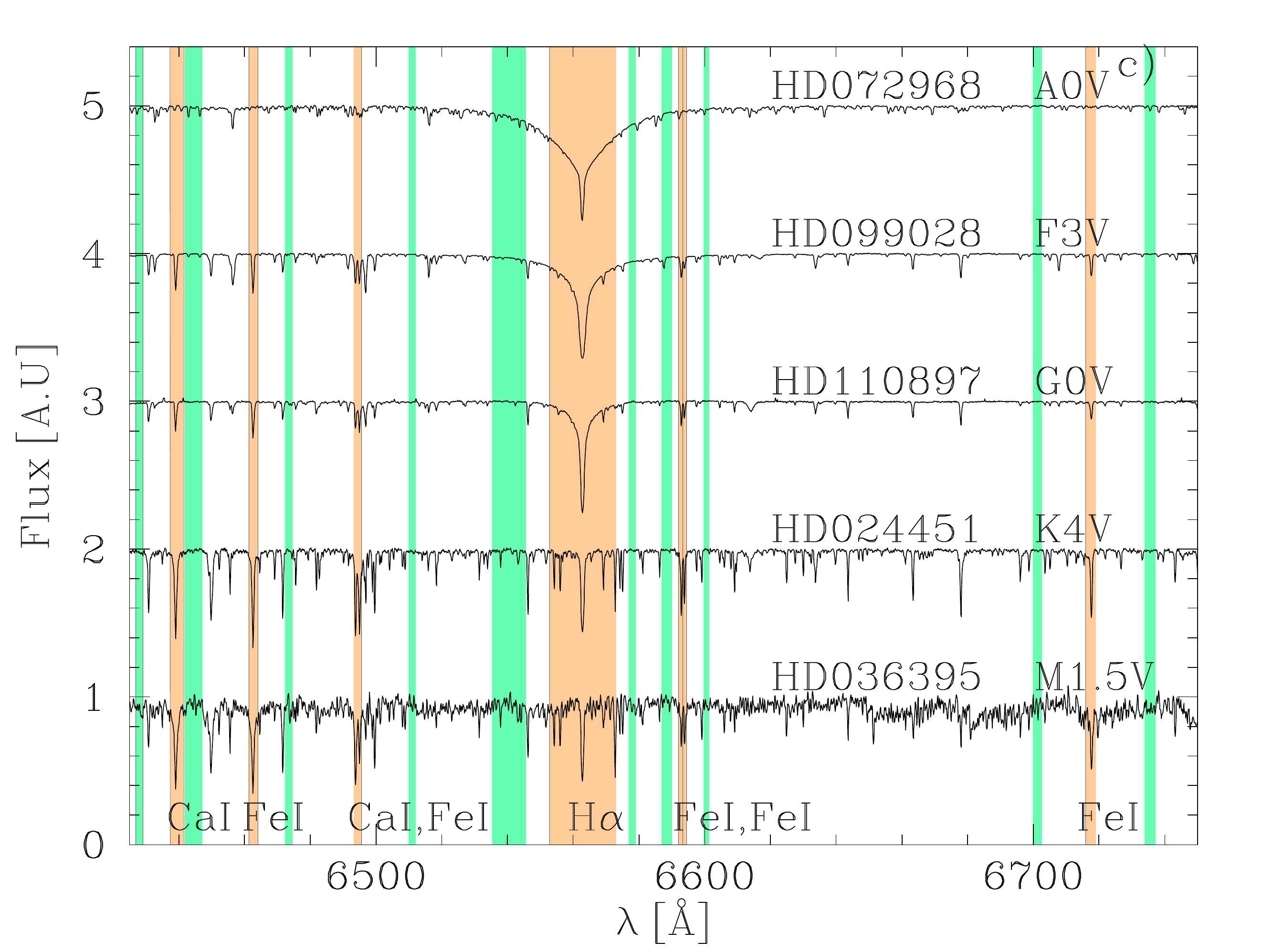}
\includegraphics[width=0.3\textwidth,angle=-90]{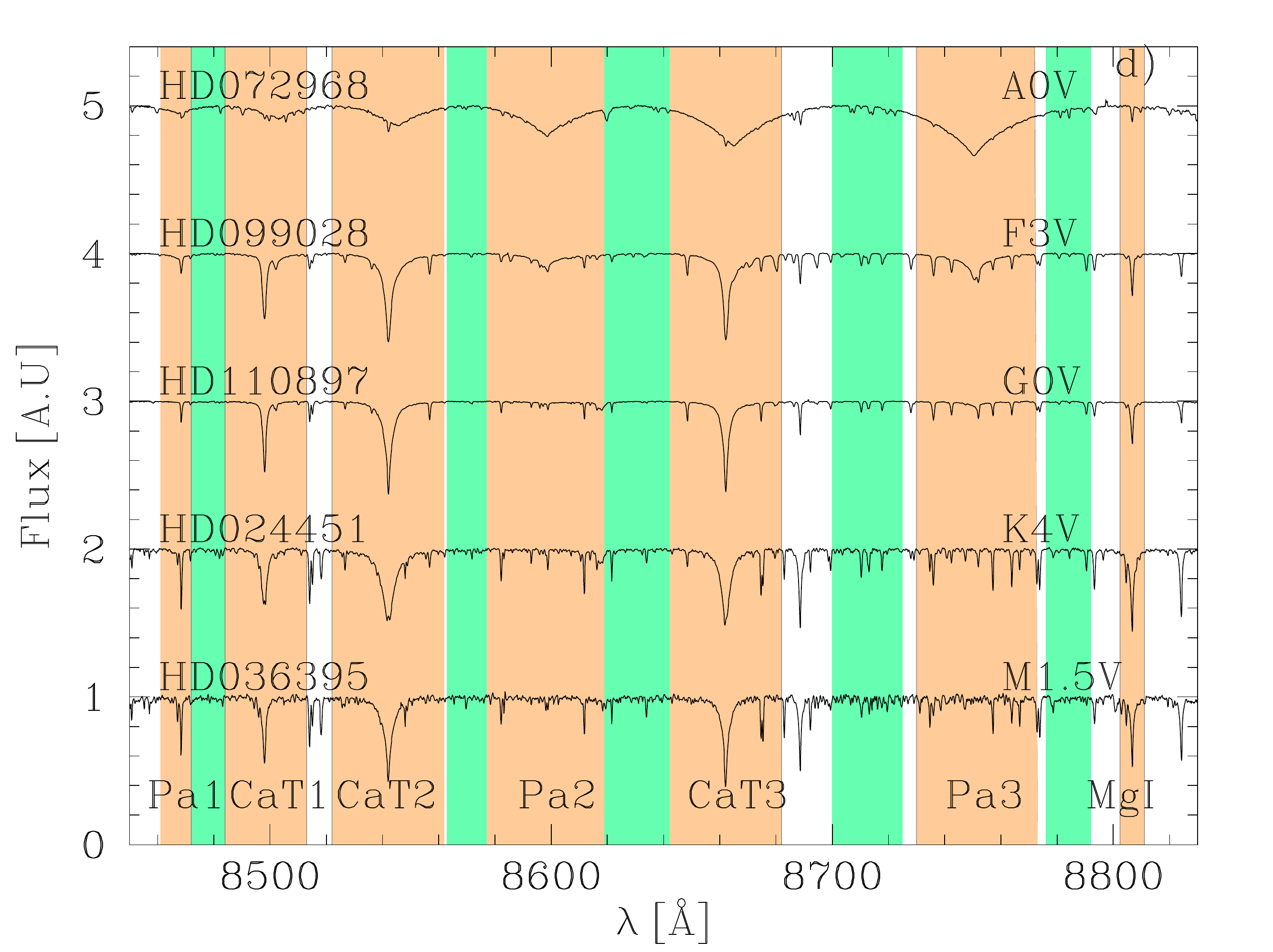}
\caption{Line bandpass (orange shade) and  continuum bandpasses (green shade) for each of the measured line or index in HR-R (left) and HR-I (right) in our {\sc MEGASTAR} spectra. Top panels (a and b) allow to explore the differences for stars with the same spectral type following a three-stars sequence in luminosity class (KIab, KIII and KV for a supergiant, giant and main sequence K star) while bottom panels (c and d) show a five-stars main sequence spectra series with different spectral types, from AOV (top) to M1.5V (bottom).}
\label{fig:indices2}
\end{figure*}

Fig.~\ref{fig:indices2} shows the same indices and line definitions with \mbox{HR-R} at the left and \mbox{HR-I} at  the right for some spectra. The top panels allow us  to see differences for stars with the same spectral type following a sequence in luminosity class (KIab, KIII and KV for a supergiant, giant and main sequence K star, respectively), while the bottom panels show the spectra for a main sequence series with different spectral types, from AOV to M1.5V.

\begin{table}
\centering
\caption{Some spectral lines identified in MEGASTAR spectra.}
\label{Table:lines}
\resizebox{7cm}{!}{
\begin{tabular}{ccc}
\hline
Spectral stellar type & Ion & Wavelength \\
                                 &       & [\AA] \\
                                 \hline
                                 \multicolumn{3}{c}{HR-R lines}\\
                   \hline                                 
Hot stars & \ion{He}{i} & 6678.15 \\
    B3 to A0 &.                 & 6867.48\\
                & \ion{He}{ii} & 6560.10\\
                & \ion{H}{i}/H$_{\alpha}$ & 6562.76\\
                &                                      & 6562.71\\
                &                                      & 6562.72\\  
                &                                      &\\      
Cool Star s& H$_{\alpha}$&  6562.76\\
Later than A & \ion{Fe}{i}   & 6430.85, 6469.19, 6475.62\\
                   &			& 6481.87, 6495.74, 6496.47\\
                   &			& 6498.94, 6518.37, 6533.93\\
                   &			& 6546.24, 6574.23, 6581.21\\
                   &			& 6591.31, 6592.91, 6593.87\\
                   &			& 6597.54, 6608.02, 6609.11\\
                   &			& 6627.54, 6633.41, 6633.75\\
                   &			& 6703.57, 6710.32, 6713.74\\
                   &			& 6716.24, 6725.36, 6750.15\\ 
                   &			& 6752.71\\
                   & \ion{Ca}{i}    & 6439.08, 6449.81, 6455.60\\
                   &			& 6471.66, 6493.78, 6499.65\\
                   &			& 6508.85, 6572.78, 6717.68\\
                   &			& 6798.48 \\      
                   &  \ion{Al}{i}    & 6696.02, 6698.67\\
                   & \ion{Si}{i}     & 6721.85, 6741.63\\
                   & \ion{Ti}{i}      & 6497.68, 6554.22, 6599.10\\
                   &                     & 6743.12\\
                   & \ion{Co}{i}     &6454.99, 6771.03\\
                   & \ion{Ni}{i}      & 6482.80, 6586.31, 6598.60\\
                   &                      & 6635.12, 6643.63, 6767.77 \\
                   &                      & 6772.31\\
                   &  \ion{V}{i}      & 6504.16\\
                   &  \ion{Cr}{i}     & 6537.92, 6630.01, 6537.92\\
                   &                       & 6630.01\\
                   &   \ion{Th}{i}    & 6457.28, 6462.61, 6531.34\\
                   &                       & 6989.65\\
                   & \ion{Fe}{ii}     & 6516.08, 6432.68 and 6456.38\\
                   & \ion{Ti}{ii}      & 6491.57\\
                   &  \ion{Sc}{ii}    & 6604.60\\
                   & \ion{Mg}{ii}    & 6545.97\\
                   \hline
                                 \multicolumn{3}{c}{HR-I lines} \\    
                   \hline                                
 Hot stars     &\ion{H}{i}         &Pa19 8413.33, Pa18 8437.96, Pa17 8467.27\\
  B3 to A0                  &                       &Pa16 8502.50, Pa15 8545.39, Pa14 8598.40\\
                    &                       & Pa13 8665.03, Pa 12 8750.47\\         
                    & \ion{He}{i}      &  8444.44, 8444.46, 8444.65\\
                    &                       & 8480.67, 8480.68, 8480.88\\
                    &                        & 8518.04, 8531, 8530.93\\
                    &                        &8532.10, 8532.11, 8532.13\\
                    &                        & 8582.51, 8582.52\\
                    &                        &8632.71, 8632.73, 8632.93\\
                    &			    & 8776.65, 8776.67, 8776.88\\
                    &                       &8849.16, 8849.37\\ 
                &                                      &\\                          
 Cool Stars& \ion{Ca}{ii}  & 8498.03, 8542.09, 8662.14\\
  Later than A & \ion{Mg}{i}     & 8806.76\\
                     \ion{Fe}{i}         & 8468.41, 8514.07, 8611.80\\
                     &                      & 8661.90, 8674.75, 8688.62\\
                     &                      & 8757.19, 8763.97, 8793.34 \\
                     &                       & 8824.22,  8838.43\\
                    & \ion{Th}{i}       &8416.73, 8421.22, 8446.51\\
                    &                        & 8478.36, 8748.03, 8758.24\\
                    & \ion{Ti}{i}         & 8412.36, 8426.50, 8434.96\\
                    &                        & 8435.65, 8675.37\\
                    & \ion{Na}{i} 	     & 8649.93, 8650.90, 8793.08\\
\hline
\end{tabular}
}
\end{table}

\begin{table}
\begin{center}
\caption{Definition of absorption spectral lines indices. The columns show: (1) title or definition of the line/index, (2) central wavelength of the {\it line} window, (3) wavelength ranges for {\it line}, (4) continuum bandpasses and (5) the reference in case the index had been previously defined by other authors.}
\label{Table:definition}
\resizebox{7cm}{!}{
\begin{tabular}{ccccc}
\hline
Line & $\rm \lambda_c$ & line & continuum & Ref.\\
     &                 & bandpass & bandpass & \\
    & (\AA) &   (\AA)  & (\AA) \\
    (1) & (2) & (3) & (4) & (5) \\
    \hline
\multicolumn{5}{c}{HR-R}\\
\hline    
\ion{Ca}{i} & 6439 & 6438.55--6439.75 & 6427.0 -- 6429.0	& TW \\
  &                 &                 & 6442.0 -- 6447.0    &   \\
  &                 &                 & 6472.5 -- 6474.5    &   \\
  &                 &                 & 6510.0 -- 6512.0    &   \\
  &                 &                 & 6535.5 -- 6545.5    &   \\
  &                 &                 & 6577.0 -- 6579.0    &   \\
  &                 &                 & 6587.0 -- 6590.0    &   \\
  &                 &                 & 6599.8 -- 6601.3    &   \\
  &                 &                 & 6700.0 -- 6702.5    &   \\
  &                 &                 & 6734.0 -- 6737.0    &   \\
\ion{Ca}{i}$_{wide}$ & 6439 &  6437.50 -- 6441.50 & as \ion{Ca}{i} & TW \\
\ion{Fe}{i} & 6463 & 6462.15 --6463.60 & as \ion{Ca}{i} & TW \\
\ion{Fe}{i}$_{wide}$ & 6463 & 6461.50 -- 6464.00 & as \ion{Ca}{i} & TW \\
\ion{Ca}{i} & 6494 & 6493.30 -- 6494.30 & as \ion{Ca}{i} & TW \\
\ion{Fe}{i} & 6495 & 6494.25 -- 6495.55 & as \ion{Ca}{i} & TW \\
\ion{Ca}{i}$+$\ion{Fe}{i} & 6494+6495 & 6493.30 -- 6495.55 & as \ion{Ca}{i} & TW \\
H$_\alpha$ & 6562.79 & 6553.0 -- 6573.0 &
6506.0 -- 6514.0 & GON05 \\
& & & 6612.0 -- 6620.0 & \\
\ion{Fe}{i} & 6593 & 6592.20 -- 6593.40 & as \ion{Ca}{i} & TW \\
\ion{Fe}{i} & 6594 & 6593.40 -- 6594.40 & as \ion{Ca}{i} & TW \\
\ion{Fe}{i} & 6593$+$6594 & 6592.00 -- 6594.60 & as \ion{Ca}{i} & TW \\
\ion{Fe}{i} & 6717 & 6716.00 -- 6719.00 & as \ion{Ca}{i} & TW \\
\hline    
\multicolumn{5}{c}{HR-I}  \\
\hline
Pa1 & 8467.90 & 8461.0 -- 8474.0 & 
8472.0 -- 8484.0 & CEN01 \\
& & & 8563.0 -- 8577.0 & \\
& & & 8619.0 -- 8642.0 &\\
& & & 8700.0 -- 8725.0 & \\
& & & 8776.0 -- 8792.0 & \\
Pa2 & 8599.04 & 8577.0 -- 8619.0 & as Pa1 & CEN01 \\
Pa3 & 8751.13 & 8730.0 -- 8772.0 & as Pa1  & CEN01 \\
CaT1$_{\rm CEN}$ & 8498.00 & 8484.0 -- 8513.0 & as Pa1  & CEN01 \\
CaT2$_{\rm CEN}$ & 8542.09 & 8522.0 -- 8562.0 & as Pa1  & CEN01 \\
CaT3$_{\rm CEN}$ & 8662.14 & 8642.0 -- 8682.0 & as Pa1  & CEN01 \\
\ion{Mg}{i} & 8807.00 & 8802.5 -- 8811.0 & 
as Pa1  & CEN09 \\ 
CaT1$_{\rm MEG}$ & 8498.00 & 8482.0 -- 8512.0 & 8450.0 -- 8460.0 & TW \\
& & & 8565.5 -- 8575.0 & \\
CaT2$_{\rm MEG}$ & 8542.09 & 8531.0 -- 8554.0 & 450.0 -- 8460.0 & TW \\
& & & 8565.5 -- 8575.0 & \\
CaT3$_{\rm MEG}$ & 8662.14 & 8650.0 -- 8673.0 & 8619.5 -- 8642.5 & TW \\
& & & 8700.5 -- 8710.0 & \\
\hline
\end{tabular}
}
\end{center}
\footnotesize{References: CEN01: \citet{cen01}; GON05: \citet{Gonzalez+2005}, CEN09: \citet{cen09}, TW: this work}
\end{table}

Table~\ref{table:indices} includes the EW  measurements and errors for all the selected spectral absorption lines and indices. The whole table is available online (few rows are given here as example). 

\begin{table*}
\centering
\caption{Measurements of the selected spectral absorption lines and indices (equivalent widths) and their errors in \AA. The whole table is available in electronic format (see Section~\ref{data}). We show a few rows here to illustrate the content. Columns (1) shows the name of the star. Columns (2) to (45) of this table give the equivalent widths along with their corresponding errors (both in \AA) for the 22 lines, defined in Table~\ref{Table:definition}.}

\label{table:indices}
\resizebox{18cm}{!}{
\begin{tabular}{lccccccccccc}
\hline
Name & $\ion{Ca}{i}_{6439}$  & $\ion{Ca}{i}_{6439,w}$ & $\ion{Fe}{i}_{6463}$  & $\ion{Fe}{i}_{6463,w}$  & $\ion{Ca}{i}_{6494}$&   $\ion{Fe}{i}_{6495}$   & $\ion{Ca}{i}_{6494}+\ion{Fe}{i}_{6495}$  & $\ion{Fe}{i}_{6593}$ & $\ion{Fe}{i}_{6594}$ & $\ion{Fe}{i}_{6593+6594}$ & $\ion{Fe}{i}_{6717}$ \\
& (\AA) & (\AA) & (\AA) & (\AA) & (\AA) & (\AA) & (\AA) & (\AA) & (\AA) & (\AA) & (\AA) \\ 
 (1) & (2)\,(3) & (4)\,(5) & (6) (7) & (8) (9) & (10) (11) & (12) (13) & (14) (15) & (16) (17) & (18) (19) & (20) (21) & (22) (23) \\
\hline
BD$+$083095 & $-$0.108 $\pm$  0.007  & $-$0.094 $\pm$ 0.031 & $-$0.133 $\pm$ 0.003  & $-$0.136 $\pm$ 0.006 & $-$0.080 $\pm$ 0.015 & $-$0.116 $\pm$ 0.008 & $-$0.196 $\pm$ 0.007 & $-$0.062 $\pm$ 0.694 & $-$0.034 $\pm$ 0.003 & -0.097 $\pm$ 0.0116  & $-$0.0727 $\pm$ 0.007 \\
BD$+$092190 & $-$0.005 $\pm$ 0.001 & $+$0.018 $\pm$ 0.001 & $-$0.005 $\pm$ 0.003  & $-$0.009$\pm$ 0.001 & $+$0.003$\pm$ 0.001 & $+$0.010 $\pm$ 0.001 &$+$ 0.012 $\pm$ 0.001 & $+$0.006 $\pm$ 0.001 & $+$0.003 $\pm$ 0.001 &  $+$0.017 $\pm$ 0.0001 & $+$0.0060  $\pm$ 0.001 \\
BD$+$130013 & $-$0.229 $\pm$ 0.001 & $-$0.285 $\pm$ 0.004 & $-$0.272 $\pm$ 0.001  & $-$0.297 $\pm$ 0.002 & $-$0.195 $\pm$ 0.001 & $-$0.261 $\pm$ 0.001 & $-$0.444 $\pm$ 0.002 & $-$0.137 $\pm$ 0.002 & $-$0.110 $\pm$ 0.003 & $-$0.243 $\pm$ 0.0017  & $-$0.171 $\pm$ 0.003 \\
BD$+$191730 &  $-$0.056 $\pm$ 0.035 & $-$0.033 $\pm$ 0.007  & $-$0.061 $\pm$ 0.003  & $-$0.050$\pm$ 0.006 & $-$0.039 $\pm$ 0.003 & $-$0.048 $\pm$ 0.007  & $-$0.086 $\pm$ 0.003 & $-$0.030 $\pm$ 0.002 & $-$0.006 $\pm$ 0.001 & $-$0.038 $\pm$ 0.0855 & $-$0.0161 $\pm$ 0.179 \\
BD$+$195116B &  $-$0.385 $\pm$ 0.005 & $-$0.542 $\pm$ 0.014 & $-$0.456 $\pm$ 0.005  & $-$0.591$\pm$ 0.008 & $-$0.346 $\pm$ 0.005 & $-$0.231 $\pm$ 0.007 & $-$0.553 $\pm$ 0.008 & $-$0.036 $\pm$ 0.009 & $-$0.108 $\pm$ 0.012 & $-$0.142 $\pm$ 0.0141  & $-$0.5009  $\pm$ 0.010 \\
\hline
\end{tabular}
}
\end{table*}

\begin{table*}
\centering
\contcaption{Equivalent widths of selected spectral absorption lines}
\resizebox{18cm}{!}{
\begin{tabular}{ccccccccccc}
\hline
H$\alpha_{\rm GON}$ & CaT1$_{\rm CEN}$ & CaT2$_{\rm CEN}$ & CaT3$_{\rm CEN}$ & $\ion{Mg}{i}_{\rm CEN}$ & Pa1$_{\rm CEN}$ & Pa2$_{\rm CEN}$ & Pa3$_{\rm CEN}$ &  CaT1$_{\rm MEG}$ & CaT2$_{\rm MEG}$ & CaT3$_{\rm MEG}$ \\
(\AA) & (\AA) & (\AA) & (\AA) & (\AA) & (\AA) & (\AA) & (\AA) & (\AA) & (\AA) & (\AA) \\ 
(24) (25) & (26) (27) & (28) (29) & (30) (31) & (32) (33) & (34) (35) & (36) (37) & (38) (39) & (40) (41) & (42) (43) & (44) (45) \\
\hline 
$-$2.705 $\pm$ 0.026 & $-$0.870 $\pm$ 0.033 & $-$2.341 $\pm$ 0.031 & $-$1.802 $\pm$  0.026 & $-$0.296 $\pm$ 0.015 & $-$0.140 $\pm$ 0.036 & $-$0.298 $\pm$ 0.035 & $-$0.488 $\pm$ 0.029 & $-$0.909 $\pm$ 0.032 & $-$2.298 $\pm$ 0.023 & $-$1.712 $\pm$ 0.020 \\
$-$3.448 $\pm$ 0.015 & $-$0.313 $\pm$ 0.066 & $-$0.654 $\pm$ 0.042 & $-$0.950 $\pm$  0.055 & $-$0.053 $\pm$ 0.010 & $-$0.137 $\pm$ 0.127 & $-$0.528 $\pm$ 0.024 & $-$0.720 $\pm$ 0.030 & $-$0.204 $\pm$ 0.023 & $-$0.615 $\pm$ 0.028 & $-$0.877 $\pm$ 0.035 \\
$-$1.458 $\pm$ 0.007 & $-$0.709 $\pm$ 0.011 & $-$2.493 $\pm$ 0.011 & $-$1.900 $\pm$  0.011 & $-$0.634 $\pm$ 0.005 & $-$0.274 $\pm$ 0.018 & $-$0.267 $\pm$ 0.365 & $-$0.266 $\pm$ 0.022 & $-$0.663 $\pm$ 0.009 & $-$2.299 $\pm$ 0.007 & $-$1.715 $\pm$ 0.007 \\
$-$3.084 $\pm$ 0.019 & $-$0.495 $\pm$ 0.036 & $-$1.527 $\pm$ 0.034 & $-$1.176 $\pm$  0.027 & $-$0.166 $\pm$ 0.019 & $-$0.026 $\pm$ 0.003 & $-$0.195 $\pm$ 0.171 & $-$0.564 $\pm$  0.031 & $-$0.524 $\pm$ 0.031 & $-$1.415 $\pm$ 0.025 & $-$1.182 $\pm$ 0.022 \\
   $+$2.747 $\pm$ 0.030 &    $+$0.055 $\pm$ 0.007 & $-$0.608 $\pm$ 0.015 & $-$0.091 $\pm$
0.002 & $-$0.549  $\pm$ 0.014 & $-$0.553 $\pm$ 0.024 & $+$0.523 $\pm$  0.015 & $-$0.788 $\pm$ 0.001 & $+$0.926 $\pm$ 0.001 & $-$0.551 $\pm$ 0.012 & $-$0.598 $\pm$ 0.011 \\
\hline
\end{tabular}
}  
\end{table*}

In the \mbox{HR-I} spectra, we have measured the most widely used indices in the near infrared defined by \citet[][hereinafter CEN01]{cen01}, and \citet[][hereinafter CEN09]{cen09}. We consider these indices as the most appropriate to be applied to composed populations studies. In Paper~I,  we have defined new indices fitted to the HR-I spectral resolution for  individual stars.  Using the definitions by CEN01, we have measured the  CaT1$_{\rm CEN}$, CaT2$_{\rm CEN}$ and CaT3$_{\rm CEN}$ indices, centred at  8498.00, 8542.09 and 8662.14\,\AA\ respectively, and  the Pa1, Pa2, Pa3 indices, centred at the series lines P17, P14 and P12.   We have also measured the \ion{Mg}{i} index as defined by CEN09. Finally, we have defined CaT indices optimized for the MEGARA spectral resolution, CaT1$_{\rm MEG}$, CaT2$_{\rm MEG}$ and CaT3$_{\rm MEG}$. The bandpasses are summarized in \ref{Table:definition}. From the individual CEN01 \ion{Ca}{ii} and Pa indices, we have derived the composed indices $\rm CaT_{\rm CEN}$ = $\rm CaT1_{\rm CEN}$ + $\rm CaT2_{\rm CEN}$ + $\rm CaT3_{\rm CEN}$ and $\rm PaT$ = $\rm Pa1 + \rm Pa2 + \rm Pa3$, which we will use for the correlations with the stellar parameters. 

\begin{figure*}
\centering
\includegraphics[width=0.24\textwidth,angle=-90]{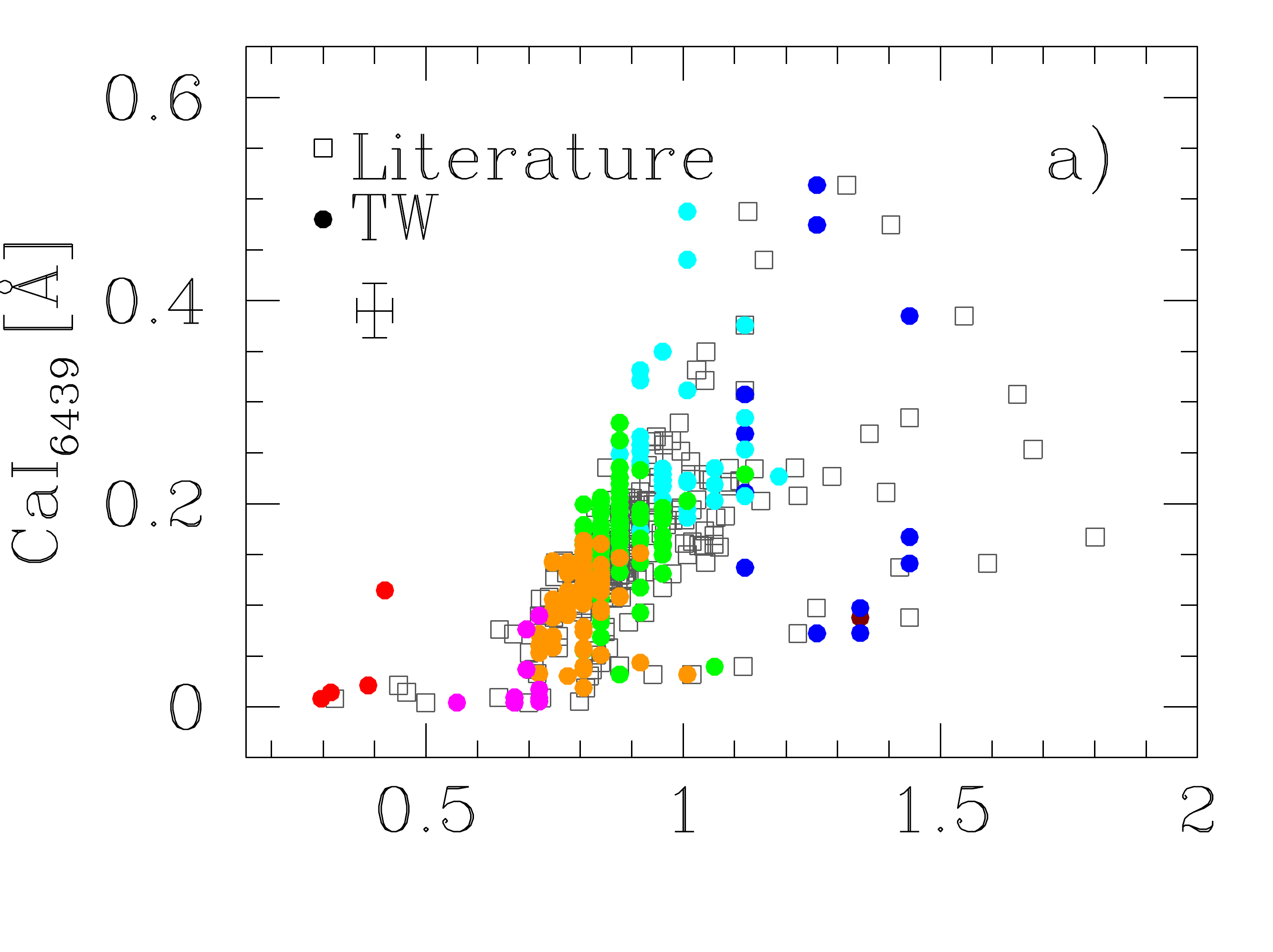}
\includegraphics[width=0.24\textwidth,angle=-90]{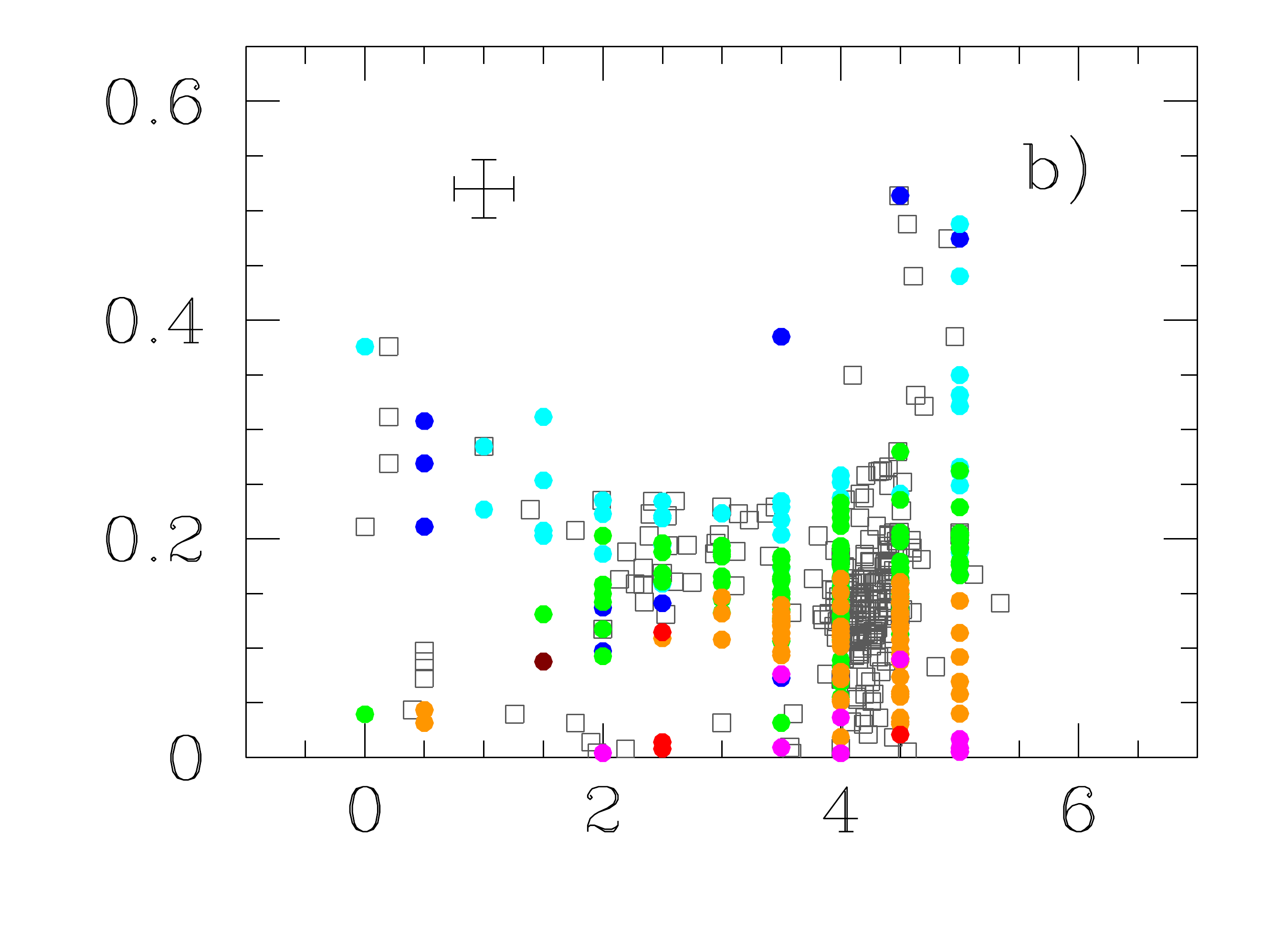}
\includegraphics[width=0.24\textwidth,angle=-90]{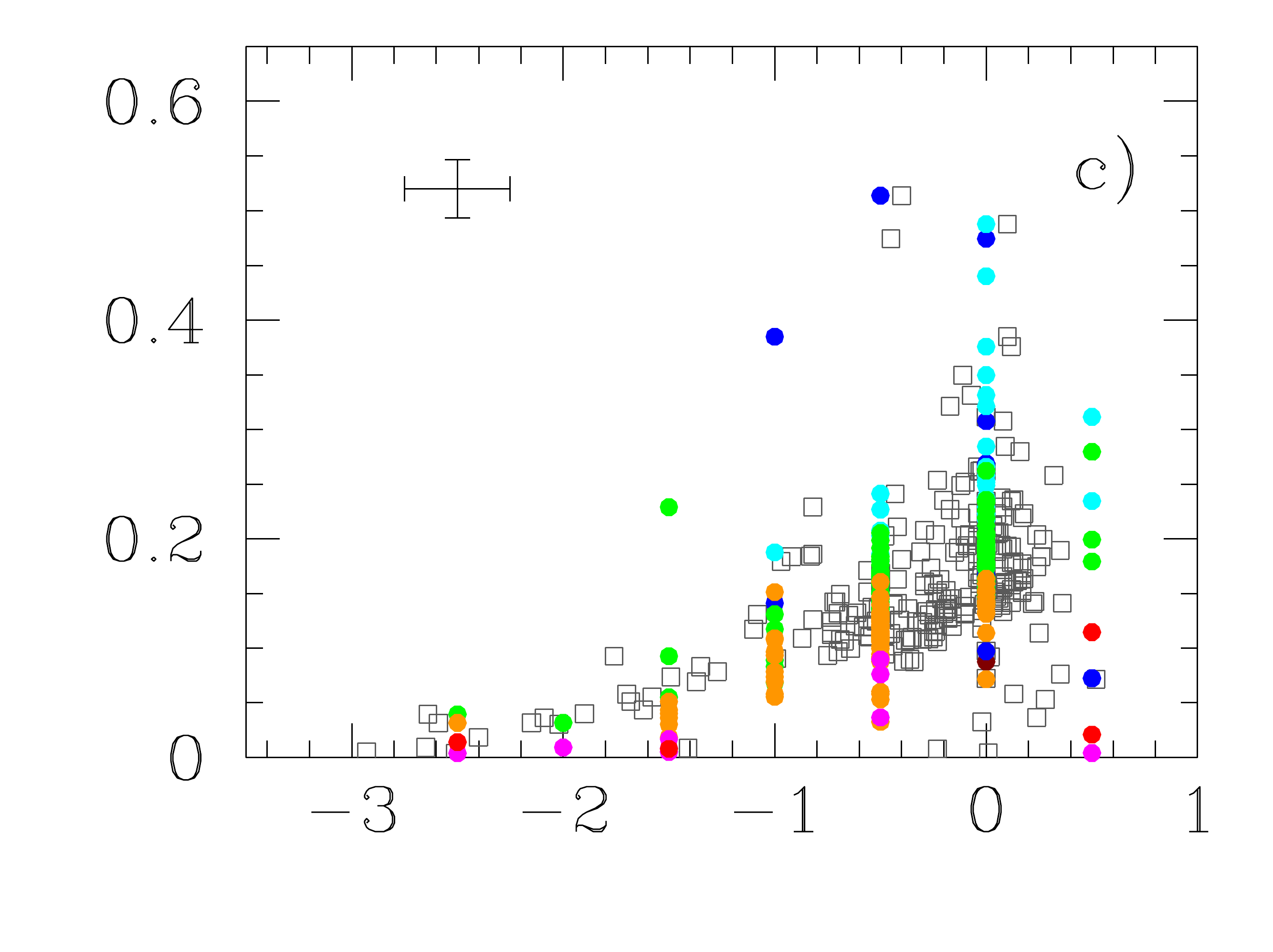}
\includegraphics[width=0.24\textwidth,angle=-90]{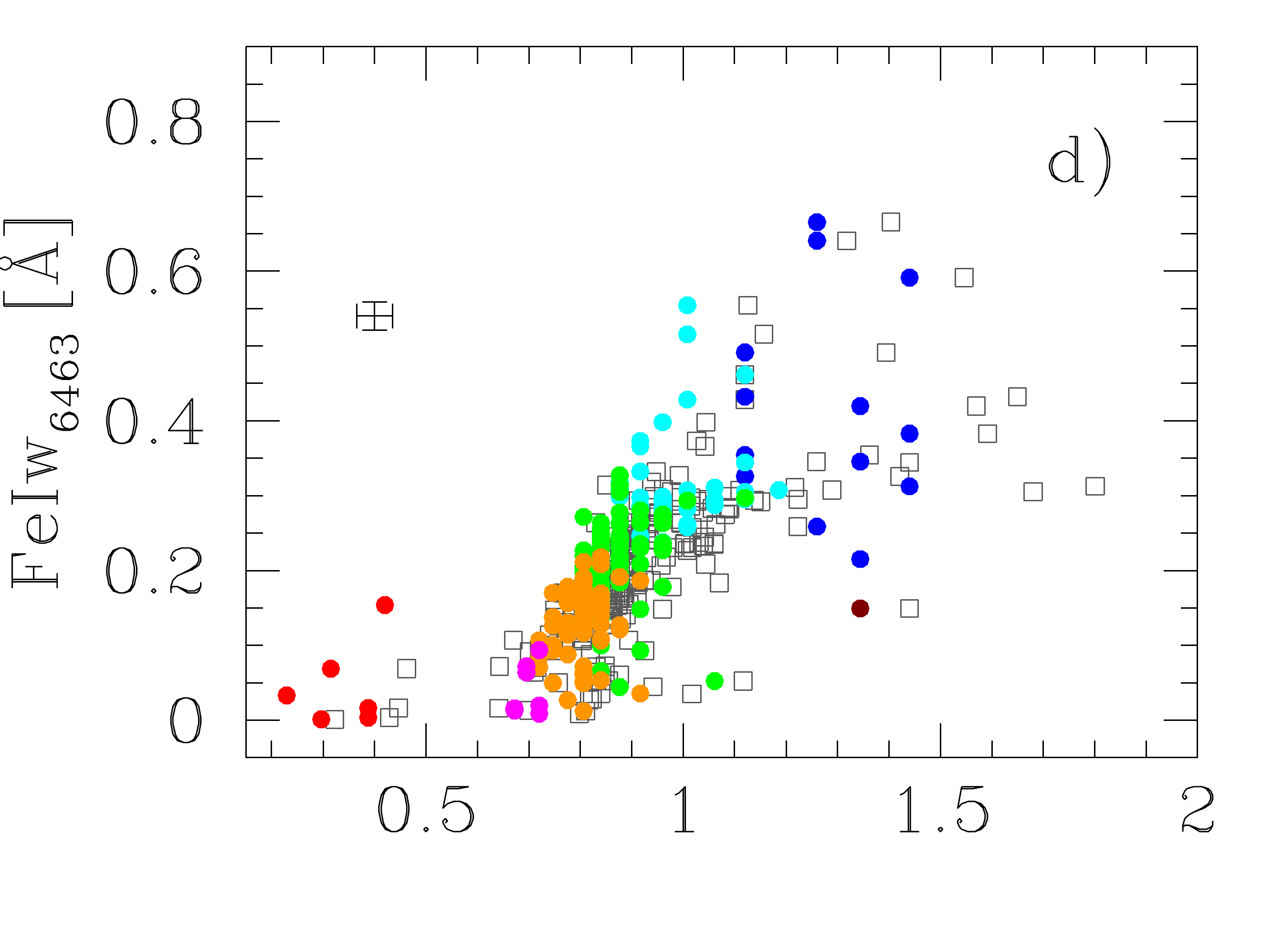}
\includegraphics[width=0.24\textwidth,angle=-90]{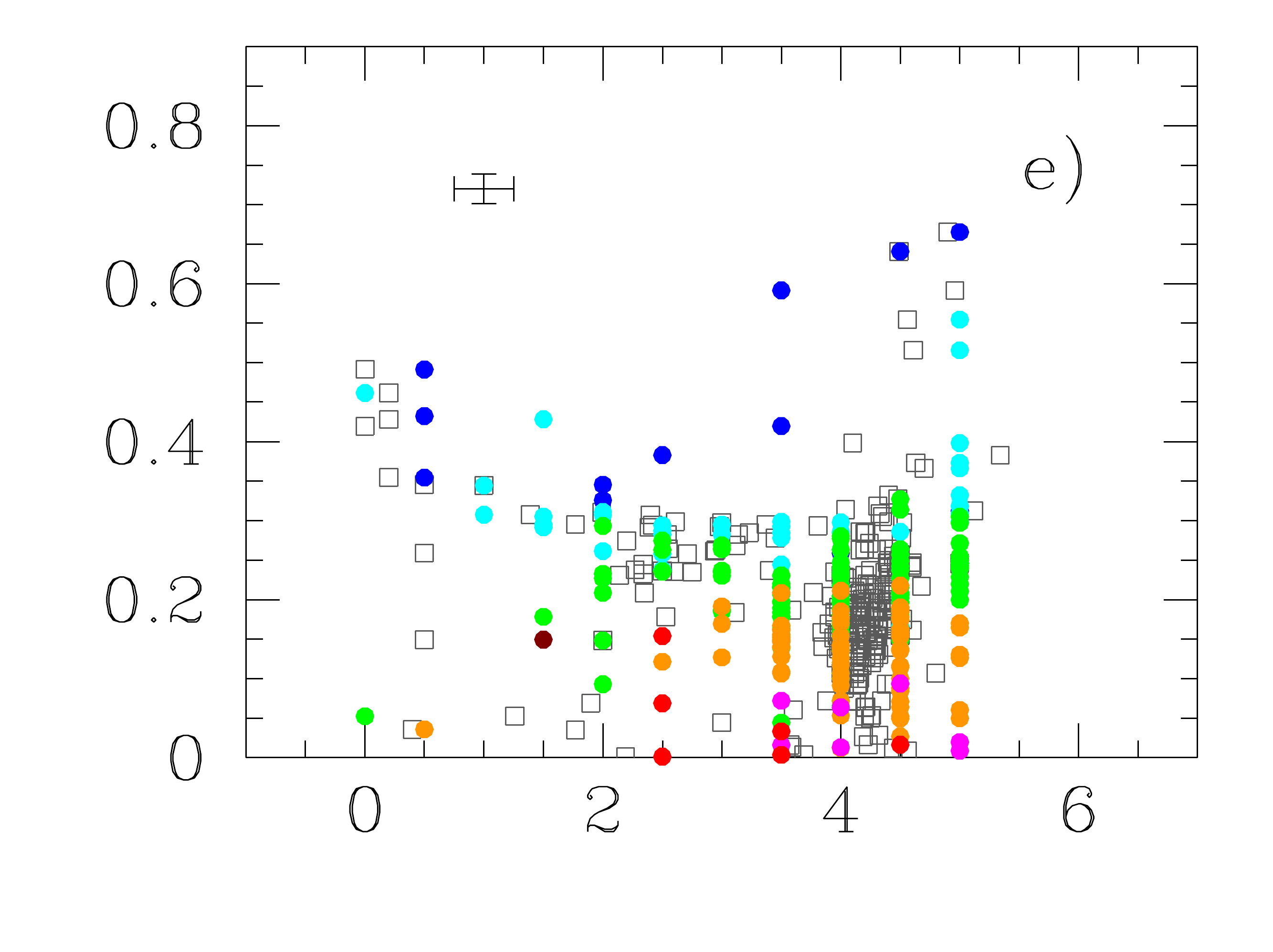}
\includegraphics[width=0.24\textwidth,angle=-90]{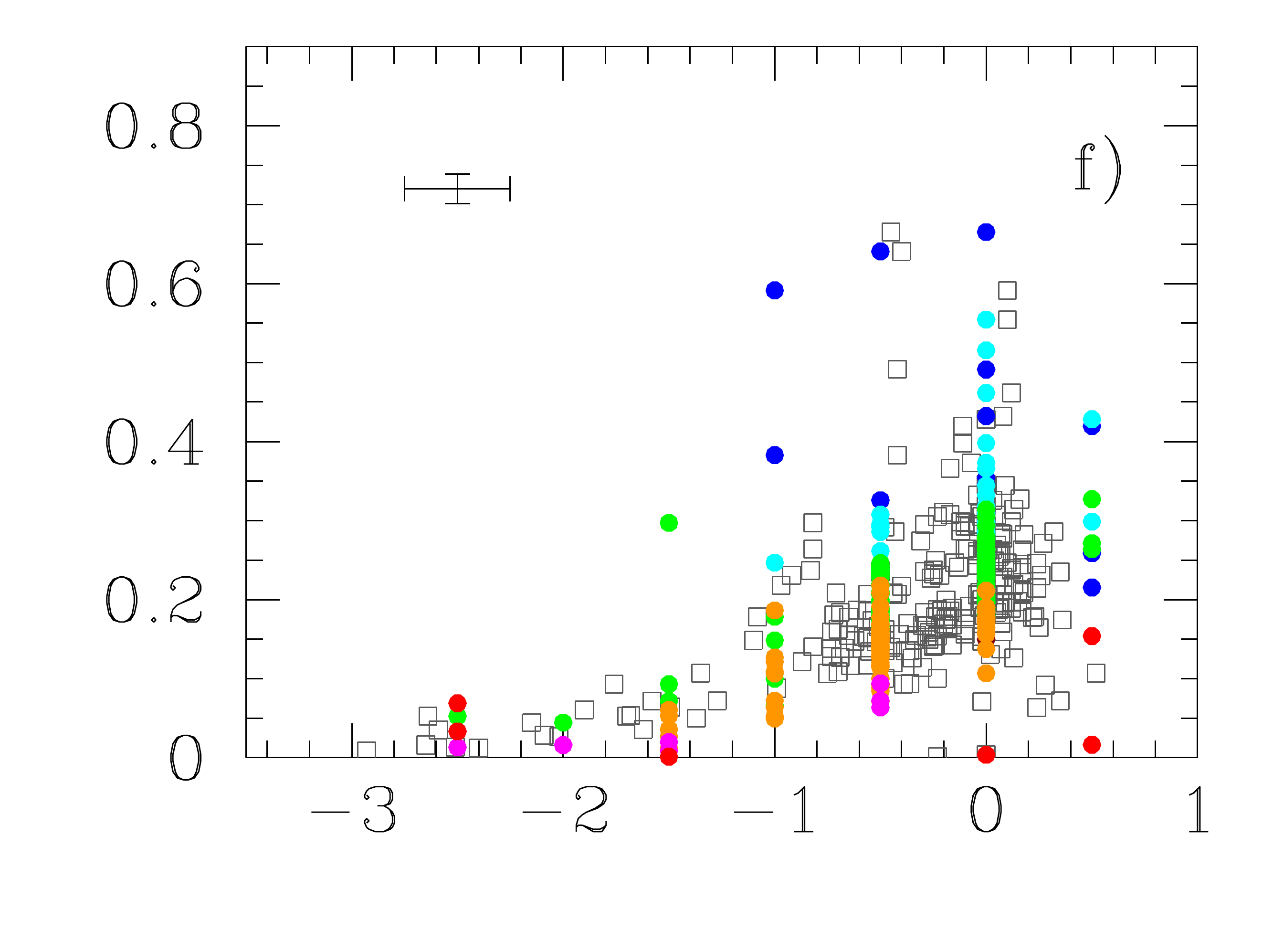}
\includegraphics[width=0.24\textwidth,angle=-90]{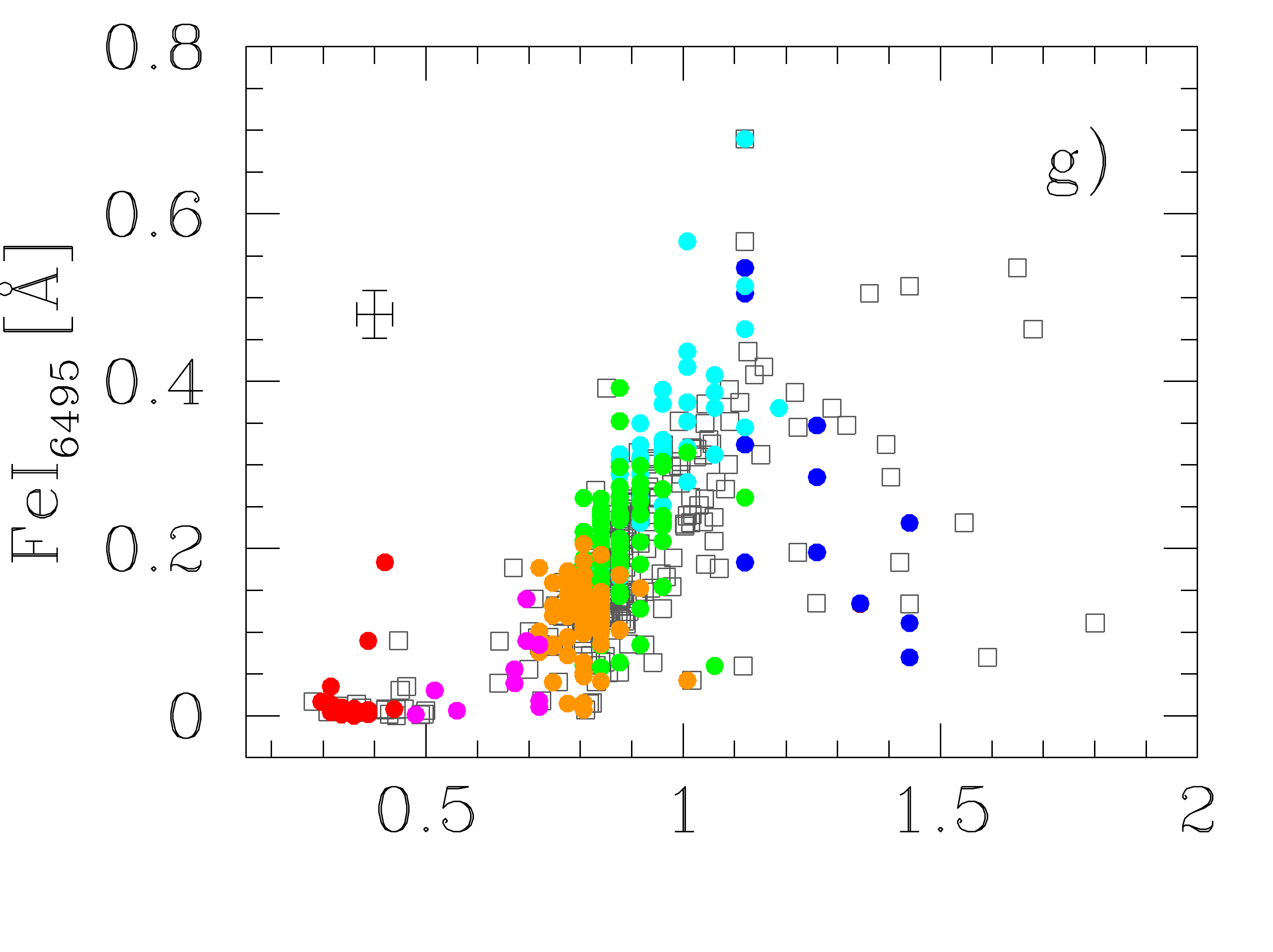}
\includegraphics[width=0.24\textwidth,angle=-90]{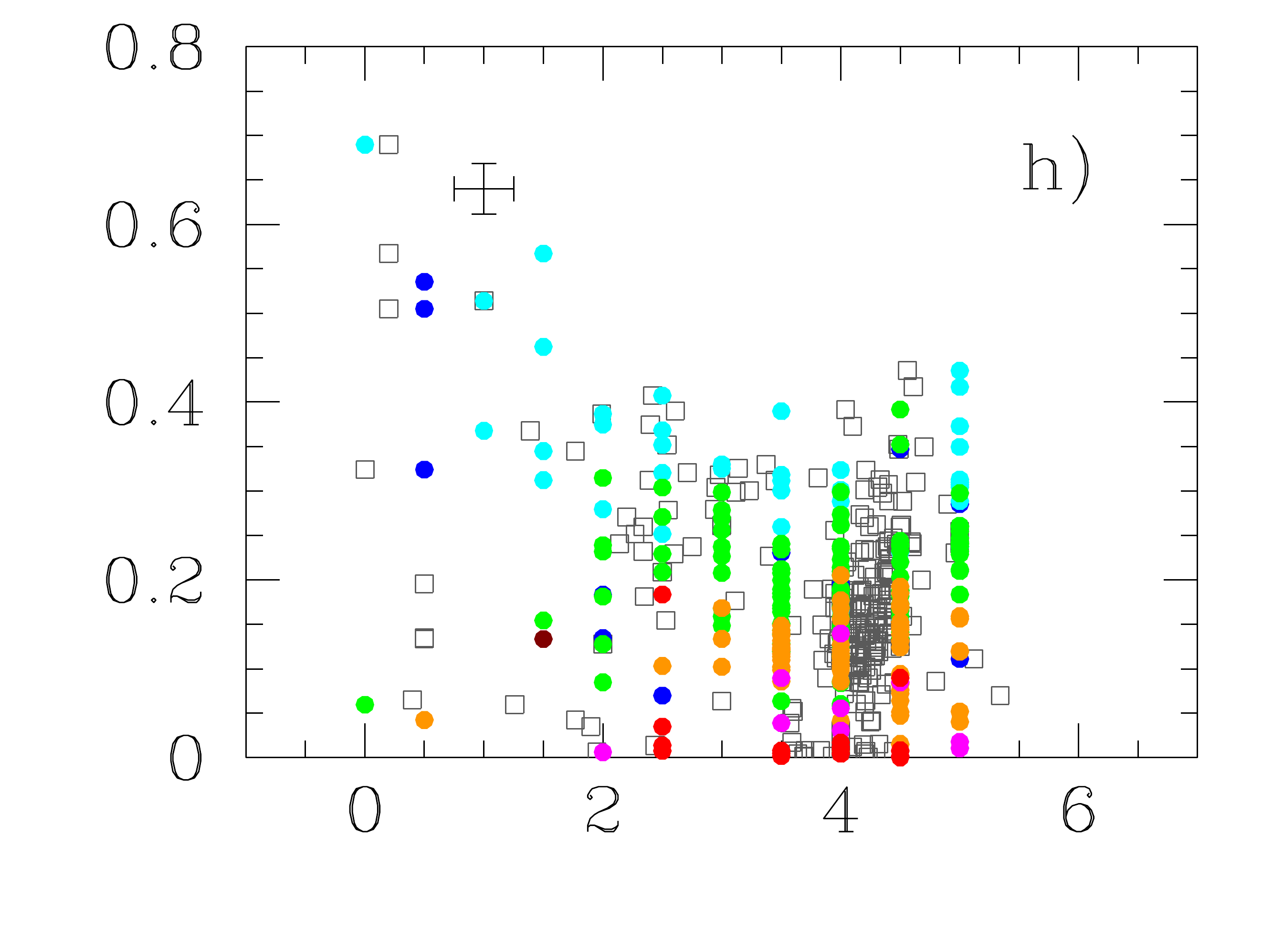}
\includegraphics[width=0.24\textwidth,angle=-90]{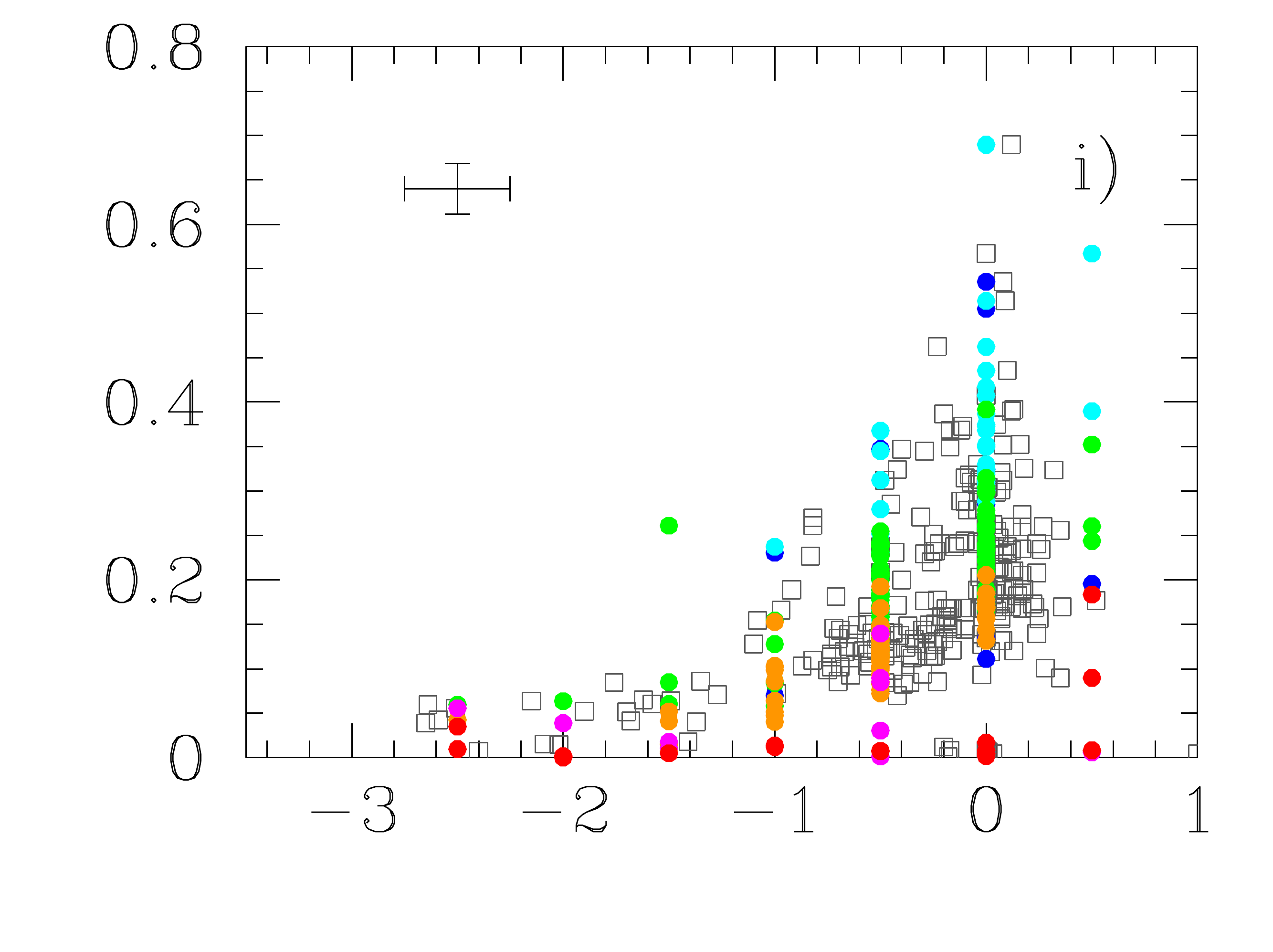}
\includegraphics[width=0.24\textwidth,angle=-90]{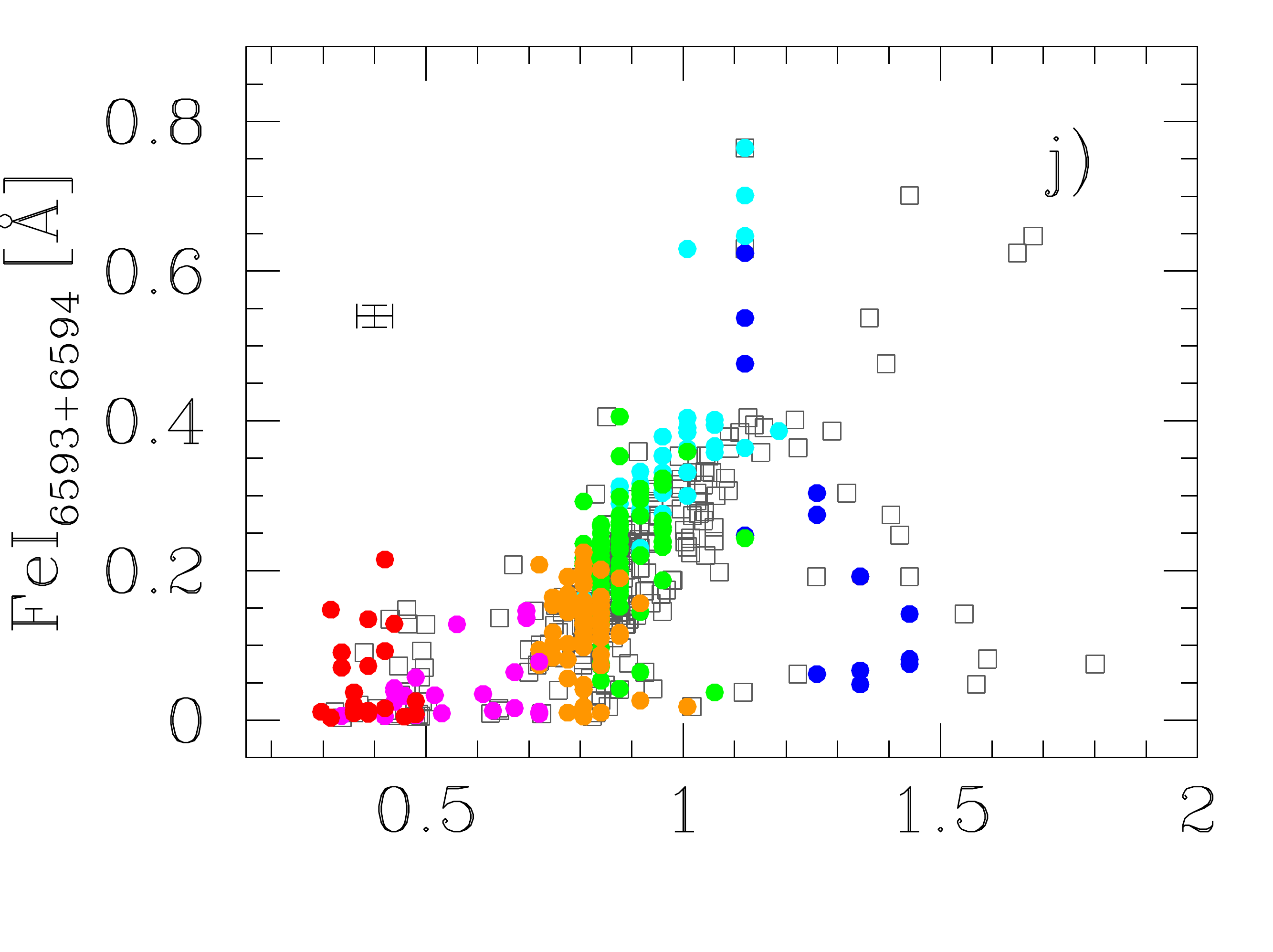}
\includegraphics[width=0.24\textwidth,angle=-90]{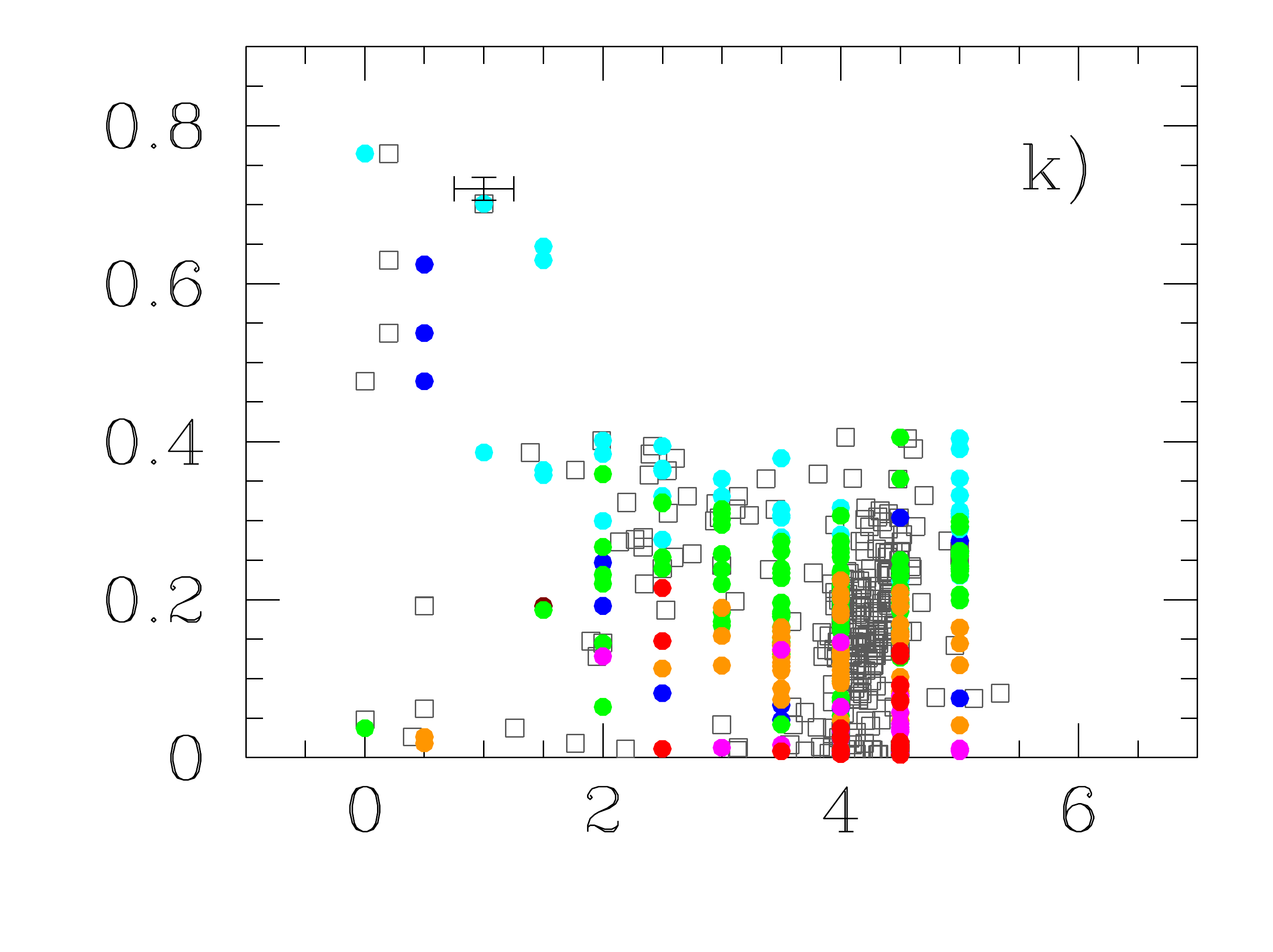}
\includegraphics[width=0.24\textwidth,angle=-90]{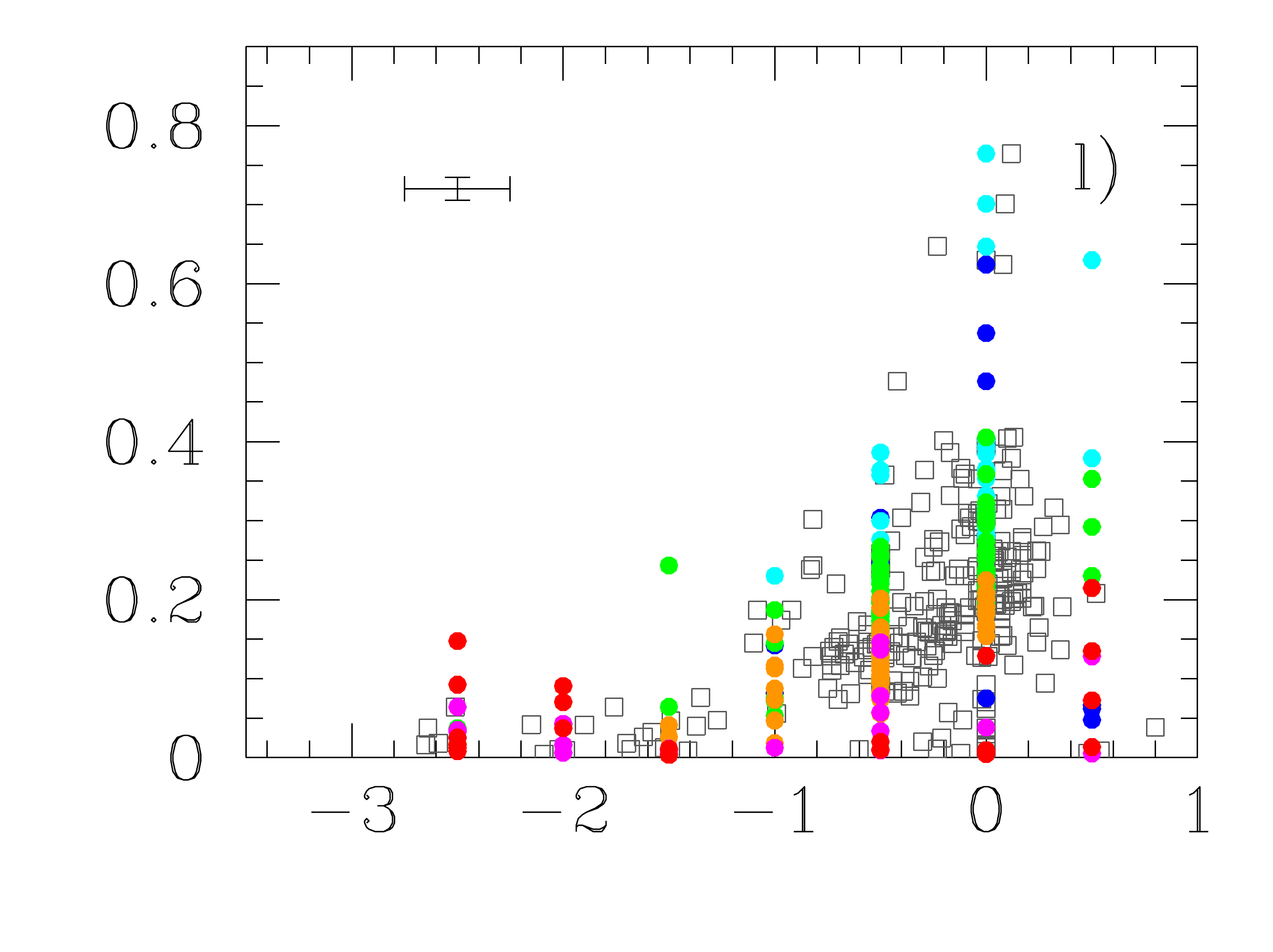}
\includegraphics[width=0.24\textwidth,angle=-90]{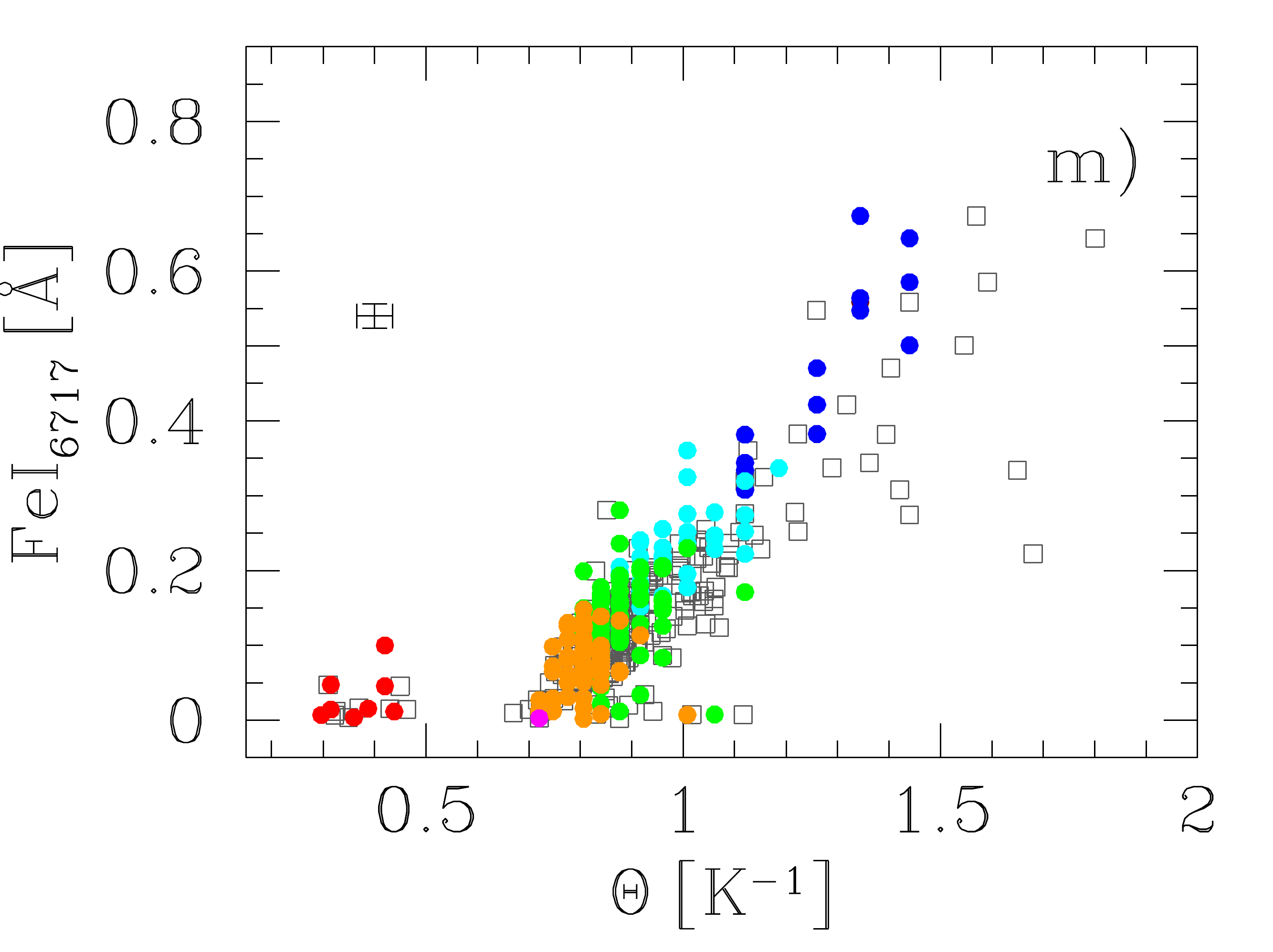}
\includegraphics[width=0.24\textwidth,angle=-90]{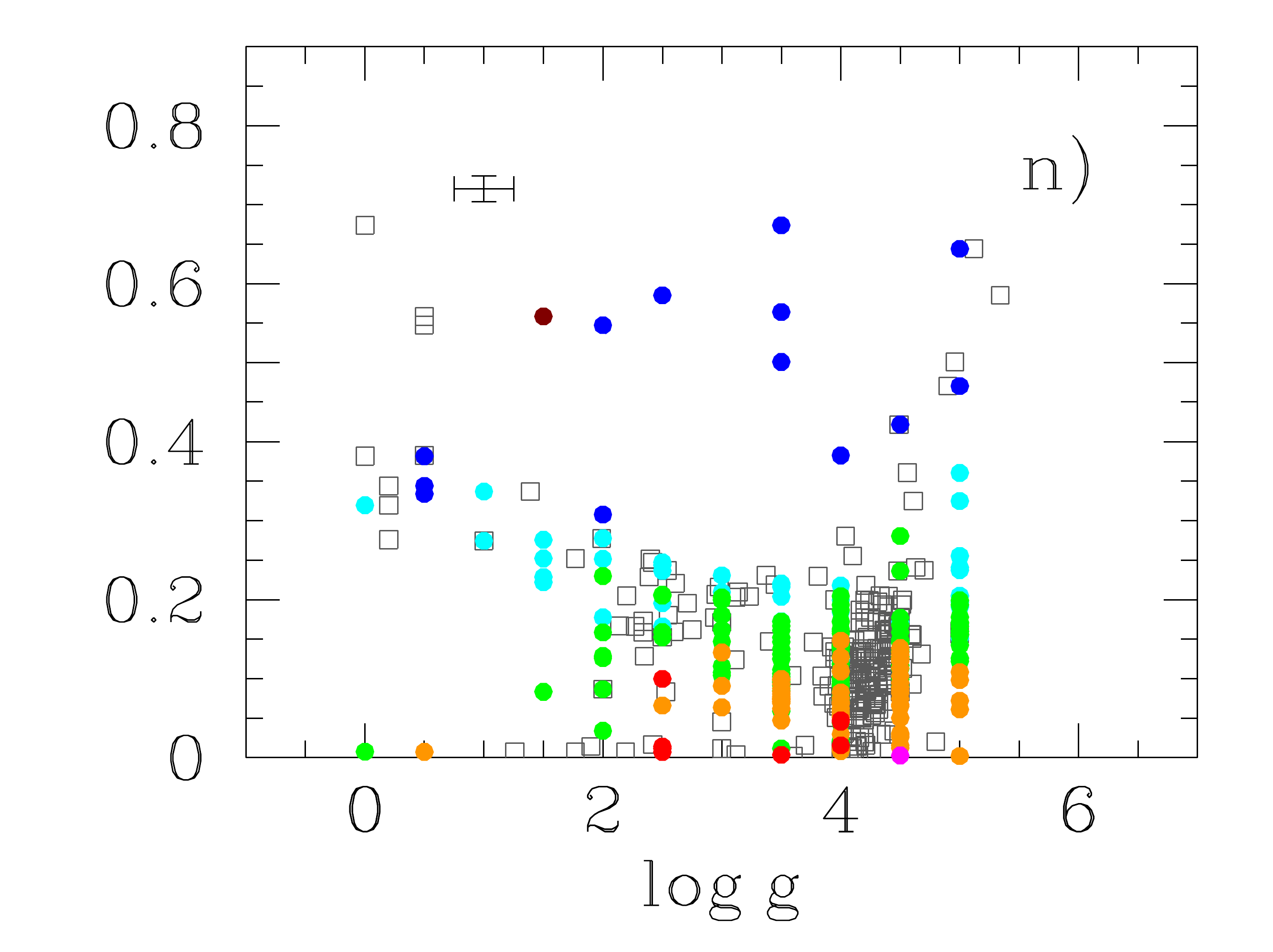}
\includegraphics[width=0.24\textwidth,angle=-90]{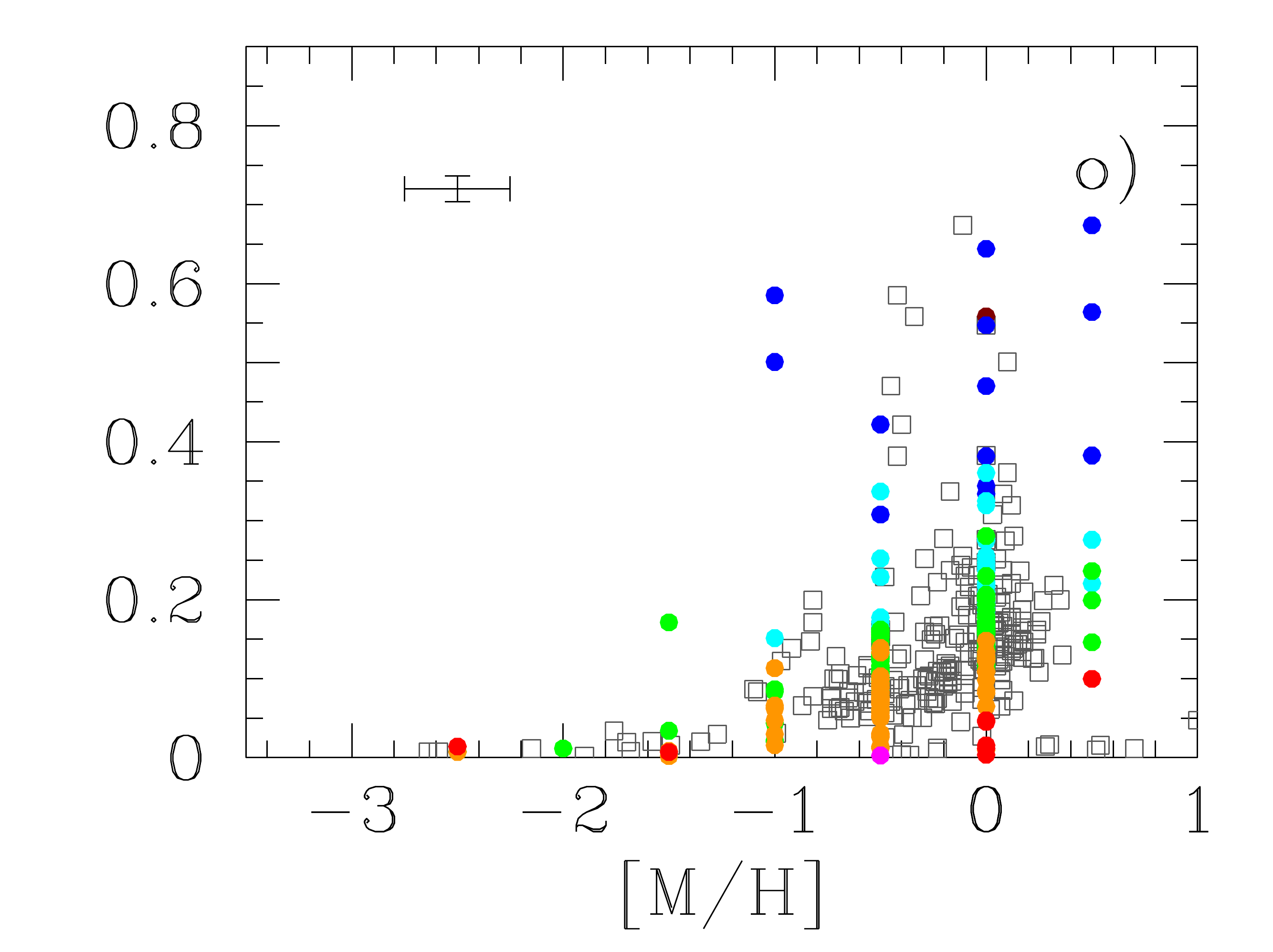}
\caption{Equivalent widths of spectral lines in the HR-R setup as a function of the stellar parameters obtained in this work,TW, shown as circles, and from the literature, shown as open squares. Left) Inverse of the $T_\mathrm{eff}$ (represented as $\theta$ = 5040/$T_\mathrm{eff}$); center) $\log{g}$; and right) [M/H]. We show from top to bottom: \ion{Ca}{i}\,6439 \AA; \ion{Fe}{i}\,6463 \AA, \ion{Fe}{i}\,6495 \AA, \ion{Fe}{i}\,6593+6594  \AA, and \ion{Fe}{i}\,6717 \AA. The red, pink, orange, green, cyan, blue, and brown circles correspond to B, A, F, G, K, M and S spectral types, respectively. The typical error bars of each measurement are given in each panel.}
\label{indices_plot1}
\end{figure*}

\begin{figure*}
\centering
\includegraphics[width=0.24\textwidth,angle=-90]{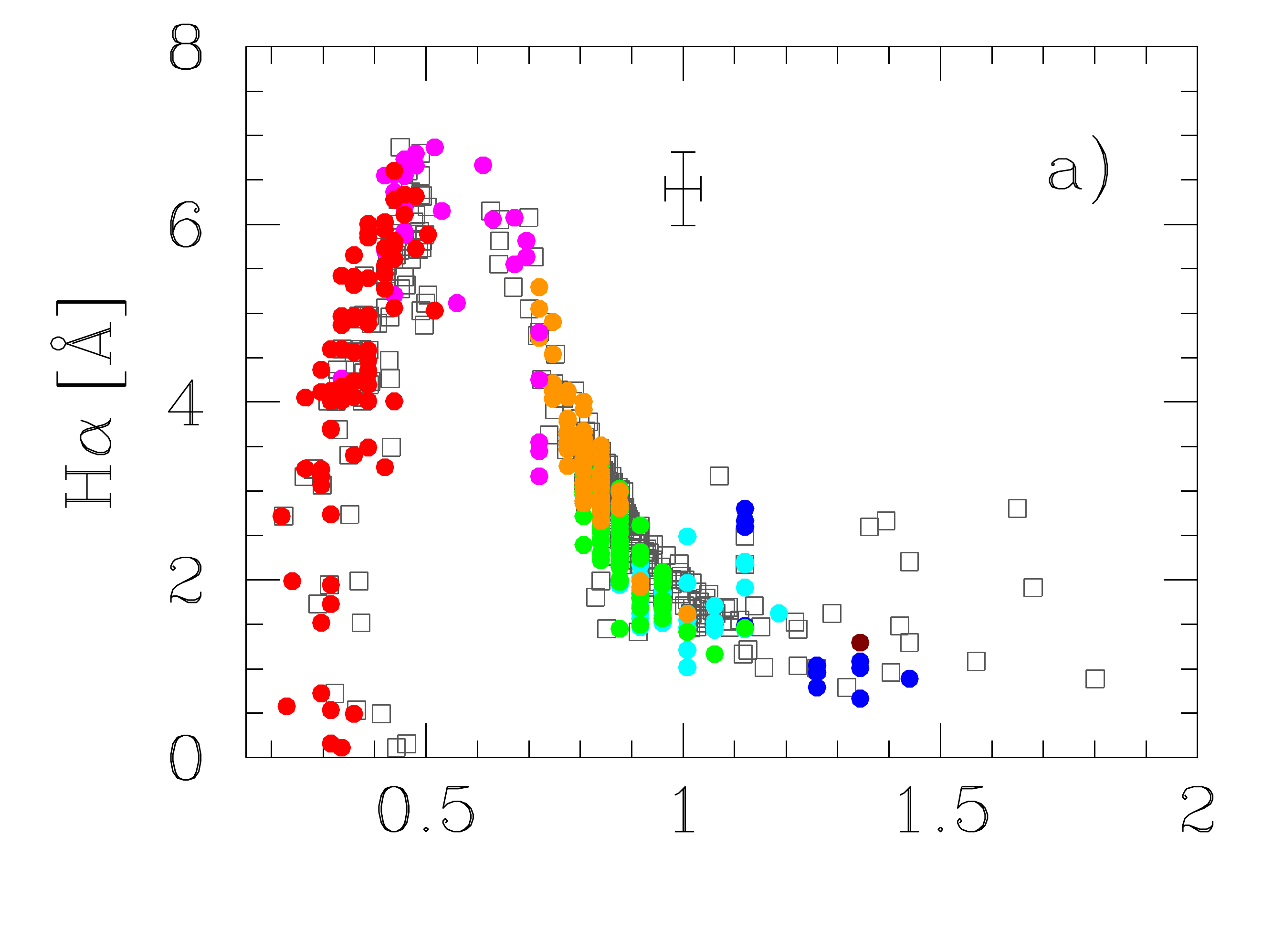}
\includegraphics[width=0.24\textwidth,angle=-90]{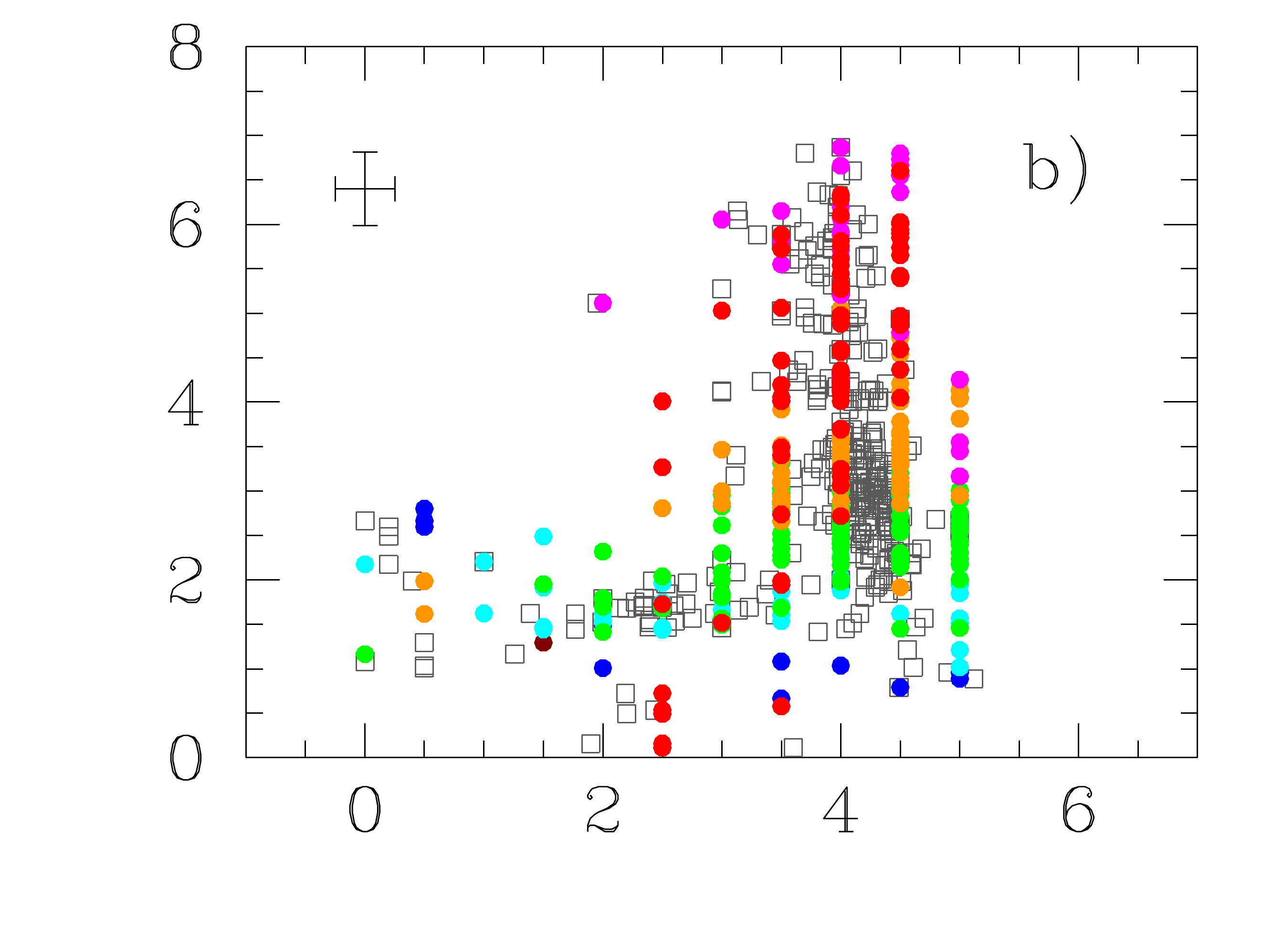}
\includegraphics[width=0.24\textwidth,angle=-90]{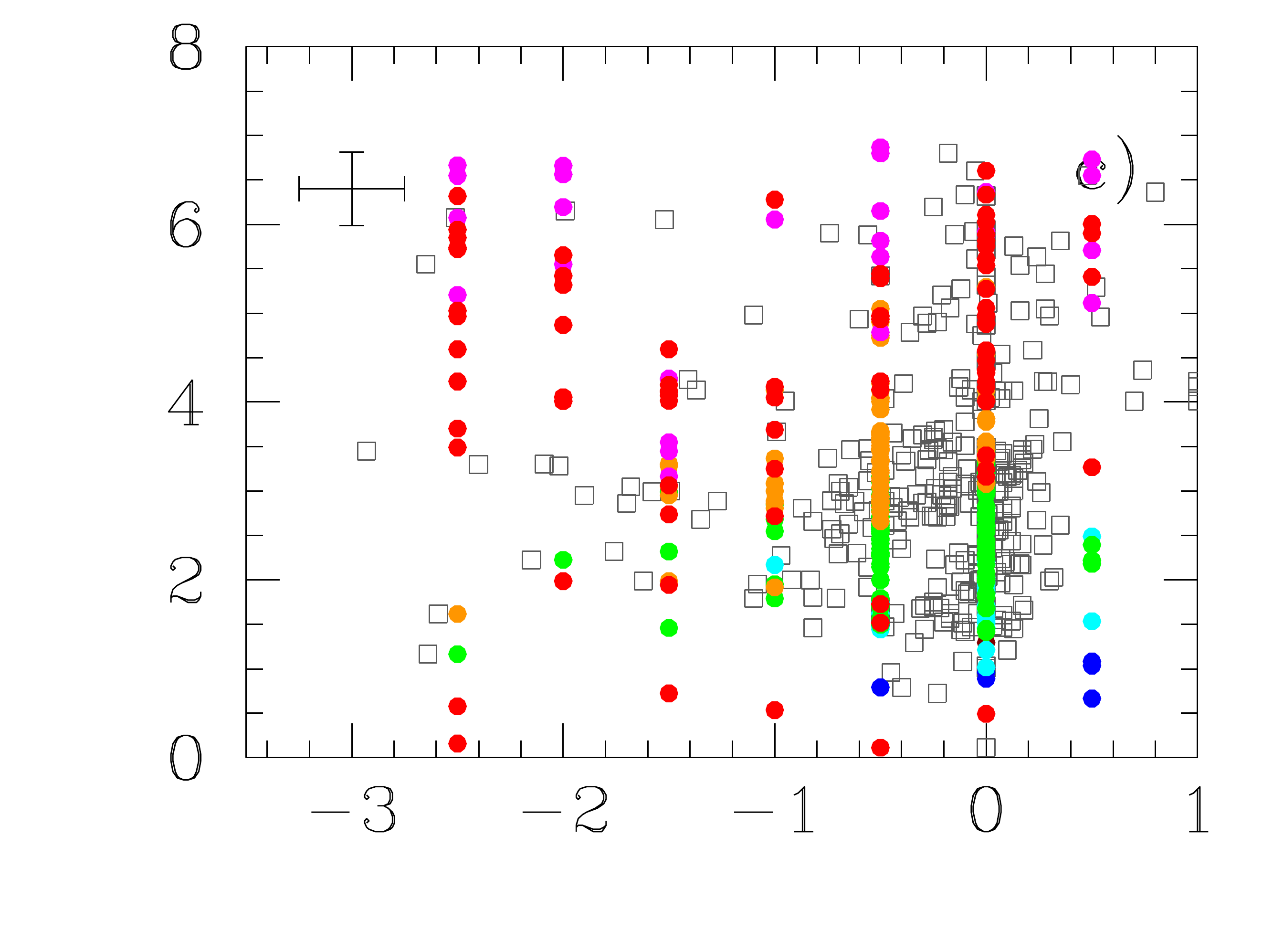}
\includegraphics[width=0.24\textwidth,angle=-90]{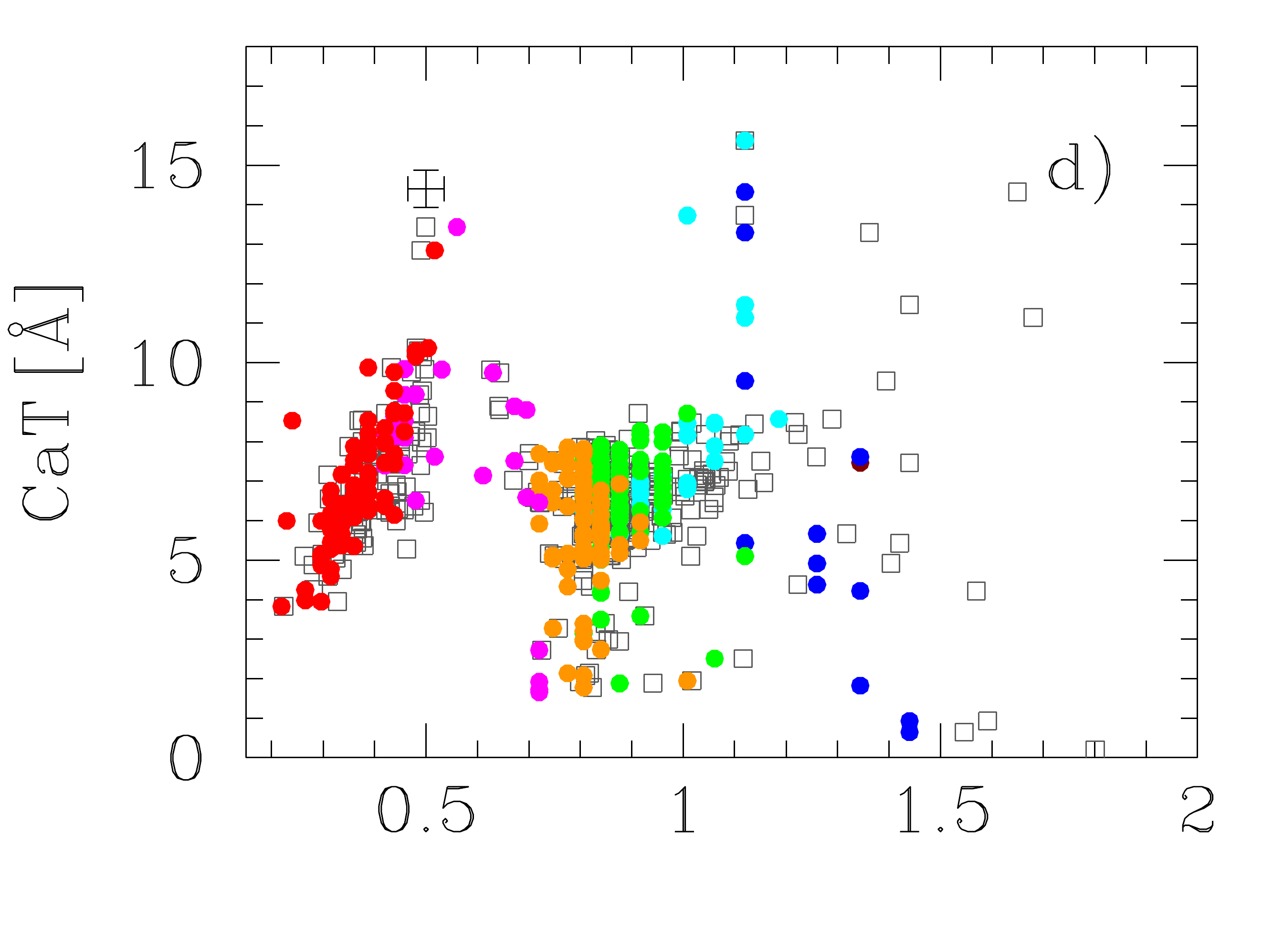}
\includegraphics[width=0.24\textwidth,angle=-90]{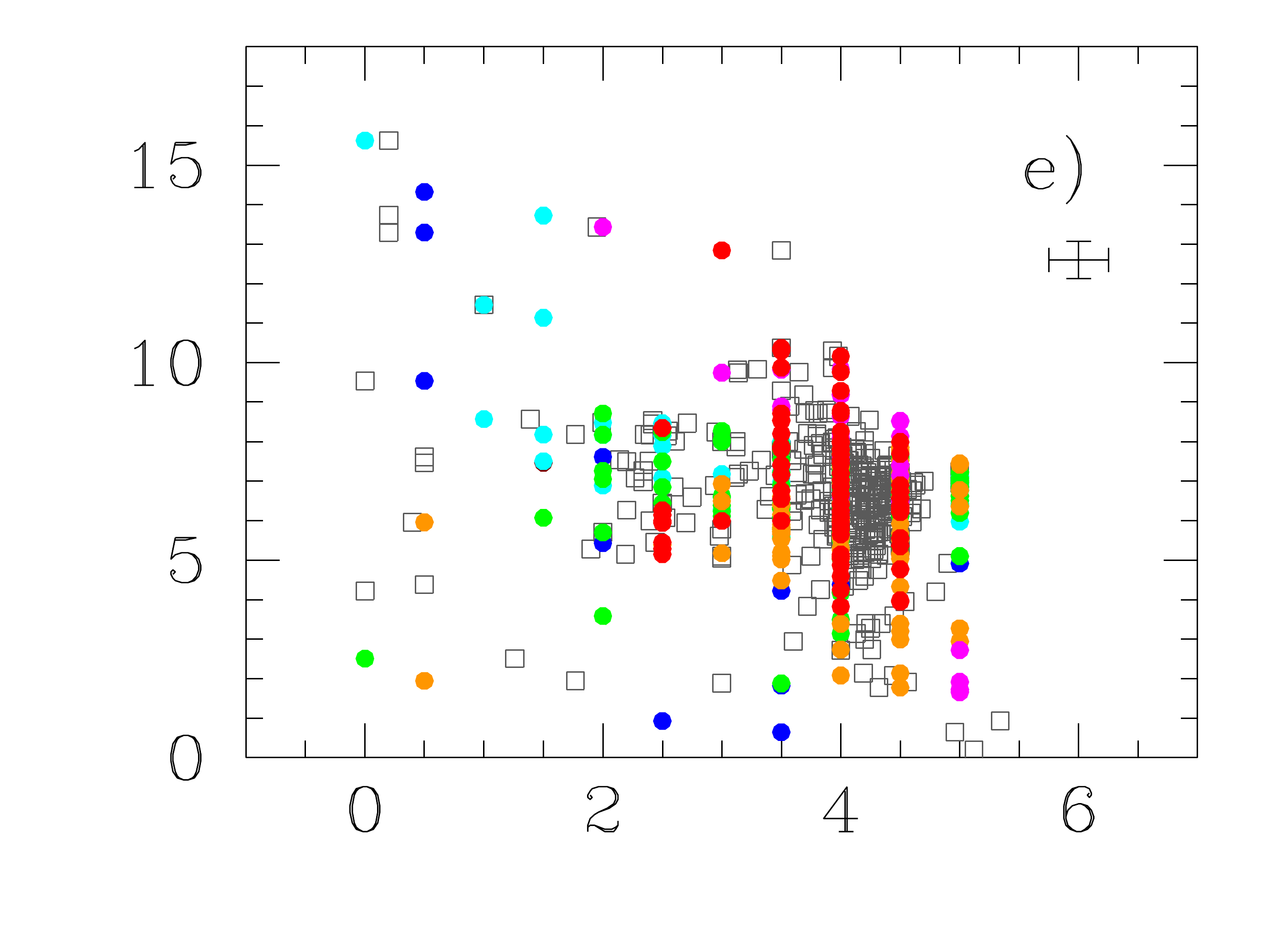}
\includegraphics[width=0.24\textwidth,angle=-90]{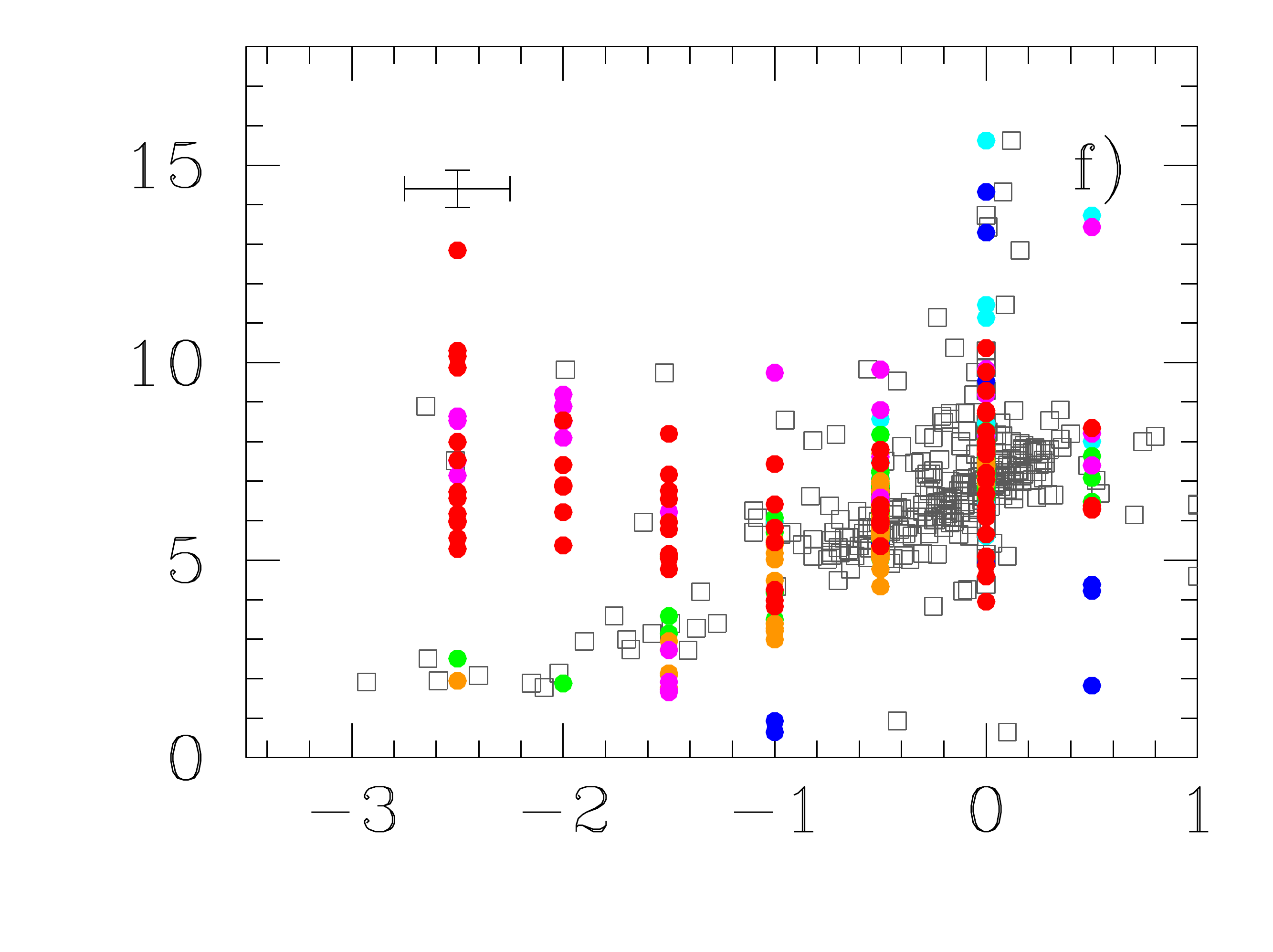}
\includegraphics[width=0.24\textwidth,angle=-90]{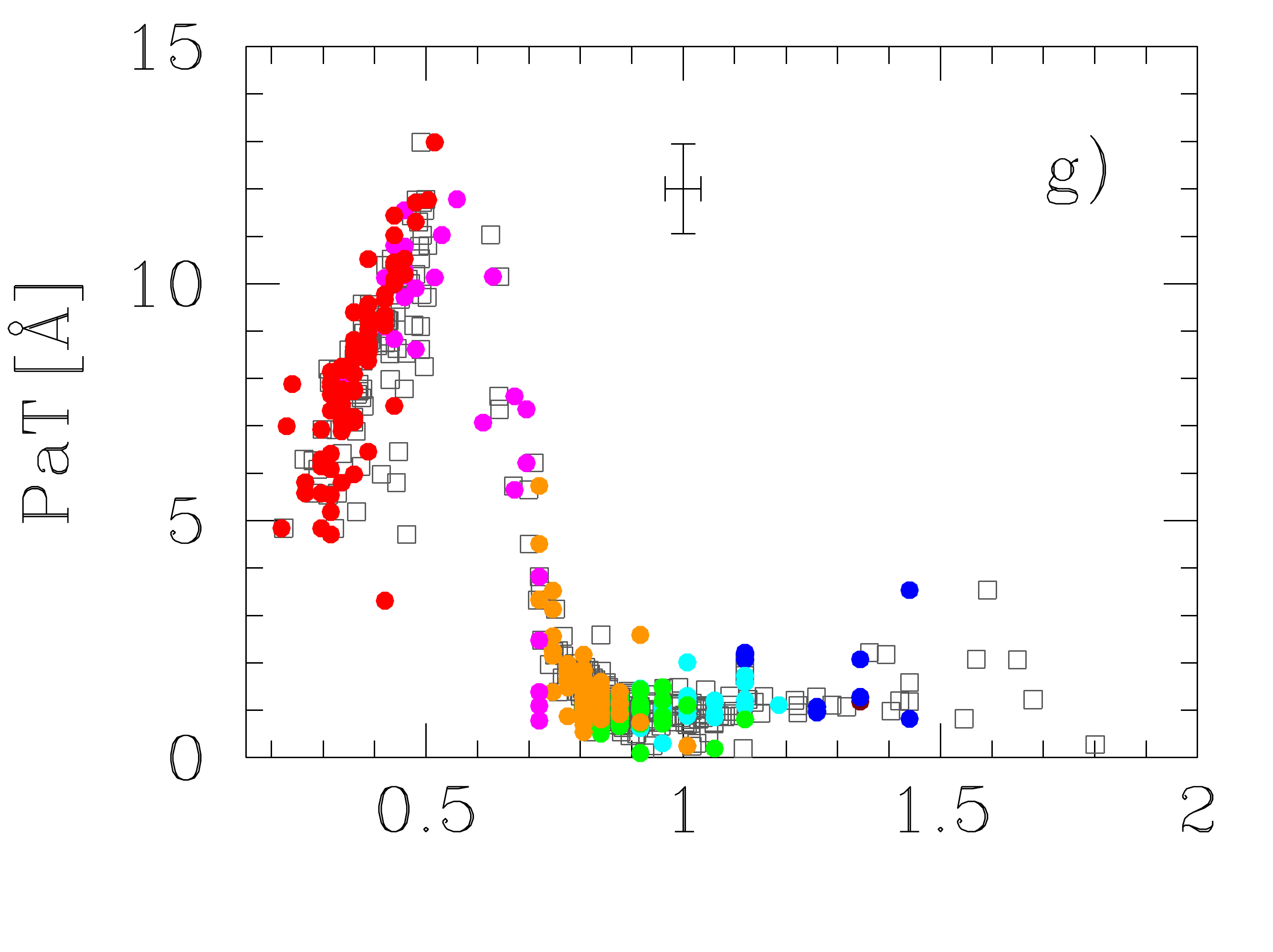}
\includegraphics[width=0.24\textwidth,angle=-90]{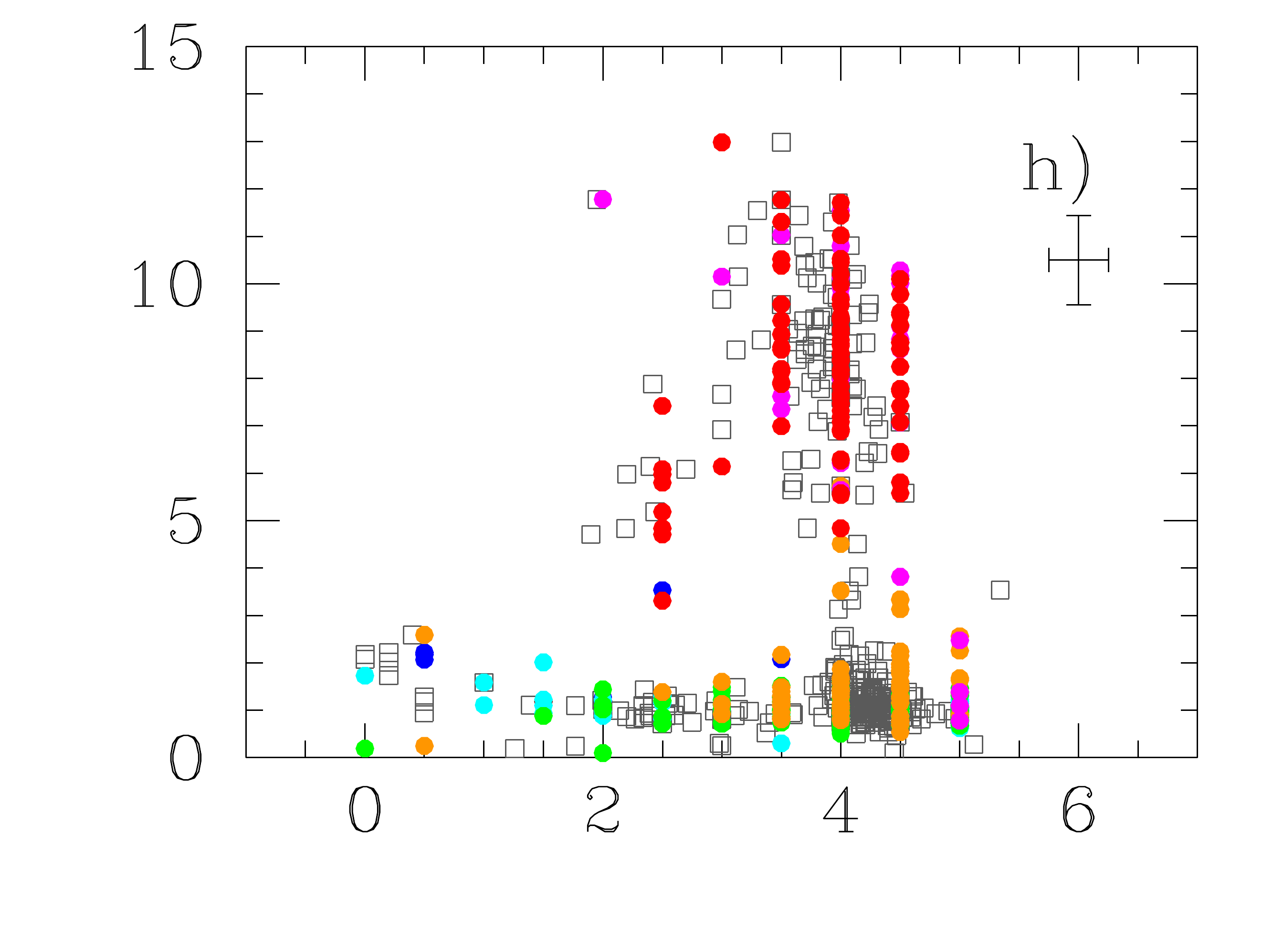}
\includegraphics[width=0.24\textwidth,angle=-90]{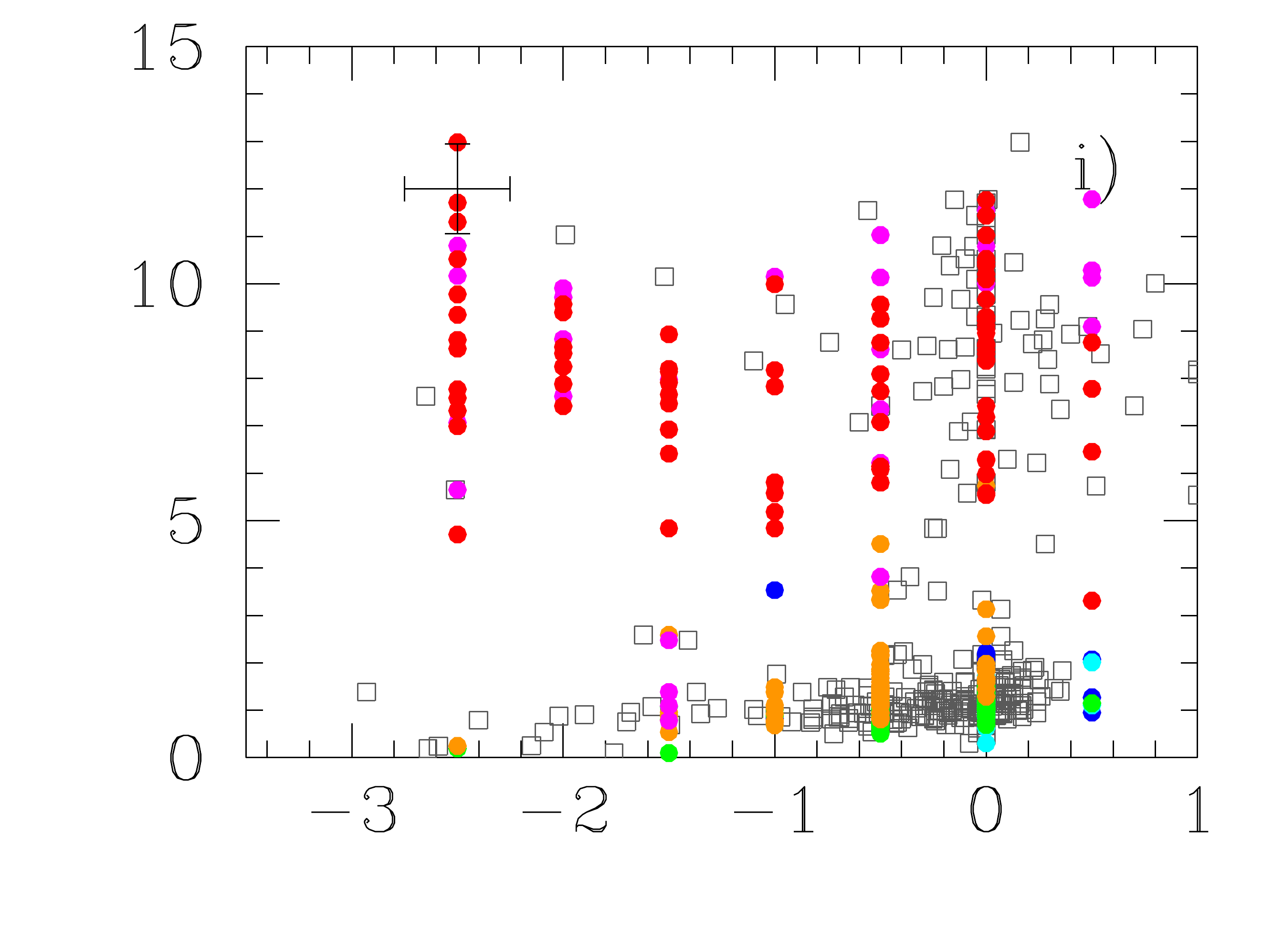}
\includegraphics[width=0.24\textwidth,angle=-90]{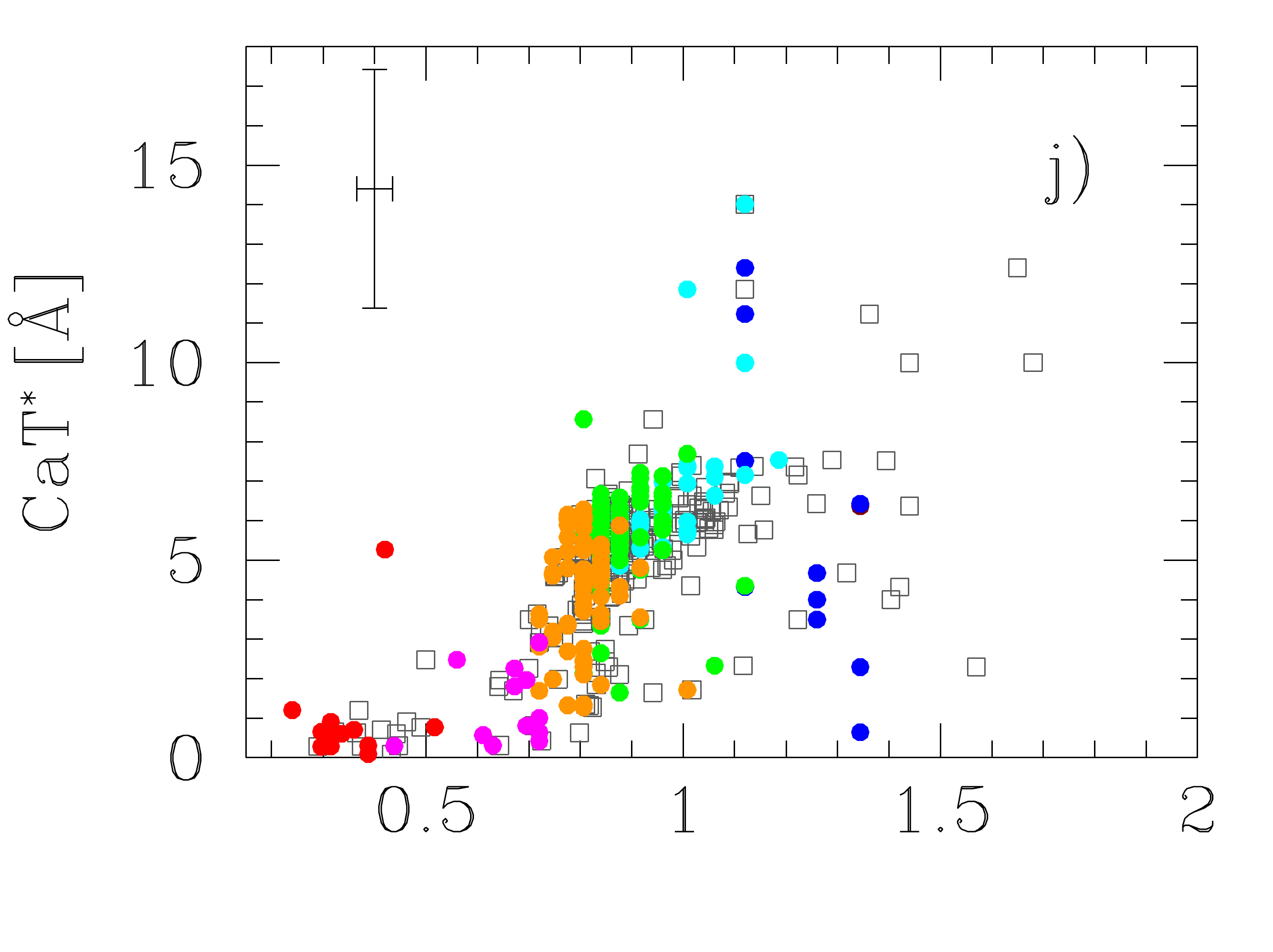}
\includegraphics[width=0.24\textwidth,angle=-90]{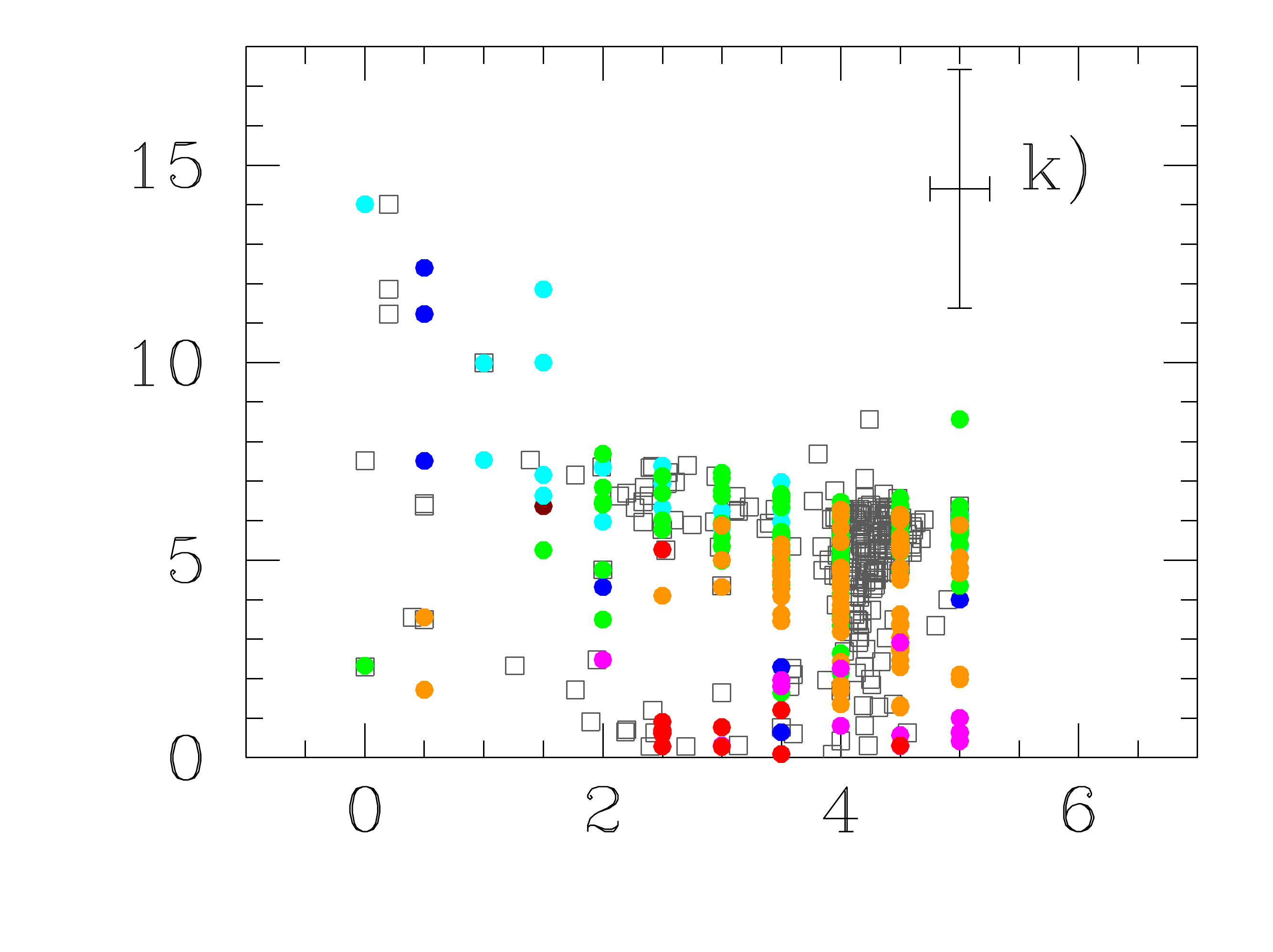}
\includegraphics[width=0.24\textwidth,angle=-90]{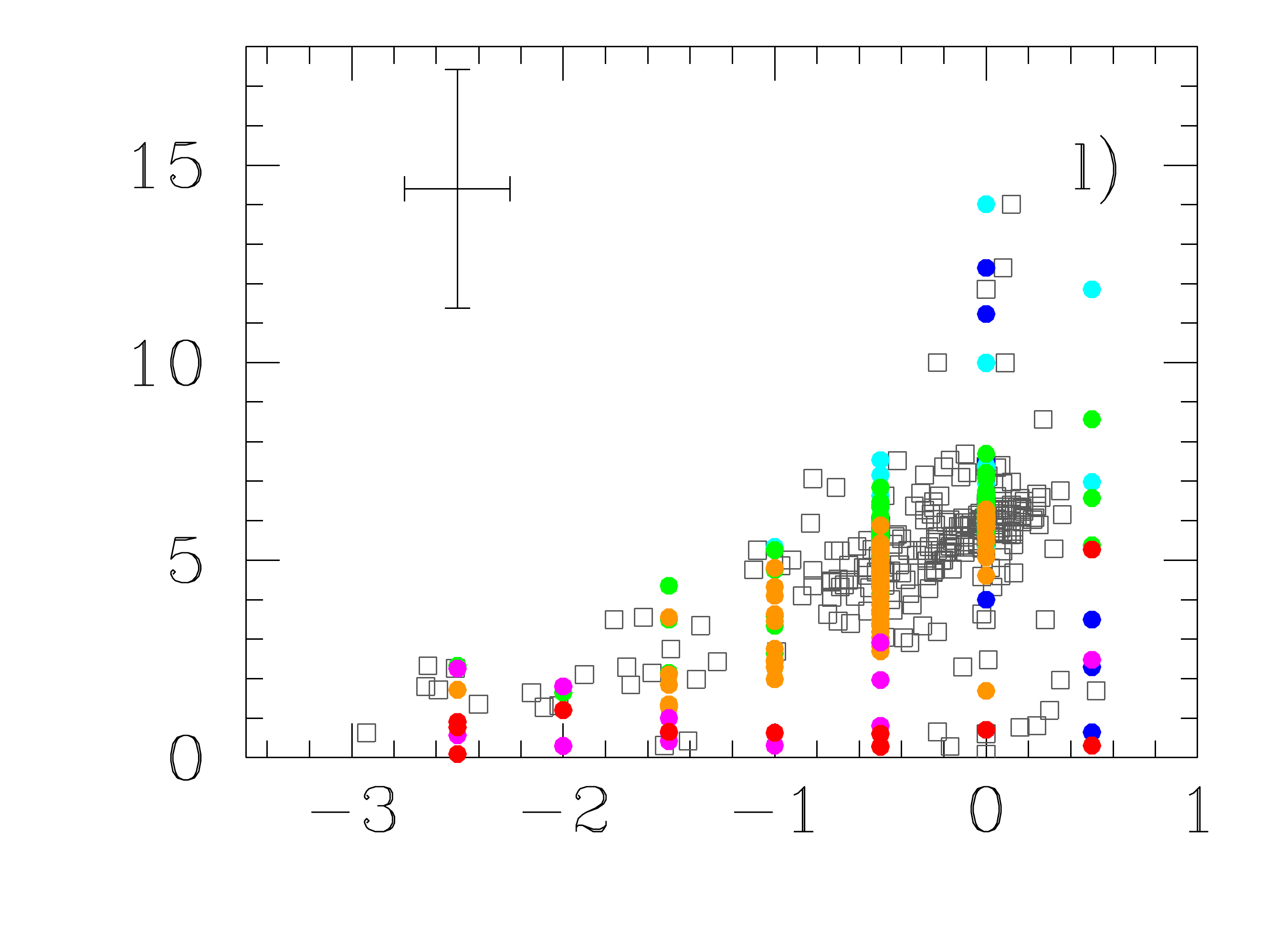}
\includegraphics[width=0.24\textwidth,angle=-90]{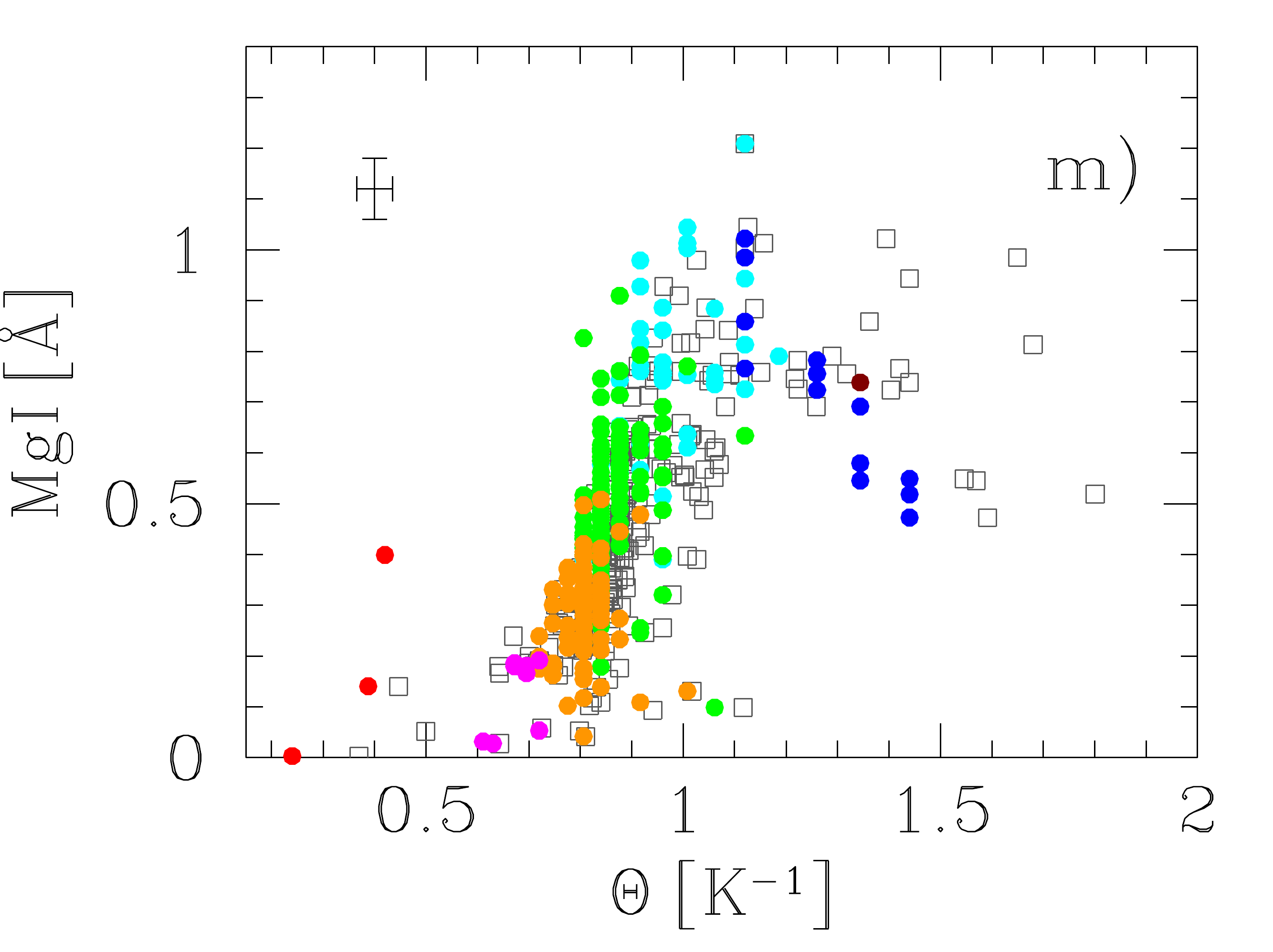}
\includegraphics[width=0.24\textwidth,angle=-90]{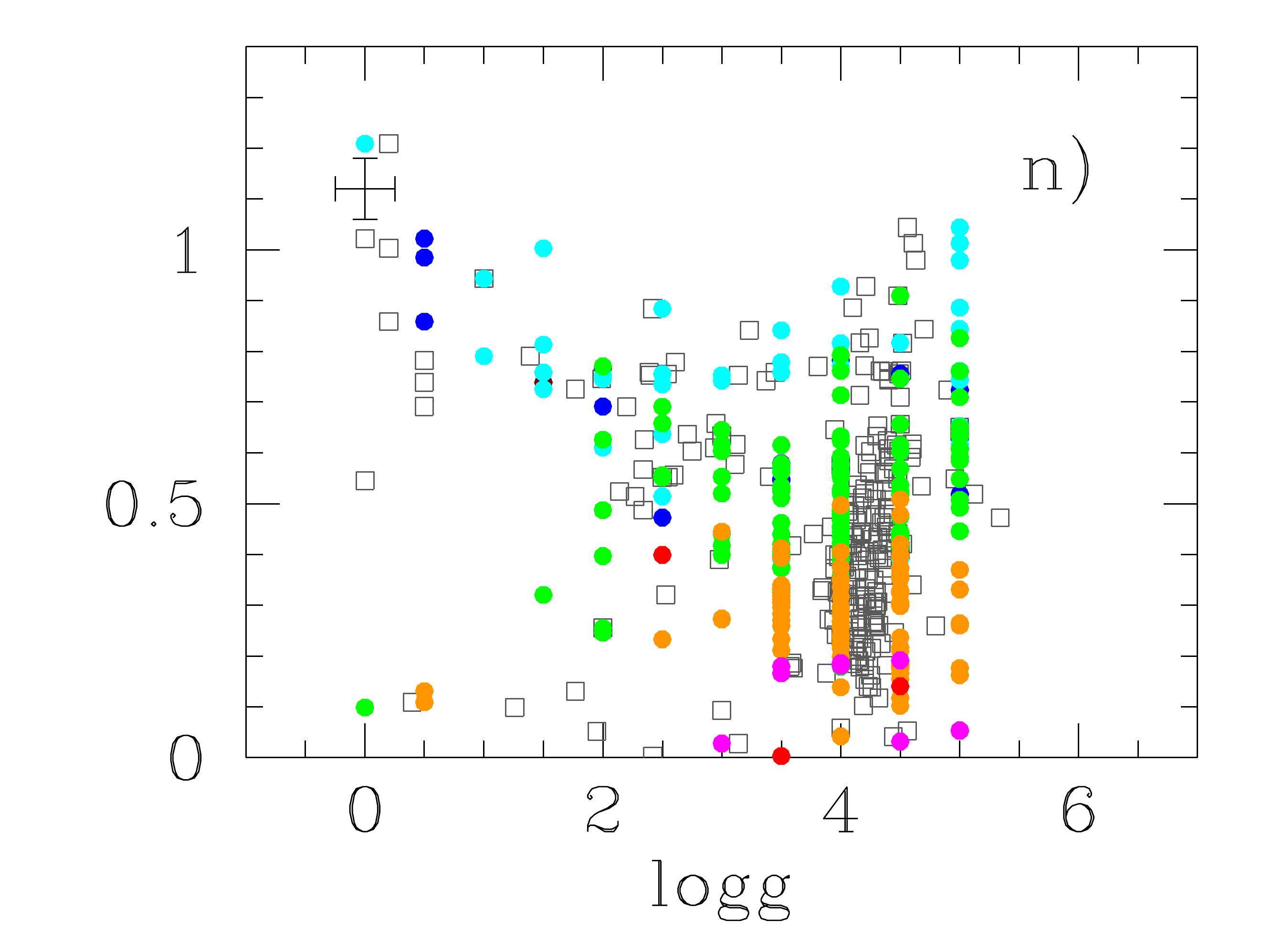}
\includegraphics[width=0.24\textwidth,angle=-90]{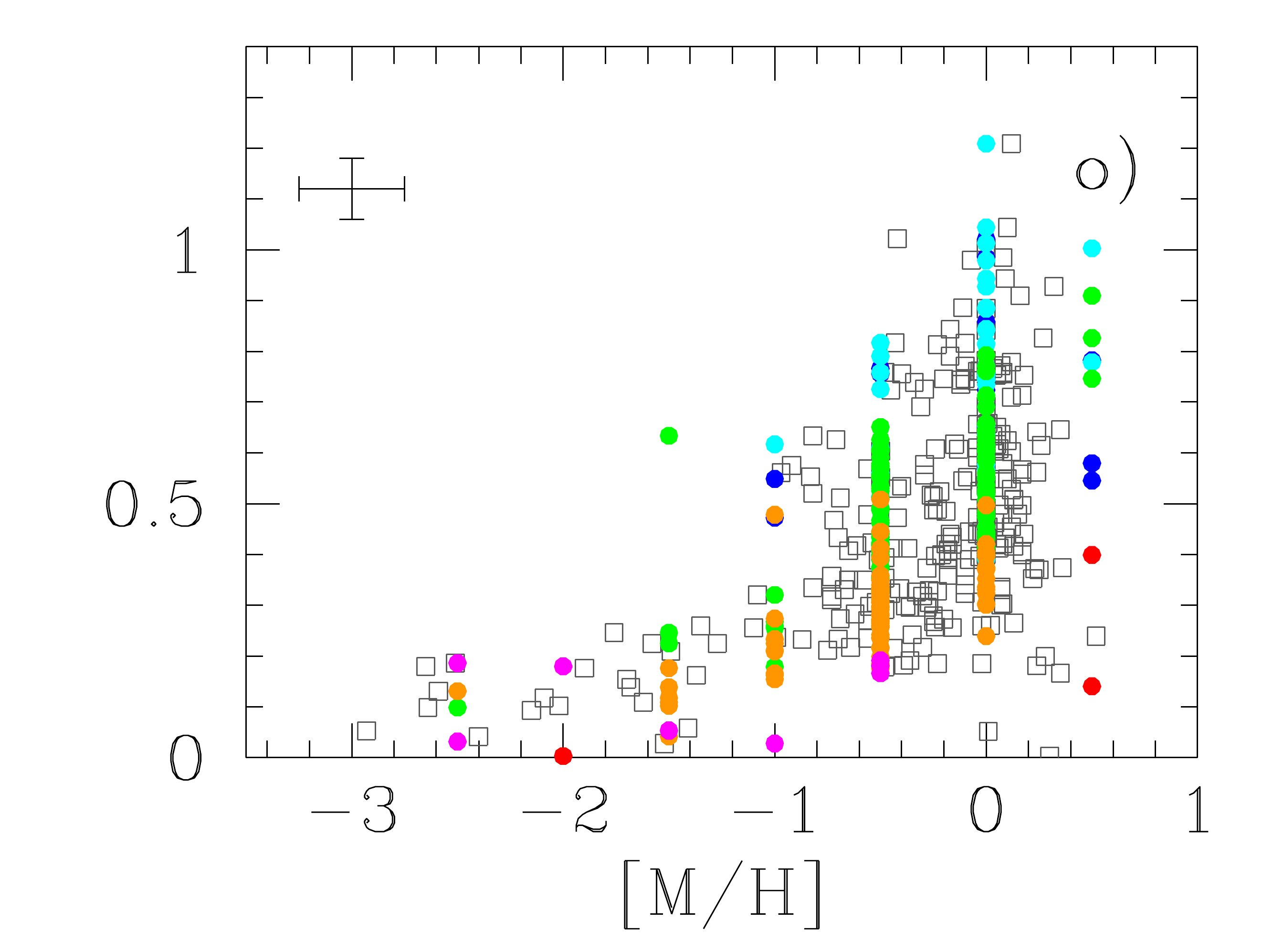}
\caption{Stellar indices as equivalent widths in \AA, as a function of the physical stellar parameters obtained in this work as dots, and from the literature as open squares. For H$\alpha$  as defined by GON05,  in HR-R. For CaT and  Pa lines  and  CaT$^{*}$, as defined in CEN01, and \ion{Mg}{i} as defined in CEN09 in HR-I. As a function of: Left) Inverse of the $T_\mathrm{eff}$ (represented as $\theta$ = 5040/$T_\mathrm{eff}$); center) $\log{g}$; and right) [M/H]. The red, pink, orange, green, cyan, blue and brown circles correspond to B, A, F, G, K, M and S spectral types, respectively. The typical error bars are given in each panel.}
\label{indices_plot2}
\end{figure*}
\normalsize

As shown in Paper~I, for hot stars, meaning in this paper those stars with spectral types from B3 to A0, the MEGARA HR-I spectral range shows lines coming from hydrogen (Paschen) and helium series (for the hottest B stars) as its strongest features. The Paschen series lines are clear and strong along the whole sequence, with maximum strength and width for giant B stars. The \ion{He}{i} lines are also identified (see Table~\ref{Table:lines}). For stars cooler than A, the CaT lines are the most prominent ones; followed in intensity by \ion{Mg}{i}. The lines from  \ion{Fe}{i} are easily identified in the spectra. The strongest \ion{Th}{i} \ion{Ti}{i} and \ion{Na}{i} lines can be also detected. 

Fig.~\ref{indices_plot1} shows the EW for spectral lines in the \mbox{HR-R} set-up as a function of  the stellar parameters: left) inverse of $T_\mathrm{eff}$ (represented as $\theta$ = 5040/T$_\mathrm{eff}$); centre) $\log{g}$; and right) [M/H]. We show from top to bottom: \ion{Ca}{i}\,6439\,\AA; \ion{Fe}{i}\,6463\,\AA, \ion{Fe}{i}\,6495\, \AA, \ion{Fe}{i}\,6593+6594\,\AA, and \ion{Fe}{i}\,6717\,\AA. In the left column plots, panels a), d), g) and j), we do not find any correlation with $T_{\mathrm{eff}}$ or the hottest stars (spectral types late-B and A). In contrast, for F, G and K stars, the EW in all the five lines studied decreases with an increase of $T_{\mathrm{eff}}$, a decrease of $\theta$). In the case of M stars, we only find a clear anti-correlation between the EW  of the  \ion{Fe}{i}\,6717\,\AA\ line and T$_\mathrm{eff}$, and, however, we see an increase of the EW of the lines \ion{Fe}{i}\,6495\,\AA\ and \ion{Fe}{i}\,6593+6594\,\AA, when $T_{\mathrm{eff}}$ increases, that is not seen in the other lines. We will confirm these results in future releases as we expect to have a larger number of M-type observed stars. In the central column plots, we observe a quite flat behaviour of the EW  with the $\log{g}$ for values of this parameter between 1 and 5, showing sequences depending on spectral types. Finally, from the plots in the last column, we see a general correlation of the EW with the metallicity, at least up to the solar value, with a possible anticorrelation for super-solar abundances, except for the \ion{Fe}{i}\,6717\,\AA\ in M stars, for which we do not observe a clear behaviour.

Similarly to Fig.~\ref{indices_plot1}, Fig.~\ref{indices_plot2} shows the behaviour of classical indices in the \mbox{HR-I} range plus H$\alpha$, for comparison purposes, as a function of the physical stellar parameters. From top to bottom we show: H$\alpha$, as defined by GON05, the sum of the EW of the three lines \ion{Ca}{ii} lines, CaT, and Pa lines, PaT, CaT$^{*}$, corrected from PaT contamination, as defined in CEN01 as CaT$^{*}= \rm{CaT} - 0.93 \rm{PaT}$, and \ion{Mg}{i} as per CEN09.  Regarding the index behaviour with $T_\mathrm{eff}$, the left panels show the \ion{H}{i} lines indices in panels a) for H$\alpha$ and panel g) for PaT. Both indices present an anti-correlation with $T_\mathrm{eff}$, reaching the maximum values around 10 000\,K, as expected, while both indices correlate with $T_\mathrm{eff}$, from this temperature, for cooler stars. This last correlation of hydrogen indices with $T_{\mathrm{eff}}$ is also observed for M-type stars (blue symbols) when using the stellar parameters derived in this paper. 

Regarding the behaviour of metal indices with $T_\mathrm{eff}$, the apparent anti-correlation of CaT (panel d) for stars hotter than 10 000\,K, corresponds to the hydrogen Pa lines dependence with $T_{\mathrm{eff}}$(panel g) due to the high contamination of CaT index with Pa lines. This trend disappears when using CaT$^{*}$ (panel j), an index corrected from Pa contamination. There are clear anti-correlations of indices CaT$^{*}$ and \ion{Mg}{i} with $T_{\mathrm{eff}}$ for stars with spectral types from A to G. In contrast, these indices correlate with $T_\mathrm{eff}$ in M stars.

Regarding gravity (central panels), we confirm the anti-correlation of CaT$^{*}$ with $\log{g}$ in stars of spectral type F and cooler, finding very large values for supergiants and decreasing after, following the sequence super-giants --giants-- main sequence ($\log{g}$ from 0 to 5). For hotter stars, we cannot obtain conclusions as the CaT$^{*}$ index depends first on $T_\mathrm{eff}$, and then on $\log{g}$. A bi-modal trend is seen in \ion{Mg}{i} for all spectral types but a dedicated study discriminating by temperature and metallicity would be needed to study this relationship of \ion{Mg}{i} with $\log{g}$. 
Finally, there is a clear correlation of the metallic indices CaT$^{*}$ and \ion{Mg I} with metal abundance, increasing their values as the metallicity increases. Of course, as [M/H] does not trace the hydrogen, we do not see any correlation in panels c, f (CaT traces Paschen lines in hot stars due to the strong contamination as mentioned before) and i. 
For the near-infrared indices, we have measured, besides the CEN01 indices with large window bands, some new indices, fitted to the MEGARA high spectral resolution, (results are also given in Table~\ref{table:indices}) which could be used for stellar parameters characterisation. MEGASTAR is a work in progress and the relationships between the indices and the stellar parameters will be revisited in the future, when we will have a larger number of observed stars and a wider range in spectral type and luminosity class. The  analysis of the whole set of  the above suggested indices is  beyond the scope of this work.

\bsp	

\label{lastpage}
\end{document}